\documentclass[preprint,aps,amsmath,nofootinbib]{revtex4-1}
\usepackage{graphicx, array,dcolumn}
\usepackage{calc,tabularx, epsfig,mathrsfs}
\usepackage{hyperref}

\usepackage{amsmath, verbatim,enumerate}
\usepackage{amssymb}
\usepackage{wasysym}
\usepackage{tikz}
\usepackage{xcolor}
\usepackage{subfigure}
\usepackage{multirow}
\usepackage{xspace}
\usepackage{mathrsfs}
\usepackage{slashed}
\usepackage{mathtools}
\usepackage{cancel}
\allowdisplaybreaks[1]
\newlength{\figurewidth}
\newcommand{\beq}{\begin{equation}}
\newcommand{\eeq}{\end{equation}}
\newcommand{\bea}{\begin{eqnarray}}
\newcommand{\eea}{\end{eqnarray}}
\newcommand{\ba}{\begin{array}}
\newcommand{\ea}{\end{array}}

\newcommand{\mn}{{\mu\nu}}

\newcommand{\pt}{\partial}

\newcommand{\Bl}{\biggl}
\newcommand{\Br}{\biggr}
\newcommand{\bl}{\bigl}
\newcommand{\br}{\bigr}

\newcommand{\ii}{\mathrm{i}}
\newcommand{\dd}{\mathrm{d}}
\newcommand{\g}{\gamma}
\newcommand{\ep}{\epsilon}

\newcommand{\lam}{\lambda}
\newcommand{\Lam}{\Lambda}

\newcommand{\nb}{\nabla}
\newcommand{\de}{\delta}
\newcommand{\D}{\Delta}

\newcommand{\om}{\omega}
\newcommand{\sg}{\sigma}

\newcommand{\eqnsizesmall}{
\fontsize{13}{12}\selectfont
}

\makeatother
\begin{document}
%
\title{
IR behaviour of one-loop complex $\mathbb{R}\times S^3$ saddles
}
\setlength{\figurewidth}{\columnwidth}
%

\author{Shubhashis Mallik} 
\email{shubhashism@iisc.ac.in}

\author{Gaurav Narain}
\email{gnarain@iisc.ac.in}

\affiliation{
Center for High Energy Physics, Indian Institute of Science,
C V Raman Road, Bangalore 560012, India.
}

\vspace{50mm}

\begin{abstract}
Gravitational path-integral over $\mathbb{R}\times S^3$ complex metrics with fluctuations is studied in 4D for Einstein-Hilbert gravity in Lorentzian signature, with the aim to investigate the IR behaviour of complex saddles for various boundary choices. General covariance doesn't allow arbitrary boundary choices for the background and fluctuations. In the ADM-decomposition, while imposing ``no-boundary'' condition at the initial boundary, two scenarios are considered for the final boundary: Dirichlet and fixed extrinsic curvature. The universe undergoes a transition from a Euclidean to a Lorentzian phase in either scenario, where the dominant saddle in Euclidean phase corresponds to a Euclidean metric, while the Lorentzian phase has two dominant complex saddles that superimpose. One-loop lapse action is computed using Hurwitz-Zeta regularization. UV-divergences canceled by suitable counterterms lead to a renormalized lapse action. One-loop Hartle-Hawking wave-function is computed using the Picard-Lefschetz (PL) and WKB methods, where the contributions from the metric fluctuations show secularly growing infrared divergences as the Universe expands. This is compared with the situation in Lorentzian dS, corresponding to a Universe transitioning from an initial state of vanishing conjugate momenta to a final state of fixed extrinsic curvature. The PL method alone is not sufficient to overcome the technical hurdles in the one-loop computation, which needs to be supplemented by an $i\ep$-prescription, achieved via slight complexification of the cosmological constant $\Lam$. The renormalized one-loop dS wavefunction is seen to have qualitatively the same leading IR behaviour as the no-boundary wavefunction, though the coefficients are seen to be sensitive to the UV boundary choices. Interestingly, for all boundary choices considered, the saddles remain KSW-allowed.

\end{abstract}

\maketitle

\tableofcontents



\section{Introduction}
The gravitational path integral is an important tool in understanding various aspects of quantum gravity. Over the years, the path integral framework has been applied in numerous studies involving black holes and cosmology, yielding physically sensible results. As proposed by Gibbons and Hawking long ago, the semiclassical properties of black hole thermodynamics can be described elegantly within the path integral framework \cite{Gibbons:1976ue}. In particular, loop corrections and contributions from non-trivial topologies can be easily captured using the gravitational path integral. In the context of cosmology with a positive cosmological constant, the object of great importance is the wave function of a closed Universe ($\Psi_{\rm H-H}$), as advocated by Hartle and Hawking in the seminal work \cite{Hartle:1983ai}. Such a state is defined by summing over all the regular and compact geometries. In this paper, we study the wave function to one-loop, restricting ourselves to summing over all possible complex geometries that respect $\mathbb{R}\times S^3$ in the Lorentzian path integral.

The (Lorentzian) gravitational path integral, on a manifold ${\cal M}$ with spacelike boundary $\pt{\cal M}$, is defined by
\beq
\label{eq:PI_LI_g}
\mathcal{Z}[{\rm Bd}_f,{\rm Bd}_i] = \int_{{\cal M} + \pt {\cal M}}{\cal D} g_\mn \, \exp\bl[i\mathsf{S}(g_\mn)/\hbar\br] \, ,
\eeq
where $\mathsf{S}[g_\mn]$ is the gravity action and ${\rm Bd}_i$ and ${\rm Bd}_f$ are the initial and final boundary conditions on $\pt {\cal M}$, respectively. We work in the Lorentzian signature to avoid the conformal factor problem in the Euclidean formulation \cite{Gibbons:1978ac}. The path integral in Eq. (\ref{eq:PI_LI_g}) includes all the metric-configurations respecting the boundary conditions having a fixed topology, which in our case is fixed to be $\mathbb{R}\times S^3$. The gravitational action, within the effective field theory, is given by
\bea
\label{eq:grav_act}
\mathsf{S}_{\rm grav}[g_\mn] = \frac{1}{16 \pi G} \int d^4x \sqrt{-g}\bigl[
-2\Lambda 
+ R \bigr]
+ \mathsf{S}_{\rm bd},
\eea
where $G$ is Newton's constant, $\Lam$ is the cosmological constant, and $\mathsf{S}_{\rm bd}$ is the appropriate boundary action, must be supplemented to ensure variational consistency. The gravitational path-integral stated in eq. (\ref{eq:PI_LI_g}) requires appropriate UV-renormalization, regularization and gauge-fixing \cite{tHooft:1974toh,Deser:1974nb,Goroff:1985th,Salam:1978fd,Narain:2011gs}. In the present work, we will renormalize by adding suitable counterterms, while regularizing via generalized zeta-function, which is a well-understood and often utilized procedure for computing one-loop functional determinants \cite{Monin:2016bwf, Elizalde:1994gf, Hawking:1976ja, Vassilevich:2003xt}. Gauge-fixing is implemented via the Faddeev-Popov procedure, leading to appropriate ghost contributions. In addition to these, two other key ingredients play significant roles in the gravitational path integral: the choice of boundary conditions and the contour of integration. Both of these must be carefully specified to ensure convergence of the Lorentzian path integral.

To define the wave function of the Universe $\Psi[{\rm Bd}_f]$, which depends only on the final boundary-configuration, the no-boundary (NB) condition is implemented at the initial-boundary (${\rm Bd}_i$) \cite{Hartle:1983ai}. The Hartle-Hawking (H-H) wave function is given by
\begin{equation}
\label{eq:wave_function}
\Psi_{\rm H-H}[{\rm Bd}_f]\equiv\mathcal{Z}[{\rm Bd}_f, {\rm NB}]=\int_{\rm NB}^{\rm Bd_f}{\cal D} g_\mn \, e^{i\mathsf{S}[g_\mn]/\hbar}\, ,
\end{equation}
where $\mathcal{Z}[{\rm Bd}_f, {\rm NB}]$ is the path-integral mentioned in eq. (\ref{eq:PI_LI_g}) with ${\rm Bd_i}$ to be a no-boundary (NB) (see \cite{Lehners:2023yrj, Maldacena:2024uhs} for a review on the no-boundary universe). 

\begin{figure}[hbtp]
    \centering
    \includegraphics[width=0.4\linewidth]{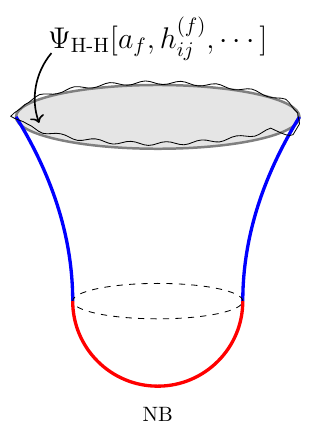}
    \caption{The H-H no boundary wave function of the universe ($\Psi_{\rm H-H}$), where the wiggle on the final hypersurface depicts small metric fluctuations. $a_f,h_{ij}^{(1)},\cdots$ are the parameters (${\rm Bd_f}$) on the final hypersurface where $\Psi_{\rm H-H}$ is defined.}
    \label{fig:no-boundary}
\end{figure}

Focusing on geometries which are spatially homogeneous and isotropic but have small metric fluctuations in the spatial part (see figure \ref{fig:no-boundary}),
such geometries in four spacetime dimensions are specified by the line element 
\beq
\label{eq:frwmet}
\begin{split}
&{\rm d}s^2 = - N_p^2(t_p) {\rm d} t_p^2 
+ 
\underbrace{
a^2(t_p) \left[
\rho_{ij}+h_{ij}+\frac{\varepsilon}{2}h_{ik}h^k{}_j+\cdots
\right]
}_{\g_{ij}(t_p,{\bf x})} {\rm d}x^i \, {\rm d}x^j\, ,\quad \varepsilon=\{0,1\}\\
&\text{where},\quad\varepsilon=0 \Rightarrow \text{Linear split},\quad \text{and} \quad \varepsilon=1 \Rightarrow \text{Exponential parametrization}
\end{split}
\eeq
where $\rho_{ij}$ corresponds to the metric on $S^3$. The non-diagonal terms in the metric fluctuation ($h_{0i}$) have been gauge-fixed to zero. The dots include higher-order terms in $h_{ij}$, which are relevant beyond the one-loop computation. In eq. (\ref{eq:frwmet}) we have introduced the parametrization $\varepsilon$ for fluctuations, which although can be arbitrary; 
we will consider it to be either $0$ or $1$. 
The former $\varepsilon=0$ is known as linear-split while the later $\varepsilon=1$ corresponds to exponential parametrization. In the later case, the series in eq. (\ref{eq:frwmet}) can be expressed as
\begin{equation}
\label{eq:exp_para}
\g_{ij}(t_p,{\bf x})\biggr|_{\varepsilon=1}=a^2\rho_{ik}(e^{h})^k_j.
\end{equation} 
While both parametrizations are used in the one-loop computations in gravity, the exponential parametrization is often found to be useful and is commonly used in the studies of expanding cosmology and EFTs in de-Sitter space \cite{Maldacena:2011nz, Maldacena:2002vr, Giddings:2010nc, Giddings:2011zd}. In this parametrization, the determinant of the fluctuation metric is preserved in the physical gauge (transverse-traceless graviton). The presence of $\varepsilon$ further helps us to investigate parametrization dependence in the wave function.

The geometries we consider in eq. (\ref{eq:frwmet}) contains lapse $N_p(t_p)$, background scale factor $a(t_p)$ and metric fluctuation $h_{ij}(t_p, x_m)$ on which the path integral needs to be performed. Imposing the proper-time gauge $N'=0$ condition \cite{Teitelboim:1981ua}, where the prime ($'$) denotes derivative w.r.t $t_p$, the path integral over $N_p$ reduces to the ordinary integration over $N_p$ with the ghost being constant (see, Batalin-Fradkin-Vilkovisky (BFV) formalism \cite{Batalin:1977pb, Feldbrugge:2017kzv, Teitelboim:1983fk,Teitelboim:1983fk,Teitelboim:1981ua}). Performing the path integral over the remaining variables eq. (\ref{eq:PI_LI_g}), we get,
\beq
\label{eq:Gform_sch}
\mathcal{Z}[{\rm Bd}_f, {\rm Bd}_i]
= \int_{\cal C} {\rm d}N_p
\int_{{\rm Bd}_i}^{{\rm Bd}_f} 
{\cal D} a {\cal D} h_{ij} {\cal D} \bar{c}_i {\cal D} c^i\,\,
e^{i (\mathsf{S}_{\rm grav}[a, N_p, h_{ij}] +  \mathsf{S}_{\rm gf} +  \mathsf{S}_{\rm gh})/\hbar} \, ,
\eeq
where ${\cal C}$ refers to the contour of integration, which we discuss below. The $\mathsf{S}_{\rm gf}$ and $\mathsf{S}_{\rm gh}$ are the gauge fixing and ghost action, which prevent overcounting of the gauge orbits. It accounts for the spatial diffeomorphisms of the $h_{ij}$ fluctuation. These can be done systematically using the Faddeev-Popov procedure \cite{Faddeev:1967fc}, where the ghost fields $c_i,\bar{c}_i$ are ``living'' on $S^3$, as they arise when the metric fluctuations $h_{ij}$, which are defined to be perturbations over $\rho_{ij}$, are gauge-fixed.

The goal of this work is to compute the path integral in eq. (\ref{eq:Gform_sch}) and understand the one-loop IR behaviour across different boundary conditions imposed on two hypersurfaces. The analysis of \cite{Ailiga:2024wdx} found evidence of the IR divergence in the path integral for a specific choice of boundary condition. In this paper, we first ask how robust the IR behaviour is to changes in the geometries and/or boundary conditions. Specifically, we consider two physically motivated scenarios: the Hartle-Hawking no-boundary condition and the pure de Sitter boundary condition. Both geometries share the same asymptotics in the future (IR), whereas in the past (UV), they have different geometries and the boundary conditions (${\rm Bd}_i$). We wish to compare the one-loop IR behaviour in these two scenarios and determine which features of this behaviour remain the same and which are sensitive to the choice of initial boundary condition. Hence, this study allows us to determine the dependence of the one-loop IR behaviour on the initial/UV boundary condition. This is the second objective of our paper.

Past studies reveal that the appropriate boundary conditions implementing NB and leading to stable evolution are the Neumann and Robin conditions for the scale factor \cite{ Narain:2021bff, Feldbrugge:2017mbc, Narain:2022msz, Ailiga:2025fny, DiTucci:2019dji, Ailiga:2023wzl, Ailiga:2024mmt, Yamada:2025rld} at the initial hypersurface. On the final hypersurface, we either fix the size (Dirichlet) or the extrinsic curvature (conformal) of the universe \cite{York:1972sj, Bousso:1998na, Witten:2018lgb, DiTucci:2019bui, Galante:2025emz, Abdalla:2026mxn, Draper:2025kcr, Krishnan:2017bte, Krishnan:2016mcj}. Fixing the curvature corresponds to fixing the Hubble radius and is physically more relevant, but leads to a non-linear boundary condition, in contrast to the others, which are linear. On the other hand, the de Sitter case corresponds to the boundary condition where the momentum conjugate to the scale factor vanishes on the initial hypersurface. Once we fix the boundary condition for the background $a(t_p)$, it automatically fixes the boundary conditions for $h_{ij}$ fluctuations due to the requirements from general covariance \cite{Brizuela:2023vmb, Ailiga:2024wdx}. Parametrization ($\varepsilon$) brings further constraints on the boundary conditions for the fluctuations. In this paper, we study this dependence of the parametrization on the allowed boundary choices, extending the analysis of \cite{Brizuela:2023vmb, Ailiga:2024wdx}.

Once the boundary conditions are determined, the path integrals for both the background and the fluctuations can be computed, yielding an effective action for lapse integration. The path integral over $a(t_p)$ and $h_{ij}(t,{\bf x})$ decouples in the special case of vanishing fluctuations at the boundaries. The former path integral can be computed exactly for linear boundary conditions, while the latter can be computed up to one loop for arbitrary $\varepsilon$. The one-loop effective action contains UV divergences and requires renormalization and regularization. UV divergences are carefully extracted where infinite summations are performed using the Hurwitz-zeta function \cite{Monin:2016bwf, Elizalde:1994gf, Hawking:1976ja}. UV divergences are cropped 
by adding suitable counterterms. Given the contour $\mathcal{C}$, the integral over lapse is evaluated using the Picard-Lefschetz method \cite{Witten:2010cx, Tanizaki:2014xba, Feldbrugge:2017kzv,Shoji:2025riv, Chou:2024sgk, Ailiga:2023wzl,Ailiga:2024mmt,Ailiga:2024wdx,DiazDorronsoro:2017hti} and WKB approximation. This concludes the computation of the wave function introduced in eq. (\ref{eq:Gform_sch}) for both the no-boundary and de-sitter boundary conditions.

The outline of the paper is as follows: In section \ref{sec:expansion}, we perform the action variation in arbitrary parametrization, respecting the covariance. In sec. \ref{sec:bound_choice_par_dep}, we 
explore how the covariance and parametrization constrain the boundary conditions for background and fluctuations. We consider the general linear boundary condition and non-linear fixed extrinsic condition for the background. 
In Sec. \ref{trans_amp}, we compute the path-integrals for background and fluctuations up to one loop using the Gel'fand-Yanglom method, yielding the effective action for lapse integration including the ghost contribution. In sec. \ref{sec:lapseNc}, we perform a one-loop expansion and analyze the lapse integration using the Picard-Lefschetz method. We analyze the semiclassical stability of the complex saddles and discuss their KSW allowability. In sec. \ref{NB_sad_cor}, we evaluate the one-loop renormlized lapse action at the no-boundary saddles. In sec. \ref{sec:wave_fun_PL_method}, we compute the wave function for both the fixed extrinsic curvature and fixed final size. We analyze its asymptotic behaviour in the deep Lorentzian regime (IR). In sec. \ref{sec:comparison_de_sitter}, we compute the wave function for Lorentzian de-Sitter and compare its IR behaviour with the no-boundary saddles. We further elaborate on the dependence of the one-loop IR behaviour on the choice of initial/UV boundary condition in sec. \ref{sec:uv_dependence}. Finally, we summarize our findings in sec. \ref{sec:conclusion}.

\section{Quadratic expansion of action}
\label{sec:expansion}

We substitute the metric (\ref{eq:frwmet}) in the gravity action mentioned in Eq. (\ref{eq:grav_act}) and retain the terms up to second order in $h_{ij}$. These terms are only relevant in the one-loop computation.  Following the Arnowitt–Deser–Misner (ADM) decompositions \cite{Arnowitt:1962hi}, the four-dimensional Ricci scalar ($R$) can be written in terms of the hypersurface variable as
\bea
\label{eq:3Rgamma}
R = {\cal R} + K_{ij} K^{ij} + K^2 + 2K^\prime/N_p \, ,
\eea
where $({}^\prime)$ denotes the derivative w.r.t $t_p$ and ${\cal R} = \g^{ij} \,\, {\cal R}_{ij}$ is the Ricci-scalar of the 3-hypersurface, $K_{ij}$ is the extrinsic curvature and $K$ is its trace defined on $\gamma_{ij}$. The determinant $\sqrt{-g} = N_p \sqrt{\g}$, where $g = \det g_{\mn}$ and $\g = \det \g_{ij}$. The gravity action mentioned in Eq. (\ref{eq:grav_act}) becomes the following \cite{Arnowitt:1962hi, Ailiga:2024wdx}:
\bea
\label{eq:EHact_ADM}
\mathsf{S}_{\rm grav}[g_\mn]
= \frac{1}{16 \pi G} \int_{\cal M} {\rm d}^4x \sqrt{-g} 
\bigl[
-2 \Lam + {\cal R} + K_{ij} K^{ij} - K^2 
\bigr] 
+ \frac{2}{16 \pi G} \int_{\pt {\cal M}} {\rm d} {\bf x} \sqrt{\g} K + \mathsf{S}_{\rm bd} \, .
\eea
Utilizing the expansion as given in appendix \ref{sec:curvature_expansion}, the above action can be computed up to quadratic order. To fix the gauge redundancy, we consider the transverse-traceless gauge fixing condition $\bar{D}^ih_{ij}=h^i_i=0$, where $\bar{D}$ is the covariant derivative on the hypersurface. In this gauge, the graviton has only two physical degrees of freedom; see appendix \ref{sec:gauge-fixing_ghost} for details of diffeomorphism breaking, gauge fixing, and ghost computation.

\subsection{
Action variation, covariance and parametrization}
\label{boundTR} 
In this section, we focus our attention on computing the boundary conditions required to evaluate the path integral in eq. (\ref{eq:PI_LI_g}). This is achieved by ensuring the proper variation of the gravity action. Studies in \cite{Brizuela:2023vmb, Ailiga:2024wdx} reveal that general covariance puts a strong constraint on the allowed boundary conditions. While the past literature focuses on the linear split, in this section, we allow the general parametrization ($\varepsilon$) and compute the allowed boundary conditions. In particular, we will be interested in seeing whether any specific parametrization makes the allowed boundary conditions ``simpler" than those previously found for the linear split.  

To set the stage, we make the following transformation for the scale factor ($a(t_p)$) and lapse ($N_p(t_p)$): $N_p(t_p) {\rm d} t_p = (N(t)/a(t)) {\rm d} t$ and $q(t) = a^2(t)$.
This transformation recasts the action in a quadratic form, allowing exact computation \cite{Feldbrugge:2017kzv, Narain:2022msz, Ailiga:2023wzl}. The line element with this transformation is given by
\beq
\label{eq:frwmet_changed}
{\rm d}s^2 = - \frac{N^2}{q(t)} {\rm d} t^2 
+ \underbrace{
q(t) \left[
\rho_{ij} + h_{ij}+\frac{\varepsilon}{2}h_{ik}h^{k}{}_j
\right]
}_{\g_{ij}(t,{\bf x})} {\rm d}x^i \, {\rm d}x^j \, .
\eeq
In the above metric, we choose the proper-time gauge ($\dot{N}=0, N=N_c$(constant), $(^.)$ being the derivative w.r.t ``$t$"). Utilizing the transformation, the gravitational action in eq. (\ref{eq:grav_act}), relevant for one-loop computation, is given by
\bea
\label{eq:EHact_exp}
&&
S_{\rm grav}[g_\mn] 
= S_{\rm grav}^{(q)} + S_{\rm grav}^{(h)} = 
\frac{1}{16\pi G}\int_{\cal M} {\rm d} t {\rm d} {\bf x} \sqrt{\rho}
\biggl[
(6k -2 \Lam q )N_c - \frac{3\dot{q}^2}{2N_c} 
\biggr]  \notag \\
&&
+ \frac{1}{16\pi G}\int_{\cal M} {\rm d} t {\rm d} {\bf x}\sqrt{\rho}
\biggl[ 
\frac{q^2}{4N_c}\dot{h}_{ij}\dot{h}^{ij} -\frac{N_c}{2}h_{ij}h^{ij}
+\frac{N_c}{4}h^{ij}\bar{D}^2h_{ij}+(\varepsilon-1)\biggl\{\frac{\dot{q}^2}{8 N_c^2}+\frac{q\ddot{q}}{2N_c^2} \notag \\
&&+\frac{1-\Lambda q }{2}\biggr\}N_c h_{ij}h^{ij}
\biggr] 
+ \frac{1}{16\pi G} \int_{\pt {\cal M}} {\rm d} {\bf x} 
\sqrt{\rho} \biggl[\left(\frac{3q \dot{q}}{N_c}\right) +(\varepsilon-1)\frac{q^2}{N_c}\dot{h}_{ij}h^{ij}\notag \\
&&+(\varepsilon-1)\frac{q\dot{q}}{4N_c}h_{ij}h^{ij}\biggr]\biggr \rvert_0^1
+ S_{\rm bd}\, ,
\eea
where we imposed the transverse-traceless gauge fixing condition $\bar{D}^ih_{ij}=h^i_i=0$ and use a different font for the action to convey the same. Here, $\pt {\cal M}$ refers to boundaries at $t=0$ and $t=1$ along with the $S^3$. The surface terms, which are the total spatial derivative, vanish and are ignored in eq. (\ref{eq:EHact_exp}). We observe that while the kinetic term is parametrization independent, the ``mass" term depends on it. For $\varepsilon=1$, both the bulk and boundary action get simplified.  The surface term $S_{\rm bd}$ needs to be computed to achieve the proper variation and has two parts: $S_{\rm bd} = S_{\rm bd}^{(q)} + S_{\rm bd}^{(h)}$.
In this section, we compute $S_{\rm bd}$ explicitly for a few physically relevant boundary choices. Respecting the general covariance, we take the following ansatz for $S_{\rm bd}$
\beq
\label{eq:Sbd_genform}
S_{\rm bd} = \frac{1}{8\pi G} \int_{\pt {\cal M}} {\rm d} {\bf x}
\sqrt{\g} \,\, F_{\pt {\cal M}}({\cal R}, K) \biggr \rvert_0^1\, ,
\eeq
where the subscript ${\pt {\cal M}}$ implies that the function $F$ can be different for different boundaries. We perform an expansion of the $S_{\rm bd}$ in powers of $h_{ij}$. Keeping terms up to second order in $h_{ij}$ we get
\begin{equation}
\label{eq:expand_F}
F_{\pt {\cal M}}({\cal R}, K)=F_{\pt {\cal M}}({\cal R}_b, K_b)+\frac{\partial F_{\pt {\cal M}}({\cal R}, K)}{\partial {\cal R}}\biggr|_{({\cal R}_b, K_b)}\mathbf{\Delta}{\cal R}+\frac{\partial F_{\pt {\cal M}}({\cal R}, K)}{\partial K}\biggr|_{({\cal R}_b, K_b)}\mathbf{\Delta} K,
\end{equation}
where $K_b$ and ${\cal R}_b$ are defined on background. $\mathbf{\Delta}{\cal R}$ and $\mathbf{\Delta} K$ are $({\cal R}-{\cal R}_b)$ and $(K-K_b)$ respectively and can be read out from the appendix \ref{sec:curvature_expansion}. In the gauge considered here, neither $K$ nor ${\cal R}$ has a linear term in $h_{ij}$. In addition, when exponential parametrization ($\varepsilon=1$) is considered, we have $\mathbf{\Delta} K=0$ \footnote{In fact, it is true for all orders in $h_{ij}$. Extrinsic curvature ($K$) doesn't change when parameterizing the fluctuations exponentially, see appendix \ref{sec:curvature_expansion}.}.
This makes the last term in eq. (\ref{eq:expand_F}) vanishing and consequently $\partial_KF_{\pt {\cal M}}$ doesn't enter into the further calculation. This simplification happens only in the exponential parametrization.
Defining the background quantities 
${\cal R}_b=\bar{{\cal R}}/q = {\cal R}_{h=0}$, 
$K_b=3\dot{q}/(2N_c \sqrt{q}) = K_{h=0}$, we get, 
\bea
\label{eq:Fexp}
S_{\rm bd}
= &&
\frac{1}{8\pi G} \int_{\pt {\cal M}} {\rm d} {\bf x}\sqrt{\rho}
\biggl[
q^{3/2} F({\cal R}_b, K_b)
+ \sqrt{q} \pt_{\cal R} F({\cal R}_b, K_b)
\left\{\frac{1-2\varepsilon}{2} h_{ij} h^{ij}\right.
\notag \\
&&
\left.+ \frac{1}{4} h_{ij} \bar{D}^2 h^{ij}
\right\} + \pt_{K} F({\cal R}_b, K_b)\biggl\{(\varepsilon-1)\frac{q^2}{2N_c}\dot{h}_{ij}h^{ij}\biggr\}
\biggr]\times\left(1+\frac{\varepsilon-1}{4}h_{ij}h^{ij}\right) \biggr \rvert_0^1\, ,
\eea
where we imposed the gauge fixing condition on $h_{ij}$.
The above boundary action consists of $S_{\rm bd}^{(q)}$ and $ S_{\rm bd}^{(h)}$. Considering $S_{\rm bd}^{(q)}$ and performing the background variation $q(t) = \bar{q}(t) + \de q(t)$, we get in the linear order of $\de q(t)$:
\beq
\label{eq:Sexp_qvar}
\de S^{(q)}_{\rm grav} = \frac{V_3}{16\pi G} \int_{0}^{1} {\rm d}t \biggl[
\left(-2 \Lam N_c + \frac{3 \ddot{\bar{q}}}{N_c} \right) \de q
\biggr]
+ \frac{V_3}{16\pi G} \biggl[
\frac{3\bar{q} \de \dot{q}}{N_c} 
+ 3 \sqrt{\bar{q}} \bar{F} \de q 
+ 2 \bar{q}^{3/2} \de F({\cal R}_b, K_b)
\biggr] \biggr\rvert_0^1\, ,
\eeq
where $\bar{F} = F(\bar{\cal R}_b, \bar{K}_b)$,
$\bar{\cal R}_b = \bar{\cal R}/\bar{q}$,
$\bar{K}_b = 3 \dot{\bar q}/(2N_c \sqrt{\bar q})$ and $V_3=\int d{\bf x}\sqrt{\rho}$. Once all the surface terms are canceled by choosing an appropriate $F$, 
the equation of motion for $q$ is
\beq
\label{eq:dyn_q_eq}
\ddot{\bar{q}} - \frac{2}{3} \Lam N_c^2 =0 \, .
\eeq
In the above, we assume no ``back-reaction" implying that $h_{ij}$ doesn't affect the variation of background ($\bar{q}$) \cite{Feldbrugge:2017mbc, Barvinsky:1992dz}. Now performing the variation for $S_{\rm grav}^{(h)}$, we get
\bea
\label{eq:Sexp_hvar}
\de S^{(h)}_{\rm grav} = 
&&
\int {\rm d}{\bf x} {\rm d}t \frac{\sqrt{\rho}}{16\pi G}
\biggl[
-\frac{\bar{q}^2 \ddot{\bar h}_{ij}}{2N_c} 
- \frac{\bar{q} \dot{\bar{q}} \dot{\bar h}_{ij}}{N_c}
+ \frac{N_c}{2} \bar{D}^2 \bar{h}_{ij} 
- N_c \bar{h}_{ij}+(\varepsilon-1)\biggl\{\frac{\dot{q}^2}{4 N_c^2}\notag \\
&&+\frac{q\ddot{q}}{N_c^2}+1-\Lambda q \biggr\}N_ch_{ij}
\biggr] \de h^{ij}
+ \frac{1}{16\pi G} \int_{\pt {\cal M}} {\rm d}{\bf x} \sqrt{\rho}
\bigl[
A_{ij}^\varepsilon \de h^{ij}+B_{ij}^\varepsilon\dot{\delta h^{ij}}
\bigr] \biggr \rvert_0^1\, ,
\eea
where $A_{ij}^\varepsilon$ and $B_{ij}^\varepsilon$ are given by
\bea
\label{eq:ABexp}
&&
A_{ij}^\varepsilon = 
\biggl[\frac{\bar{q}^2(2\varepsilon-1)}{2N_c}+\frac{\bar{q}^2(\varepsilon-1)}{N_c}\partial_{ K}F\biggr] \dot{\bar h}_{ij}+\biggl[\frac{\bar{q}\dot{\bar{q}}(\varepsilon-1)}{2N_c}+\bar{q}{\sqrt{\bar{q}}}\bar{F}(\varepsilon-1)+\sqrt{\bar{q}}\partial_{\cal R}\bar{F}\bar{\nabla}^2 \, ,
\notag \\
&& +2\sqrt{\bar{q}}(1-2\varepsilon)\partial_{\cal R}\bar{F}\biggr]\bar{h}_{ij}\, ,
\notag \\
&& B_{ij}^\varepsilon=\frac{\bar{q}^2(\varepsilon-1)}{N_c}\left(1+\partial_{ K }\bar{F}\right)\bar{h}_{ij}
\eea
where $\pt_{\cal R} {\bar F} = \pt_{\cal R}F(\bar{\cal R}_b, \bar{K}_b)$,
$\pt_K {\bar F} = \pt_KF(\bar{\cal R}_b, \bar{K}_b)$ and 
$\bar{h}_{ij}$ satisfies the following equation of motion 
\beq
\label{eq:hijEQM}
\bar{q}^2 \ddot{\bar h}_{ij}
+2\bar{q} \dot{\bar{q}} \dot{\bar h}_{ij}
- N_c^2 \bar{D}^2 \bar{h}_{ij} 
+ 2N_c^2\bar{h}_{ij}-(\varepsilon-1)\biggl\{\frac{\dot{q}^2}{2 N_c^2}+\frac{2q\ddot{q}}{N_c^2}+2-2\Lambda q \biggr\}N_c^2 h_{ij} = 0
 \, .
\eeq
Clearly, the on-shell equation depends on the parametrization ($\varepsilon$). In special cases when the expression inside the curly bracket vanishes, it becomes independent of $\varepsilon$.
To ensure a consistent variational problem, one requires 
\beq
\label{eq:BD_h_gen}
A_{ij}^\varepsilon \de h^{ij}+B_{ij}^\varepsilon\dot{\delta h^{ij}} \biggr \rvert_0^1= 0 \, ,
\eeq
which automatically fixes the boundary condition for $h_{ij}$! So, what we obtain is that once we choose a boundary condition for $q(t)$, i.e., some specific $F$, eq. (\ref{eq:ABexp}) and (\ref{eq:BD_h_gen}) constraint the allowed boundary condition for $h_{ij}$. Clearly, as expected, the boundary conditions depend on the chosen parametrization. Specifically, we observe that when $\varepsilon=1$, $B^\varepsilon_{ij}=0$ and the allowed boundary choices become
\begin{equation}
\label{eq:allowed_bc_exp_para}
h_{ij}={\rm const.},\hspace{5mm}{\rm and}\hspace{5mm}\frac{\bar{q}^2}{2N_c}\dot{\bar h}_{ij}+\biggl[\sqrt{\bar{q}}\partial_{\cal R}\bar{F}\bar{\nabla}^2 -2\sqrt{\bar{q}}\partial_{\cal R}\bar{F}\biggr]\bar{h}_{ij}=0\, .
\end{equation} 
The above boundary conditions are linear! To compare, when we considered a linear split, for example, allowed boundary choices were found to be both linear and non-linear \cite{Brizuela:2023vmb, Ailiga:2024wdx}. This linearity in the space of allowed boundary conditions appears only for $\varepsilon=1$. Our analysis suggests that when parameterizing the fluctuation as an exponential, the allowed boundary choices become ``simpler".

\subsection{On shell action}
\label{EQM_qh} 
After finding the boundary conditions, we proceed to solve the equations of motion for 
$q(t)$ and $h_{ij}(t, {\bf x})$ as given in Eqs. (\ref{eq:dyn_q_eq}) 
and (\ref{eq:hijEQM}) respectively.
The e.o.m for $q(t)$ is solved by
\beq
\label{eq:qsol_gen}
\bar{q}(t) = \frac{\Lam N_c^2}{3} t^2 + c_1 t + c_2 
= \frac{\Lam N_c^2}{3} (t-r_1)(t-r_2)\, ,
\eeq
where $c_{1,2}$ are constants depending on the boundary conditions, while $r_{1,2}$ are the (complex) roots of $q(t)=0$. To solve the e.o.m for $h_{ij}$, we perform a mode decomposition over the spherical harmonics on $S^3$. For the choice of gauge, where $h_{ij}$ being transverse-traceless (TT), one can write \cite{Gerlach:1978gy,Higuchi:1986wu}
\bea
\label{eq:h_ij_Sph_Har}
&&
h_{ij}(t, {\bf x}) 
=\sum_{l=2}^\infty 
\sum_{n=2}^{l} \sum_{m=-n}^{n}
h^{l}_{nm}(t) (G_{ij})^l_{nm} ({\bf x}) \,, 
\notag \\
{\rm where \,\,\,\,}
&&
\bar{D}^2 (G_{ij})^{l}_{mn} ({\bf x}) = - [l(l+2) -2] (G_{ij})^{l}_{mn} ({\bf x})
\hspace{5mm} l \geq 2
\, .
\eea
As the action for fluctuation $h_{ij}$ mentioned in Eq. (\ref{eq:EHact_exp}) respects isotropy ($S^3$ symmetry), it is therefore independent of $n$ and $m$. So, we suppress the $n$ and $m$ indices and write $h^{l}_{nm}(t) \equiv h_l(t)$ with the degeneracy for each $l$-mode $g_l =2 (l+3)(l-1)$.
The factor $2$ appearing in $g_l$ reflects the two degrees of freedom of the physical graviton in TT gauge. Using the mode decomposition, we get that each $h_l(t)$ coefficient satisfies the following e.o.m: 
\beq
\label{eq:htL_eqm}
\ddot{\bar h}_l + \frac{2 \dot{\bar{q}}}{\bar{q}} \dot{\bar h}_l
+\biggl[
l(l+2)-(\varepsilon-1)\biggl\{\frac{\dot{q}^2}{2 N_c^2}+\frac{2q\ddot{q}}{N_c^2}+2-2\Lambda q \biggr\}\biggr]\frac{N_c^2}{\bar{q}^2}
\bar{h}_l = 0 \, .
\eeq

 We write the general solution as \footnote{ As $\mathbb{P}_1^{\xi_l}$ becomes ill-defined for certain (positive) integer degeneracies in $\xi_l$, we choose $\mathbb{P}_1^{-\xi_l}$ to be one of the independent solution. As $\xi_l$ can never be negative integer, $\mathbb{P}_1^{-\xi_l}$ and $\mathbb{Q}_1^{\xi_l}$ serve as a good independent solution for all $N_c$ (i.e., for all values of $\xi_l$ and $\tau_t$), with well-defined Wronskian (see appendix \ref{sec:def_mode_fun} for definitions). Another way to
 regulate these integer degeneracies, as taken in \cite{Barvinsky:1992dz, Ailiga:2024wdx}, would be to replace $1$ by $1-i\epsilon$ and then taking the limit $\epsilon\to 0$ at the end of the computation. 
}
\beq
\label{eq:hl_sol_gen}
\begin{split}
&\hspace{20mm}\bar{h}_l(t) = \frac{1}{\sqrt{\bar{q}}} 
\bigl[
d^{(l)}_{p} \mathbb{P}_1^{-\xi_l} (\tau_t) + d^{(l)}_{q} \mathbb{Q}_1^{\xi_l} (\tau_t) 
\bigr] \, ,\\
&\xi_l = \sqrt{2\varepsilon-1
-\frac{36 l(l+2) - 72(\varepsilon-1)}{N_c^2 \Lam^2 (r_1 - r_2)^2}} \, ,
\hspace{5mm}
\tau_t = \frac{2t - (r_2+r_1)}{r_1 - r_2},
\end{split}
\eeq
where $\mathbb{P}_1^{\xi_l}$ and $\mathbb{Q}_1^{\xi_l}$ are Legendre-P
and Legendre-Q functions, respectively.
While $\xi_l$ depends on the parametrization ($\varepsilon$), $\tau_t$ is independent of it.
The constants $d^{(l)}_{p}$ and $d^{(l)}_q$ are fixed based on the choice of
boundary conditions and depend on $\varepsilon$. 
For later purposes, we define: 
\beq
\label{eq:W_l10_func}
W_l^\varepsilon(t,0)= \mathbb{P}_1^{-\xi_l}[\tau_t] \mathbb{Q}_1^{\xi_l}[\tau_0] 
- \mathbb{P}_1^{-\xi_l}[\tau_0] \mathbb{Q}_1^{\xi_l}[\tau_t] \, .
\eeq
Using the on-shell equations for $\bar{q}(t)$ and $\bar{h}_{ij}(t)$ mentioned in eqs. (\ref{eq:qsol_gen}) and (\ref{eq:hijEQM}), the on-shell action in eq. (\ref{eq:EHact_exp}) evaluates to 
\bea
\label{eq:EHact_exp_onshell}
&&
S_{\rm grav}^{\rm on-shell} 
= S_{\rm grav}^{(\bar{q})} + S_{\rm grav}^{(\bar{h})} = 
\frac{V_3}{16\pi G} \Bl[
\frac{2 \Lam^2 N_c^3}{9} + (6k + \Lam c_1) N_c + \frac{3c_1^2}{2N_c}
\Br]
\notag \\
&&
+ \frac{1}{16\pi G}\int_{\pt {\cal M}} {\rm d} {\bf x} \sqrt{\rho}\biggl[
(4\varepsilon-3)\frac{\bar{q}^2}{4N_c}\dot{\bar{h}}_{ij}\bar{h}^{ij}+(\varepsilon-1)\frac{\bar{q}\dot{\bar{q}}}{4N_c}\bar{h}_{ij}\bar{h}^{ij}\biggr]\biggr \rvert_0^1
+ S_{\rm bd} \, .
\eea
It is interesting to note that the on-shell action for $h_{ij}$ is a purely surface term and depends on the parametrization ($\varepsilon$). In the next section, we will compute the $S_{\rm bd}$ explicitly for a few special boundary choices.

\section{Boundary choices and parametrization dependence}
\label{sec:bound_choice_par_dep}
In this section, we compute $S_{\rm bd}$ explicitly for a few boundary conditions, which could be either linear or non-linear. The linear boundary conditions are Dirichlet, Neumann and Robin, while the non-linear boundary condition we consider is the fixed extrinsic curvature ($K$)/conformal. We explore the dependencies of the boundary choices for fluctuations, given these as background boundary conditions. We investigate the parametrization dependence of the boundary conditions and the on-shell action, extending the analysis of \cite{Ailiga:2024wdx} for the case of the linear split with linear boundary conditions. Expressing the conjugate momentum corresponding to $q(t)$ as $\pi_q = -3\dot{q}/(2 N_c)$, we get $K_b=-\pi_q/\sqrt{q}$.
For variational consistency, the 
boundary term appearing in Eq. (\ref{eq:Sexp_qvar}) vanishes leading to
\beq
\label{eq:BDpart_qexp}
\biggl[
\biggl(
3 \sqrt{\bar{q}} \bar{F} 
- \frac{2\bar{\cal R}}{\sqrt{\bar{q}}} \pt_{\cal R} \bar{F} 
+ \bar{\pi}_q \pt_K \bar{F} 
\biggr) \de q
- 2 \bar{q} \biggl(
1+ \pt_K \bar{F}
\biggr) \de \pi_q 
\biggr] \biggr \rvert_0^1= 0 \, , 
\eeq
where $\bar{F} = F(\bar{\cal R}_b, \bar{K}_b)$.

\subsection{
Linear boundary condition}
\label{DBC_bg} 
In this section, we consider the linear boundary conditions on the background. These are either Dirichlet, Neumann or Robin boundary conditions. Such boundary conditions have been extensively studied in the past; for example, see \cite{Ailiga:2024wdx}. Those studied had been performed for the special case when $\varepsilon=0$, i.e., linear split. Here, we will briefly discuss the case when fluctuation is parametrized through exponential, $\varepsilon=1$.  
We start by imposing the Dirichlet boundary condition at the endpoints, which fixes $q$ and hence $\delta q=0$. 
The covariant surface term becomes \cite{Brizuela:2023vmb,Ailiga:2024wdx} 
\begin{equation}
    \label{eq:cov_sur_term}
    S_{\rm bd}^{\rm DBC}=\frac{1}{8\pi G}\int_{\partial M}d^3x\sqrt{\gamma}\,\,[-K+f({\cal R})],
\end{equation}
where $f({\cal R})$ is any arbitrary function. 
For the Dirichlet condition, the on-shell action is given by
\bea
\label{eq:onsh_dbc_qbar}
S_{\rm grav}^{(\bar{q})} \br \rvert_{\rm DBC}
= \frac{V_3}{16\pi G}
\Bl[\frac{\Lam^2}{18}N_{c}^3 
+ \{6k - \Lam(q_f + q_i)\} N_c - \frac{3 (q_f - q_i)^2}{2 N_c}\Br] \, .
\eea
Given $F$, one can compute the boundary condition for the fluctuation $h_{ij}$ from eqs. (\ref{eq:ABexp}) and (\ref{eq:BD_h_gen}) depending on parametrization ($\varepsilon$). 
In this paper, we will focus on the 
Dirichlet choices: $\de h_{ij} \rvert_0^1 =0$
($h_{ij}$ is fixed at the endpoints), which is independent of $\varepsilon$. 
The on-shell solution for $h_l(t)$ is given
in eq. (\ref{eq:hl_sol_gen}). For the boundary 
condition $h_l(t=1)=h_1^{(l)}$ and $h_l(t=0)=0$, $d^{(l)}_{p}$ and $d^{(l)}_{q}$ are given by: $d^{(l)}_{p}W_l^\varepsilon(1,0)=h_1^{(l)}\sqrt{q_f} \mathbb{Q}_1^{\xi_l}[\tau_0]$, $d^{(l)}_{q}W_l^\varepsilon(1,0)=-h_1^{(l)}\sqrt{q_f} \mathbb{P}_1^{-\xi_l}[\tau_0]$.
Evaluating the on-shell action including the surface terms, as mentioned in eq. (\ref{eq:EHact_exp_onshell}), for each $l$ mode, we get
\bea
\label{eq:hh_onSH_DBC}
&&
S^{(l)}_{\rm grav} [\bar{q}, \bar{h}_l, N_c] \Br \rvert_{\rm DBC} 
=\frac{1}{16\pi G}\Bl[
\frac{\bar{q}^2}{4N_c}\bar{h}_l\dot{\bar{h}}_l
+(1-\varepsilon)\frac{\bar{q}\dot{\bar{q}}}{2N_c} \bar{h}_l^2
\Br]\Br \rvert_{0}^{1} \, ,
\notag \\
&&
= \frac{q_f \bl(h^{(l)}_1\br)^2}{(16\pi G) (24N_c)}
\Bl[
\{3(q_f-q_i) + \Lam N_c^2\}(3-4\varepsilon) +6q_f \bl\{ \ln W_l^\varepsilon(t,0) \br\}^\prime \br\rvert_{t=1}
\Br] \, ,
\eea
where $({}^\prime)$ denotes the derivative with respect to $t$. Clearly, the on-shell action depends on the chosen parametrization ($\varepsilon$). Since the on-shell action is a purely boundary term, this dependence is only at the boundaries.


We now consider imposing the Robin boundary condition 
for $q(t)$ at one of the endpoints. 
In this case, we fix a 
linear combination of the scale factor $q$ and the
conjugate momentum $\pi_q$ at the end point. 
This means $\pi_q + \boldsymbol{\beta}\, q = P_i= {\rm fixed}$ and hence, $\de \pi_q + \boldsymbol{\beta} \de q = 0$.
In the limit $\boldsymbol{\beta}\rightarrow 0$, it reduces to the Neumann boundary condition, where one fixes the momentum $(\pi_q=P_i,\,\delta \pi_q=0)$.
The covariant boundary action is found to be \cite{Ailiga:2024wdx}
\beq
\label{eq:Sbd_rbc_cov}
S^{\rm RBC}_{\rm bd}
= 
\frac{1}{8\pi G} \int_{\pt {\cal M}} {\rm d} {\bf x} \sqrt{\g} 
\biggl[
c  \biggl(2\boldsymbol{\beta} \sqrt{\frac{6k}{\cal R}} - K \biggr)^{2\om}
{\cal R}^{3/2-\om} - \boldsymbol{\beta} \sqrt{\frac{3k}{2 {\cal R}}}
\biggr]
\hspace{5mm}
\forall c,\om \in \mathbb{R} \, ,
\eeq
with the simplest choice being $c=0$. $S^{\rm RBC}_{\rm bd}$ is an infinite set of covariant actions, each of which constrains the boundary choices for fluctuation from eqs. (\ref{eq:ABexp}) and (\ref{eq:BD_h_gen}) depending on parametrization ($\varepsilon$). In the limit
$\boldsymbol{\beta}\to0$, this boundary action goes to the action for the Neumann condition. For example, when $\boldsymbol{\beta}=0$ and $\varepsilon=1$, the boundary choices are given by eq. (\ref{eq:allowed_bc_exp_para}) with the given $F$. In this paper, we stick to the simplest choice ($c=0$).
The on-shell action $S_{\rm grav}^{(\bar{q})}$ with the Robin BC at the initial 
boundary and the Dirichlet condition at the final boundary is given by \cite{Ailiga:2024wdx, Ailiga:2024mmt}
\bea
\label{eq:stot_onsh_rbc}
S_{\rm grav}^{\rm (\bar{q})}[N_c] &=& 
\frac{V_3}{16\pi G}\biggl\{\frac{1}{(27 + 18 N_c \boldsymbol{\beta})} \biggl[
\boldsymbol{\beta} \Lam^2 N_c^4 + 6 \Lam^2 N_c^3 +
N_c^2 \{108 \boldsymbol{\beta}  k-18 \Lam  (P_i+\boldsymbol{\beta} q_f)\}
\notag \\
&&
+18 N_c \left\{9
k+P_i^2 -  3 q_f \Lambda \right\}
+ 54 P_i q_f - 27 \boldsymbol{\beta} q_f^2
\biggr]\biggr\} \, .
\eea
We consider the Dirichlet boundary condition for fluctuation on both hypersurfaces. The on-shell action for each $l$ mode is given by
\begin{equation}
\label{eq:hh_onSH_RBC}
\begin{split}
&
S^{(l)}_{\rm grav} [\bar{q}, \bar{h}_l, N_c] \Br \rvert_{\rm RBC} 
= \frac{1}{16\pi G} 
\Bl[
\Bl(\frac{\bar{q}^2}{4 N_c} \bar{h}_{l} \dot{\bar{h}}_{l} 
+(1-\varepsilon) \frac{\bar{q} \dot{\bar{q}}}{2 N_c} \bar{h}^2_{l} \Br) \Br \rvert_{1} 
 \\
& 
+ \Bl\{(3-4\varepsilon)\frac{\bar{q}^2}{4 N_c} \bar{h}_{l} \dot{\bar{h}}_{l} 
+ (1-\varepsilon)\frac{\bar{q} \dot{\bar q}}{4 N_c} \bar{h}^2_{l} 
- \boldsymbol{\beta} \bar{q}^2 \bar{h}^2_{l}\Bl( \frac{7-8\varepsilon}{24}
- \frac{l(l+2)-2}{48}  
\Br)\Br\}\Br\rvert_{0} \Br] 
 \\
 &= 
\frac{q_f (h_{1}^{(l)})^2}{(16\pi G)(24 N_c)}
\Bl[\frac{6 \Lam N_c^2 - 6 N_c P_i 
+ \boldsymbol{\beta} (6 q_f N_c + 2 \Lam N_c^3 )}{(3 + 2 N_c \boldsymbol{\beta})} (3-4\varepsilon)
+ 6 q_f \bl\{\ln{W}_l^\varepsilon(t,0)\br\}'\big|_{t=1}\Br] \, .
\end{split}
\end{equation}
In the limit $\boldsymbol{\beta}\rightarrow 0$, we obtain the on-shell action for the Neumann boundary condition. Similar to the DBC case, on- shell action is purely a boundary term and depends on parametrization ($\varepsilon$).
\subsection{ Fixed extrinsic curvature
} 
\label{sec:fix-k}
So far, we have focused only on linear boundary choices. However, there is one physically well-motivated nonlinear boundary condition worth studying. It fixes the trace of the extrinsic curvature ($K$), which is the same as fixing the Hubble rate ($H_1$)
\begin{equation}
    \label{eq:fixing_K}
    K=-\frac{\pi_q}{\sqrt{q}}=3H_1=\text{fixed},
\end{equation}
where we have $H_1\leq H$ with $H=\sqrt{\Lam/3}$ due to the closed slicing of de-Sitter space. In the Euclidean regime $H_1=\pm i\gamma H$, and hence is purely imaginary. Such boundary choices have been considered previously in the context of no-boundary cosmology and slow-roll inflation \cite{DiTucci:2019bui, Abdalla:2026mxn, Bousso:1998na} and have been found to be preferred over the Dirichlet boundary condition, which fixes the final size of the universe \cite{DiTucci:2019bui}. Such boundary conditions, also known as conformal boundary conditions, have been studied in higher derivative gravity theories \cite{York:1972sj,Witten:2018lgb,Galante:2025emz}. In \cite{DiTucci:2019bui}, the authors studied only the background dynamics; here, we aim to extend their analysis by considering the $h_{ij}$ fluctuation. 

Taking the variation of the boundary condition in eq. (\ref{eq:fixing_K}) and combining eq. (\ref{eq:BDpart_qexp}), we get
\begin{equation}
    \label{eq:fixed_K_variation}
    \delta \pi_q+\frac{3H_1}{2\sqrt{q}}\delta q =0\hspace{3mm}\Rightarrow \hspace{3mm} 3F-2\mathcal{R}\partial_{\cal R}F+K=0 
\end{equation}
The above equation is an ordinary differential equation for $\cal R$. Solving it, we get the boundary action
\beq
\label{eq:Sbd_fix_K}
S^{{\rm fixed}\, K}_{\rm bd}
= 
\frac{1}{8\pi G} \int_{\pt {\cal M}} {\rm d} {\bf x} \sqrt{\g} 
\biggl[
-\frac{K}{3}+f(K)\,{\cal R}^{3/2}
\biggr]
\, ,
\eeq
where $f(K)$ is any arbitrary function of $K$. 
The first term in eq. \ref{eq:Sbd_fix_K} was previously known in the literature \cite{Witten:2018lgb,DiTucci:2019bui,Galante:2025emz,Krishnan:2017bte}; however, the second term is new. This term will lead to non-trivial boundary choices and dynamics for $h_{ij}$. Considering the parametrization $\varepsilon=1$, we get the boundary choices 
\begin{equation}
\label{eq:allowed_bc_exp_para_fixed_k}
h_{ij}={\rm const.},\hspace{5mm}{\rm and}\hspace{5mm}\frac{\bar{q}^2}{6N_c}\dot{\bar h}_{ij}+\sqrt{\frac{3}{2}}f(K)[\bar{\nabla}^2 -2]\bar{h}_{ij}=0.
\end{equation}
In this paper, we will stick to the simplest choice $f=0$.
The on-shell solution 
for scale factor $\bar{q}(t)$ with the Neumann BC ($\pi_q=P_i$) at the initial 
boundary and fixed $K$ at the final boundary is given by
\beq
\label{eq:qsol_RBC}
\bar{q}(t) = \frac{\Lam N_c^2}{3} (t^2-1) -\frac{2N_cP_i}{3}(t-1)+
\frac{(P_i-\Lambda  N_c)^2}{9H_1^2}\, .
\eeq

The on-shell action $S_{\rm grav}^{(\bar{q})}$ is given by
\bea
\label{eq:stot_onsh_fixed_k}
S_{\rm grav}^{\rm (\bar{q})}[N_c] &=& 
\frac{V_3}{8\pi G(27 H_1^2)}\biggl[\Lambda^2(3 H_1^2-\Lambda) N_c^3+3P_i\Lambda(\Lambda-3H_1^2)N_c^2
\notag \\
&& +\{9H_1^2(P_i^2+9)-3P_i^2\Lambda\}N_c+P_i^3
\biggr] \, .
\eea
From eqs. (\ref{eq:ABexp}) and (\ref{eq:Sbd_fix_K}), we observe that when fixing $K$, Dirichlet BC ($h_{ij}=\rm const$) is allowed only for exponential parametrization ($\varepsilon=1$). This is an example where the allowed boundary condition depends on parametrization. Moreover, when $\varepsilon=1$, such a boundary condition is allowed irrespective of the chosen boundary condition on the background (independent of $F$). However, special Dirichlet BC ($h_{ij}=0$) is always allowed irrespective of background boundary condition and/or parametrization. Since we only focus on the Dirichlet B.c for fluctuation, we consider only the case $\varepsilon=1$ in the following.

The on shell solution for $h_l(t)$ is mentioned 
in Eq. (\ref{eq:hl_sol_gen}), with $d^{(l)}_{p}$ and $d^{(l)}_{q}$ are given by
\bea
\label{eq:hl_dpdq_RBC}
d^{(l)}_{p} =  \frac{h^{(l)}_1 \sqrt{\bar{q}(1)} \mathbb{Q}_1^{\xi_l}[\tau_0] }{W_l^\varepsilon(1,0)} \, ,
\hspace{5mm}
d^{(l)}_{q} = - \frac{ h^{(l)}_1 \sqrt{\bar{q}(1)} \mathbb{P}_1^{-\xi_l}[\tau_0] }{W_l^\varepsilon(1,0)} \, ,
\hspace{5mm} \text{with},\hspace{4mm}\varepsilon=1\, ,
\eea
Note the appearance of $\sqrt{\bar{q}(1)}$ instead of $q_f$ in $d^{(l)}_{p}$ and $d^{(l)}_{q}$.
The on-shell action for each mode of fluctuation, including the boundary term, is given by (for $\varepsilon=1$)
\bea
\label{eq:hh_onSH_fix_K}
&&
S^{(l)}_{\rm grav} [\bar{q}, \bar{h}_l, N_c] \Br \rvert_{{\rm fixed}\,\, K}
= \frac{1}{16\pi G} 
\frac{\bar{q}^2}{4 N_c} \bar{h}_{l} \dot{\bar{h}}_{l}  \Br \rvert_{0}^1 
\notag \\
 =&& 
\frac{(h_{1}^{(l)})^2(P_i-N_c\Lambda)^3}{(16\pi G)(324 N_cH_1^4)}
\Bl[3H_1^2N_c
+ (P_i-N_c\Lambda)\bl\{\ln{W}_l^{\varepsilon=1}(t,0)\br\}'\big|_{t=1}\Br] \, .
\eea

It is interesting to note that when $\varepsilon=1$, the $K$ doesn't change due to the  $h_{ij}$ fluctuation and remains the same as the background ($\mathbf{\Delta} K=0$). This implies that once we fix the background $K$, the dynamics and boundary choices are completely independent of $K$. In such a scenario, intrinsic curvature ($\mathcal{R}$) serves as the only independent variable. This simplification wouldn't occur in any other parametrization ($\varepsilon\neq 1$) or with other boundary choices for background. Finally, we conclude this section by noting that, in the context of boundary choices in the path integral, the exponential parametrization is preferred, as all the allowed boundary conditions become linear in this parametrization. \\
 
\section{Path integral over $q(t)$ , $h_{ij}(t,{\bf x})$ and ghosts}
\label{trans_amp}


After finding the relevant boundary conditions for both $q$ and $h_{ij}$, we proceed to compute the path integral in eq. (\ref{eq:Gform_sch}). Utilizing the transformation as mentioned earlier in sec. (\ref{boundTR}), the path integral becomes
\beq
\label{eq:grav_path_qform}
\mathcal{Z}[{\rm Bd}_f, {\rm Bd}_i]
= \int_{\mathcal{C}} {\rm d} N_c \,[\mathbb{G}]
\int_{{\rm Bd}_i}^{ {\rm Bd}_f} {\cal D} q(t) {\cal D} h_{ij}(t, {\bf x})
\exp \left(\frac{i}{\hbar} S_{\rm grav}[q, h_{ij}, N_c] \right) \, ,
\eeq
where ${\rm Bd}_i$ and  ${\rm Bd}_f$ are the initial and final 
boundary conditions. $S_{\rm grav}[q, h_{ij}, N_c]$ is the gravity action with the gauge-fixing condition imposed and is mentioned in 
eq. (\ref{eq:EHact_exp}). $[\mathbb{G}]$ is the ghost determinant arising from Faddeev-Popov gauge-fixing procedure, which has been computed in the appendix \ref{sec:gauge-fixing_ghost}. This ghost operator can be exponentiated via introduction of anti-commuting ghost fields. In subsection \ref{sec:ghost_det} we compute the contribution of ghost-fields to the wavefunction to one-loop. 

To compute the path integral in eq. (\ref{eq:grav_path_qform}) up to one loop, we use the background field formalism \cite{Abbott:1980hw}, where the fields $q(t)$ and $h_{ij}(t,{\bf x})$ are decomposed as
\beq
\label{eq:qh_exp_bg}
q(t) = \bar{q}(t) + \mathsf{\ep}_q Q(t) \, ,
\hspace{5mm}
h_{ij} = \bar{h}_{ij} + \mathsf{\ep}_h H_{ij} \,,
\eeq
where, $\mathsf{\ep}_q, \mathsf{\ep}_h$ are introduced to keep track of perturbation expansion. $\bar{q}$ and $\bar{h}_{ij}$ are the on-shell quantities. Keeping the terms up to $\mathcal{O}(\ep^2_{q}), \mathcal{O}(\ep_{q}\ep_{h})$, and $\mathcal{O}(\ep^2_{h})$, relevant for one-loop computation, we get:
\bea
\label{eq:Sgrav_exp_QH}
&&
S_{\rm grav}[N_c,q,h_{ij}] = 
S_{\rm grav}[N_c, \bar{q}, \bar{h}_{ij}]
+ \int_{\cal M} \frac{{\rm d} t {\rm d} {\bf x} \sqrt{\rho}}{16\pi G}
\biggl[
\ep_q N_c \biggl\{
-2 \Lam Q - \frac{3 \dot{\bar q} \dot{Q}}{N_c^2}
+ \frac{Q \bar{q} \dot{\bar h}_{ij} \dot{\bar h}^{ij}}{2N_c^2}
\notag \\
&&(\varepsilon-1)\biggl(\frac{\dot{\bar{q}}\dot{Q}}{4N_c^2}+\frac{\bar{q}\ddot{Q}+\ddot{\bar{q}}Q}{2N_c^2}-\frac{\Lambda Q}{2}\biggr)N_c\bar{h}_{ij}\bar{h}^{ij}\biggr\} 
+ \ep_h N_c
\biggl\{
-  {\bar h}_{ij} H^{ij}
+ \frac{ \bar{q}^2 \dot{\bar h}_{ij} \dot{H}^{ij}}{2N_c^2}\notag \\
&&+ \frac{\bar{h}_{ij}}{2}  {\bar D}^2 H_{ij}+(\varepsilon-1)\biggl(\frac{\dot{\bar{q}}^2}{4N_c^2}+\frac{\bar{q}\ddot{\bar{q}}}{N_c^2}+1-\Lambda \bar{q}\biggr) \bar{h}_{ij}H^{ij}
\biggr\}
+ \ep_q^2 N_c\biggl\{
-\frac{3\dot{Q}^2}{2N_c^2}
+ \frac{Q^2 \dot{\bar h}_{ij} \dot{\bar h}^{ij}}{4N_c^2} \notag \\
&&
+(\varepsilon-1)\biggl(\frac{\dot{Q}^2}{8N_c^2}+\frac{Q\ddot{Q}}{N_c^2}\biggr)\bar{h}_{ij}\bar{h}^{ij}\biggl\}
+ \ep_h \ep_q N_c
\biggl\{
 \frac{\bar{q} Q \dot{\bar h}_{ij} \dot{H}^{ij}}{N_c^2}+(\varepsilon-1)\biggl(\frac{\dot{\bar{q}}\dot{Q}}{2N_c^2}+\frac{\bar{q}\ddot{Q}+\ddot{\bar{q}}Q}{N_c^2}\notag \\
&&
 -\Lambda Q\biggr) \bar{h}_{ij}H^{ij}
\biggr\}
+\ep_h^2 N_c
\biggl\{
\frac{\bar{q}^2 \dot{H}_{ij} \dot{H}^{ij}}{4N_c^2} 
-\frac{1}{2}H_{ij}H^{ij}+ \frac{H_{ij}{\bar D}^2 H^{ij}}{4}  +(\varepsilon-1)\biggl(\frac{\dot{\bar{q}}^2}{8N_c^2}+\frac{\bar{q}\ddot{\bar{q}}}{2N_c^2}\notag \\
&&+\frac{1-\Lambda \bar{q}}{2}\biggr)H_{ij}H^{ij}
\biggr\}
\biggr] + \cdots \, ,
\eea
where $S_{\rm grav}[N_c, \bar{q}, \bar{h}_{ij}]$ refers to on shell action. The on-shell equation for $q$ and $h_{ij}$ are obtained from vanishing the linear order terms $\mathcal{O}(\ep_{q}),\mathcal{O}(\ep_{h})$. As the gravity action in eq. (\ref{eq:EHact_exp}) is non-linear with the mixing terms between $q$ and $h_{ij}$, these on-shell equations are coupled. This makes it difficult to solve these equations. However, assuming there is no `` back reaction" ($h_{ij}$ doesn't affect the background $q$), the on-shell equations decouple, and they are given by eqs. (\ref{eq:dyn_q_eq}) and (\ref{eq:hijEQM}). These can be solved exactly. Moreover, following the background field formalism, $Q(t)$ and $H_{ij}(t,{\bf x})$ don't affect the background fields. The remaining quadratic terms in the fluctuations are non-diagonal with non-vanishing $\mathcal{O}(\ep_{q}\ep_{h})$. In the special case when $\bar{h}_{ij}(t,\bf{x})=0$ (only quantum field $H_{ij}(t, {\bf x})$ propagates virtually), the cross-term vanishes, making it diagonal. In this case, eq. (\ref{eq:grav_path_qform}) becomes the following:
\bea
\label{eq:grav_path_qhform_1}
\mathcal{Z}[{\rm Bd}_f, {\rm Bd}_i] \Br \rvert_{\bar{h}_{ij}(t,{\bf x})=0}
= &&
\int_{\mathcal{C}} {\rm d} N_c \,[\mathbb{G}]
\int_{{\rm Bd}_i}^{{\rm Bd}_f} {\cal D} q(t) 
\exp\left(\frac{i}{\hbar} S^{(q)}_{\rm grav}[q, N_c] \right)
\notag \\
&& \times
\int_{{\rm Bd}_i}^{ {\rm Bd}_f} 
{\cal D} H_{ij}(t, {\bf x})
\exp \left(\frac{i}{\hbar} S^{(h)}_{\rm grav}[\bar{q}, H_{ij}, N_c] \right) \, ,
\eea
where 
\bea
\label{eq:act_SH_begexp}
S^{(h)}_{\rm grav}[\bar{q}, H_{ij}, N_c]
=&& 
\int_{\cal M} \frac{{\rm d} t {\rm d} {\bf x} \sqrt{\rho}}{16\pi G}
N_c
\biggl[
 \frac{\bar{q}^2 \dot{H}_{ij} \dot{H}^{ij}}{4N_c^2} 
-\frac{1}{2}H_{ij}H^{ij}+ \frac{H_{ij}{\bar D}^2 H^{ij}}{4}   +(\varepsilon-1)\biggl(\frac{\dot{\bar{q}}^2}{8N_c^2}\notag \\
&&+\frac{\bar{q}\ddot{\bar{q}}}{2N_c^2}+\frac{1-\Lambda \bar{q}}{2}\biggr)H_{ij}H^{ij}
\biggr] \, .
\eea
The path integrals can be done exactly for a variety of boundary choices, leading to a one-loop effective action for the lapse integration. The integration can be solved using the Picard-Lefschetz method and WKB approximation.

\subsection{Path integral over $q(t)$}
\label{SubS:PI_qt}

First, we compute the path integral over $q(t)$, appeared in eq. (\ref{eq:grav_path_qhform_1}). It can be computed exactly for a limited set of boundary choices, which are ``linear", including Dirichlet, Neumann and Robin. These were computed in \cite{Feldbrugge:2017kzv, Narain:2021bff, Narain:2022msz, Ailiga:2023wzl} and we just mention the results. These are given by \cite{Ailiga:2024wdx}
\bea
\label{eq:q_pathInt_BC}
\int_{{\rm Bd}_i}^{{\rm Bd}_f} {\cal D} q(t) \,
\exp\bl[\frac{i}{\hbar} S_{\rm grav}^{(q)}\br]
= \bl\{\D_q(N_c) \br\}^{-1/2} \exp\Bl[\frac{i}{\hbar} S_{\rm grav}^{(\bar{q})}(N_c) \Br] 
= e^{i{\cal A}_{q}(N_c)/\hbar}\, ,
\eea
where $\D_q^{\rm DBC}(N_c) = N_c$, $\D_q^{\rm RBC}(N_c) = 1+2N_c\boldsymbol{\beta}/3$ 
are the one-loop prefactors computed via Gel'fand-Yaglom or Van Vleck determinant method. On-shell action $S_{\rm grav}^{(\bar{q})}(N_c)$ for
fixed Robin and fixed extrinsic curvature
are mentioned in Eqs. (\ref{eq:stot_onsh_rbc}) and (\ref{eq:stot_onsh_fixed_k}) respectively. In the limit $\boldsymbol{\beta}$ goes to zero, one obtains the expressions for the Neumann condition. $\D_q$'s are computed when boundary choices at both ends are ``linear". In fact, we kept the final boundary condition as Dirichlet. In the case when we fix the extrinsic curvature at the final end, the boundary choice becomes non-linear! (see eq. (\ref{eq:fixing_K})). This makes it difficult to compute the determinant using the available methods. Technically, the differential operator doesn't become entirely a bulk term as the quadratic surface terms remain non-vanishing. One requires non-linear boundary choices to make these terms vanish. We will not explicitly compute this in the paper. As noted in \cite{Ailiga:2024wdx}, the one-loop prefactor from the background neither affects the UV nor IR understanding of the path integral at the saddles. Hence, the prefactor is not relevant for the issues addressed in this paper. We exponentiate $\D_q(N_c)$ to obtain the quantum-corrected action for the lapse
$N_c$. This is given by
\beq
\label{eq:quant_Nc_act_q}
{\cal A}_{q}(N_c) = S_{\rm grav}^{(\bar{q})}(N_c)
+ \frac{i \hbar}{2} \ln \D_q(N_c) \, . 
\eeq
%

\subsection{Path integral over $H_{ij}$}
\label{SubS:PI_hij}
Having computed the path integral over background, we now proceed to compute the path integral over metric fluctuations ($h_{ij}$). As explained earlier to avoid mixing in the non-diagonal terms, we choose the on-shell fluctuations to vanish. In this approximation, the path integral over $h_{ij}$ reduces to the path integral over $H_{ij}$ as given in eq. (\ref{eq:grav_path_qhform_1}). We proceed by performing a mode decomposition using eq. (\ref{eq:h_ij_Sph_Har}), under which the path integral measure becomes
\beq
\label{eq:PI_hij_mea}
[{\cal D} H_{ij}]=\prod_{l=2}^\infty \,\, \prod_{n=2}^l \,\, \prod_{m=-n}^n \,\, 
{\cal D} H^l_{nm}(t).
\eeq
The action mentioned in eq. (\ref{eq:act_SH_begexp}) becomes the sum over action for individual modes ($l,m,n$):
\bea
\label{eq:mode_hL_act}
S^{(h)}_{\rm grav}
= \frac{1}{16\pi G} \sum_{l=2}^\infty \,\, \sum_{n=2}^l \,\,\sum_{m=-n}^n &&\int_0^1 {\rm d}t
\Bl[
\frac{\bar{q}^2}{4N_c}\bl(\dot{H}^l_{mn}\br)^2
-
\biggl\{\frac{1}{4}l(l+2)-(\varepsilon-1)\biggl(\frac{\dot{\bar{q}}^2}{8N_c^2}\\
\notag
&&+\frac{\bar{q}\ddot{\bar{q}}}{2N_c^2}+\frac{1-\Lambda \bar{q}}{2}\biggr)\biggr\}N_c\bl(H^l_{mn}\br)^2
\Br] \, ,
\eea
Eq. (\ref{eq:mode_hL_act}) is the sum of the action of a collection of ``free" harmonic oscillators with non-trivial time-dependency. The path integral can be solved exactly for the Dirichlet-type boundary condition. We also note that in situations, when either $\varepsilon=1$ (exponential parametrization) or the quantity multiplied to $\varepsilon$ vanishes, the action resembles the action of a massless minimally coupled scalar mode on the background $\bar{q}$ \cite{Grishchuk:1974ny}. This point needs further study in future. Taking care of degeneracy ($g_l$) arising from the spherical symmetry, the path integral evaluates to
\bea
\label{eq:h_PI_mode_degen}
\int_{{\rm Bd}_i}^{ {\rm Bd}_f} 
{\cal D} H_{ij}(t, {\bf x}) 
&&
\exp \left(\frac{i}{\hbar} S^{(h)}_{\rm grav}[\bar{q}, H_{ij}, N_c] \right) 
\notag \\
&&
= \prod_{l=2}^\infty
\biggl[
\int_{{\rm Bd}_i}^{{\rm Bd}_f} 
{\cal D} H_l(t) \exp\left(\frac{i}{\hbar} S^{(l)}_{\rm grav}[\bar{q}, H_l, N_c] \right) 
\Br]^{g_l} 
= e^{i {\cal A}^\varepsilon_{h}(N_c)/\hbar}\, ,
\eea
where $H_{mn}^l(t)\equiv H_l(t)$ as the modes are independent of $m$ and $n$. ${\rm Bd}_i$ and ${\rm Bd}_f$ are $H_l(0)=0$ and $H_l(1)=0$, respectively. Eq. (\ref{eq:h_PI_mode_degen}) is the infinite product of one-loop prefactors for each harmonic oscillator. The path integral over each $l$-mode takes the following form

\beq
\label{eq:hL_PI_stru}
\int_{{\rm Bd}_i}^{{\rm Bd}_f} 
{\cal D} H_l(t) \exp\Bl(\frac{i}{\hbar} 
S^{(l)}_{\rm grav}[\bar{q}, H_l, N_c] \Br)
= \bl\{\D^{(l)}_h(N_c) \br\}^{-1/2}\, ,
\eeq
where, $\D^{(l)}_h(N_c)$ is the functional determinant of one-dimensional self-adjoint Sturm-Liouville operator ($\mathcal{D}_{h}^l$):
\begin{equation}
    \label{eq:determin}
    \D^{(l)}_h(N_c) =\det\biggl[-\frac{d}{dt}\left(\frac{\bar{q}^2}{N_c}\frac{d}{dt}\right)-N_c\biggl[l(l+2)-4(\varepsilon-1)\biggl(\frac{\dot{\bar{q}}^2}{8N_c^2}+\frac{\bar{q}\ddot{\bar{q}}}{2N_c^2} +\frac{1-\Lambda \bar{q}}{2}\biggr)\biggr].
\end{equation}

In the following section, we compute the determinant of eq. (\ref{eq:determin}) explicitly using the Gel'fand-Yaglom method. 

\subsection{One-loop determinant via Gel'fand-Yaglom method}
\label{sec:GY_method}

In this section, we will compute the functional determinant appearing in eq.(\ref{eq:determin}) using the Gel'fand-Yaglom (GY) method \cite{Gelfand:1959nq}. 
This method is based on the generalized zeta function and can be used for a generic one-dimensional self-adjoint Sturm-Liouville operator with the general ``linear" boundary conditions \cite{Dunne:2007rt,Kirsten:2003py, Kirsten:2004qv}. This technique has previously been applied in the context of minisuperspace cosmology and gravity path integral \cite{Ailiga:2024wdx, Matsui:2025guo}. According to the GY method (see appendix \ref{sec:GY_append} for details of the method), the determinant of the most general second-order Sturm-Liouville operator ($\mathcal{D}$)
\begin{equation}
\label{eq:STL_opp_main}
\mathcal{D}=-\frac{d}{dt}\left(P(t)\frac{d}{dt}\right)+Q(t) \,, \quad t\in[0,1]
\end{equation} where $t$ is normalized to be bounded between $[0,1]$ is given by
\begin{equation}
\label{eq:detMN_GY}
\det{\mathcal{D}}=\det\left[M+N\left(\begin{matrix}
u_1(1) &u_2(1)\\
v_1(1) & v_2(1)
\end{matrix}\right)\right] \, ,
\end{equation}
where, $u_1(t)$ and $u_2(t)$ are the two independent solutions to $\mathcal{D}u_i(t)=0$,
which are zero eigenvalue ($\lam=0$) solutions of Eq. (\ref{eq:eigenEQ}), with initial conditions obtained from $E_{(\lam =0)}=I_2$:
\begin{equation}
\label{eq:INC_cond_u1u2}
\begin{split}
&u_1(0)=1, \quad\quad  \dot{u}_1(0)=0\\
&u_2(0)=0, \quad\quad\dot{u}_2(0)=1/P(0) \, .
\end{split}
\end{equation}
The matrices $M$ and $N$ depend on the boundary conditions imposed on the eigenvalue problem (see appendix \ref{sec:GY_append}).

%
For the differential operator in eq. (\ref{eq:determin}) $P(t)$ and $Q(t)$ are given by 
\begin{equation}
\begin{split}
    P(t)=\frac{\bar{q}^2}{N_c},\hspace{5mm}
    Q(t)=- N_c\biggl[l(l+2)-4(\varepsilon-1)\biggl(\frac{\dot{\bar{q}}^2}{8N_c^2}+\frac{\bar{q}\ddot{\bar{q}}}{2N_c^2} +\frac{1-\Lambda \bar{q}}{2}\biggr)\biggr],
\end{split}
\end{equation}
where $\bar{q}$ mentioned in eq. (\ref{eq:qsol_gen}) for a generic boundary condition.
For the ``Dirichlet-Dirichlet" boundary condition on $H_{ij}$, $M$ and $N$ matrices are given by
\begin{equation}
\label{eq:M_and_N_dirich_matric}
M=\left(\begin{matrix}
    1&0\\
    0&0
\end{matrix}\right),\hspace{5mm}N= \left(\begin{matrix}
    0&0\\
    1&0
\end{matrix}\right) \, .
\end{equation}
Hence the determinant $\det\mathcal{D}_{h}=u_2(1)$. Solving $\mathcal{D}_{h}u_2=0$ with the initial condition mentioned, we get
\begin{equation}
\label{eq:u_2(t)}
u_2(t)=\frac{N_c\left(\mathbb{P}_1^{-\xi_l}(\tau_t)\mathbb{Q}_1^{\xi_l}(\tau_0)
-\mathbb{Q}_1^{\xi_l}(\tau_t)\mathbb{P}_1^{-\xi_l}(\tau_0)\right)}{\sqrt{q(t)}\,\bar{q}(0)^{3/2}
\left(\dot{\mathbb{P}}_1^{-\xi_l}(\tau_0)\mathbb{Q}_1^{\xi_l}(\tau_0)
-\dot{\mathbb{Q}}_1^{\xi_l}(\tau_0)\mathbb{P}_1^{-\xi_l}(\tau_0)\right)},
\end{equation}
where, $\xi_l$ and $\tau_t$ are given in eq. (\ref{eq:hl_sol_gen}). Utilizing the Wronskian identity  $\mathcal{W}\{\mathbb{P}_{\nu}^{-\mu}(x),\mathbb{Q}_{\nu}^\mu(x)\}=-e^{i\pi\mu}/(x^2-1)$ \cite{dlmf},
the denominator in eq. (\ref{eq:u_2(t)}) is simplified to
\begin{equation}
\label{eq:wronskian_2}
    \,\bar{q}(0)\left(\dot{\mathbb{P}}_1^{-\xi_l}(\tau_0)\mathbb{Q}_1^{\xi_l}(\tau_0)
    -\dot{\mathbb{Q}}_1^{\xi_l}(\tau_0)\mathbb{P}_1^{-\xi_l}(\tau_0)\right)= N_c\, \mathbb{M}_\varepsilon(\xi_{l},N_{c}), 
\end{equation}
where, $\mathbb{M}_\varepsilon(\xi_{l},N_c)= e^{i(\xi_{l}+1)\pi}\Lam N_c (r_1 - r_2)/6$.
Finally, the determinant in eq. (\ref{eq:determin}) simplifies to
\begin{equation}
\D^{(l)}_h(N_c)=u_2(1)
=\frac{W_l^{\varepsilon}(1,0)}{\mathbb{M}_\varepsilon(\xi_{l},N_c)\sqrt{\bar{q}(1)\bar{q}(0)}}\, ,
\end{equation}
where, $W_l^{\varepsilon}(1,0)$ is given in eq. (\ref{eq:W_l10_func}).
The quantum corrected lapse action, as defined in eq. (\ref{eq:h_PI_mode_degen}), for a generic boundary condition, it is given by 
\bea
\label{eq:QT_Nc_act_h_divST}
{\cal A}_{h}^\varepsilon(N_c) = \sum_{l=2}^\infty 
\Bl[
-\frac{i g_l \hbar}{4} \{\ln \bar{q}(1)+\ln\bar{q}(0)\}
+ \frac{i g_l \hbar}{2} \ln W_l^\varepsilon(1,0)-\frac{i g_l \hbar}{2} \ln \mathbb{M}_\varepsilon(\xi_{l},N_c) \Br]\, .
\eea
Eq. (\ref{eq:QT_Nc_act_h_divST}) contains the contribution from one-loop fluctuation coming from all the oscillator modes in $h_{ij}$. It depends on both the background boundary conditions (via $r_1$ and $r_2$) and parametrization ($\varepsilon$).

\subsection{Path integral over ghosts}
\label{sec:ghost_det}

In this section, we will compute the ghost determinant ($[\mathbb{G}]$) up to one-loop appearing in eq. (\ref{eq:grav_path_qhform_1}), see appendix \ref{sec:gauge-fixing_ghost} for details of the gauge fixing and ghost determinant via the Faddeev-Popov method. The one-loop ghost operator is given by
\begin{equation}
\label{eq:ghost_comp_one_loop}
    \Delta^{1-\rm loop}_{ij}=\rho_{ij}\bar{\nabla}^2+\bar{\nabla}_j\bar{\nabla}_i, \hspace{5mm} [\mathbb{G}]=\det[\Delta^{1-\rm loop}_{ij}]
\end{equation}
For the restricted set of diffeomorphisms we consider, the operator is defined on the comoving hypersurfaces on which the ghost fields live. Hence $c_i,\bar{c}_i$ are functions of ${\bf x}$ only. The $[-\Delta^{1\rm -loop}_{ij}]$ leads to positive definite eigenvalues (upon orthogonal decomposition), except for the negative and zero modes. Appearance of zero mode (related to the Killing symmetries of $S^3$)
signifies the presence of the residual gauge freedom. Constant scalar mode appears as a negative mode. In the Lorentzian signature, the determinant is given by
\begin{equation}
   \det \Delta^{1-\rm loop}_{ij}
   =\int \mathcal{D}\bar{c}_i\mathcal{D}c_ie^{ i\int d^3x\sqrt{\rho}\,\bar{c}^i (\rho_{ij}\bar{\nabla}^2+\bar{R}_{ij}+\bar{\nabla}_i\bar{\nabla}_j)c^j} \,,
\end{equation}
where we utilized the commutation relation $[\bar{\nabla}_i,\bar{\nabla}_j] c^k=\bar{R}^k_{\,\,\,nij}c^n$, where $\bar{R}^{k}_{\,\,\,nik}$ is the Riemann tensor on $S^3$.
We decompose the ghost fields as $c_i=c_i^{T}+\bar{\nabla}_ic,\,\,\,\bar{c}_i=\bar{c}^{T}_i+\bar{\nabla}_ic$, where $\bar{\nabla}^ic^{T}_i=\bar{\nabla}^i\bar{c}^{T}_i=0$.
Taking care of the Jacobian $[{\rm det} (-\bar{\nabla}^2)]^{-1}$ and noting that
\begin{equation}
    \bar{c}^i (\rho_{ij}\bar{\nabla}^2+\bar{R}_{ij}+\bar{\nabla}_i\bar{\nabla}_j)c^j=\bar{c}^{iT}\rho_{ij}(\bar{\nabla}^2+\bar{R}/3)c^{jT}-2\bar{c}(\bar{\nabla}^2)(\bar{\nabla}^2+\bar{R}/3)c,
\end{equation}
the determinant simplifies to
\begin{equation}
\begin{split}
      \det \Delta^{1-\rm loop}_{ij}&=\int \mathcal{D}\bar{c}_i^T\mathcal{D}c_i^T e^{- i\int d^3x\sqrt{\rho}\,\bar{c}^{iT} \rho_{ij}[\Delta_{(1)}]c^{jT}}\int \mathcal{D}\bar{c}\mathcal{D}c e^{-i\int d^3x\sqrt{\rho} \bar{c}[\Delta_{(0)}]c},\\
     \end{split}
\end{equation}
where, the spin-1 and spin-0 operators are $\Delta_{(1)}=\Delta_{(0)}=-\bar{\nabla}^2-2$. Performing the mode decomposition over the harmonics of $S^3$ and doing the spatial integrations, the determinant reduces to
\begin{equation}
\label{eq:one-loop_det}
\begin{split}
   \det \Delta^{1-\rm loop}_{ij}=(\text{Z.M})&\prod_{l=2}^\infty\left[\int d\bar{c}^T_l d c^T_le^{- i\bar{c}^T_l(\lambda_l^{(1)}-2)c^T_l}\right]^{d_l^{(1)}}\times\left[\int  d\bar{c}_ld c_le^{- i\bar{c}_l(\lambda_l^{(0)}-2)c_l}\right]_{l=0}^{d_l^{(0)}}\\
   &\times\prod_{l=1}^\infty\left[\int d\bar{c}_ld c_le^{- i\bar{c}_l(\lambda_l^{(0)}-2)c_l}\right]^{d_l^{(0)}},
    \end{split}
\end{equation}
where ``Z.M" is the zero mode ($l=1$) contribution for the transverse vector, and in the scalar part, we have separated the negative mode ($l=0$). The eigenvalues $\lambda_l^{(1)},\lambda_l^{(0)}$ and $d_l^{(1)},d_l^{(0)}$ are the degeneracies of the laplacian $-(\bar{\nabla}^2)$ for spin-1 and spin-0 fields, respectively. They are given by
\begin{equation}
\label{eq:eigenvalue}
    \begin{split}
&d_l^{(1)}=2l(l+2),\lambda_l^{(1)}=[l(l+2)-1],l\geq 1\\
&d_l^{(0)}=(l+1)^2,\lambda_l^{(0)}=[l(l+2)],\,\,\,\,l\geq 0.
    \end{split}
\end{equation}
Evaluating the integrals in eq. (\ref{eq:one-loop_det}) (dropping Z.M term), we get
\begin{equation}
\begin{split}
    \det \Delta^{1-\rm loop}_{ij}&=(- 2i)\prod_{l=2}^\infty\left[ i(\lambda_l^{(1)}-2)\right]^{d_l^{(1)}}\times\prod_{l=1}^\infty\left[ i(\lambda_l^{(0)}-2)\right]^{d_l^{(0)}},
    \end{split}
\end{equation}
where $- 2i$ comes from the negative eigenvalue. Remarkably, with zeta regularization, all the factors of $i$ cancel, giving
\begin{equation}
    \label{eq:phase_cancel}
    \begin{split}
    (- i)\prod_{l=2}^\infty ( i)^{d_l^{(1)}}\prod_{l=1}^\infty (i)^{d_l^{(0)}}=(- i)\exp\left(\sum_{l=2}^\infty d_l^{(1)}\ln( i)+\sum_{l=1}^\infty d_l^{(0)}\ln( i)\right)=1.
\end{split}
\end{equation}
This makes the ghost contribution real and positive. This result agrees with \cite{Polchinski:1988ua, Ivo:2025yek},  where it is suggested that one should take the absolute value of the determinant to account for the correct phase of the sphere partition function.
On writing $\det \Delta^{1-\rm loop}_{ij}=e^{i\mathcal{A}_h^{\rm ghost}/\hbar}$, we get
\begin{equation}\label{eq:ghost_Action}
\begin{split}
\mathcal{A}_h^{\rm ghost}&= 
-i\hbar\biggl[\sum_{l=2}^\infty d_l^{(1)}\ln[(l+3)(l-1)]+\sum_{l=1}^{\infty}d_l^{(0)}\ln[(l+1+\sqrt{3})(l+1-\sqrt{3})]+\ln 2\biggr],
\end{split}
\end{equation}
where we utilized the eigenvalues $\lambda_l^{(1)},\lambda_l^{(0)}$ from eq (\ref{eq:eigenvalue}). The infinite sums in the above equation are divergent and require regularization. We evaluate the finite part by employing the generalized Zeta-regularization \cite{Elizalde:1994gf, Monin:2016bwf}, which extracts the physically meaningful finite part. This can be computed exactly. Expanding the above logarithms and apply regularization for each term (see appendix \ref{sec:zeta_sum}), we get 
\begin{equation}
    \label{eq:zeta_reg_ghost}
\begin{split}
    &\mathcal{A}_h^{\rm ghost}\biggr|_{\zeta-{\rm reg}}=\frac{-i\hbar}{\pi ^2}\biggl[\zeta (3)-\pi ^2 \biggl\{\zeta'_H(-2,2-\sqrt{3})+\zeta'_H(-2,\sqrt{3}+2)\\
    &+2 \sqrt{3} \zeta'_H(-1,2-\sqrt{3})-2 \sqrt{3} \zeta'_H(-1,\sqrt{3}+2)+3 \zeta'_H(0,2-\sqrt{3})\\
    &+3 \zeta'_H(0,\sqrt{3}+2)+\log (8)-6 \log (\pi )\biggr\}\biggr]\approx -4.92469i\hbar.
    \end{split}
\end{equation}
Hence, the ghost contribution in the path integral $e^{i\mathcal{A}_h^{\rm ghost}/\hbar}$ is a real constant (independent of $N_c$) and enhancing in nature. Since, it is independent of $N_c$, it will play no role in the following Picard-Lefschetz analysis.
\section{Saddles stability, allowability and $N_c$-integral}
\label{sec:lapseNc}
After computing the path integral over the background $q(t)$, fluctuations $h_{ij}$ (up to one loop) and the ghosts, we are left with an ordinary integral over $N_c$ as mentioned in eq. (\ref{eq:grav_path_qform}). It is given by 
\beq
\label{eq:grav_path_NC_form}
\mathcal{Z}_\varepsilon[{\rm Bd}_f, {\rm Bd}_i]
= \int_{\mathcal{C}} {\rm d} N_c \,\,
\exp \bl\{i {\cal A}(N_c,\varepsilon)/\hbar \br\} \, ,
\eeq
where we introduce the $\varepsilon$ explicitly to study the parametrization dependence. The lapse effective action ${\cal A}(N_c,\varepsilon)$ contains a semiclassical piece and a $\mathcal{O}(\hbar)$ correction arising from one loop determinant from both the $q$ and $h_{ij}$-path integral. Before we consider the one-loop effects, in this section we first analyze the semiclassical part appearing in the $N_c$ integral and analyze the stability of the metric fluctuation at the saddles. 

The lapse effective action ${\cal A}(N_c,\varepsilon)$ in eq. (\ref{eq:grav_path_NC_form}) is given by
\bea
\label{eq:Nc_act_total}
&&
{\cal A}(N_c,\varepsilon) = {\cal A}_{q}(N_c) + {\cal A}_{h}^\varepsilon(N_c)+\mathcal{A}_h^{\rm ghost}
= S_{\rm grav}^{(\bar{q})}(N_c)
+  \frac{i \hbar}{2} \ln \D_q(N_c) 
\notag \\
&&
- \frac{2i \hbar}{3} \{\ln \bar{q}(1)+\ln\bar{q}(0)\} 
+ \frac{i \hbar}{2} \sum_{l=2}^\infty 
g_l  \ln W_l^\varepsilon(1,0)-\frac{i\hbar}{2} \sum_{l=2}^\infty 
g_l  \ln \mathbb{M}_\varepsilon(\xi_{l},N_c)+\mathcal{A}_h^{\rm ghost} \notag \\
&&
= {\cal A}_0(N_c) + i \hbar {\cal A}_1(N_c,\varepsilon) +\mathcal{A}_h^{\rm ghost}
\, ,
\eea
where ${\cal A}_0(N_c)$ is the classical action, ${\cal A}_1(N_c) $ is the one-loop correction and $\mathcal{A}_h^{\rm ghost}$ is the ghost contribution (independent of $N_c$) \footnote{Here, we have utilized the Zeta-regularization \cite{Hawking:1976ja,Elizalde:1994gf} to sum
\beq
\label{eq:reg_sum_gl}
\notag
\sum_{l=2}^\infty g_l=\sum_{l=2}^\infty \frac{g_l}{l^s}\biggr|_{s=0^+} = 2 \zeta(-2) + 4 \zeta(-1) - 6\zeta(0) = 8/3 \, .
\eeq
}.
The integral in eq. (\ref{eq:grav_path_NC_form}) can only be solved using the saddle-point approximation once the contour $\mathcal{C}$ is specified. We take this to be the real axis (fixed final size of the universe) or the axis parallel to it (fixed final Hubble rate) \cite{Feldbrugge:2017kzv, DiTucci:2019bui}. 
The expressions for quantum correction $\ln \D_q(N_c)$ are known in literature \cite{ Narain:2022msz, Ailiga:2023wzl, Ailiga:2024mmt}. The contribution from the $H_{ij}$ is obtained earlier for the $\varepsilon=0$ parametrization in \cite{Ailiga:2024wdx}. It is important to note that eq. (\ref{eq:Nc_act_total}) is obtained for a generic boundary choice on the background.

Similar one-loop studies in cosmological systems have been attempted in \cite{Barvinsky:1992dz}. The authors worked in the Euclidean path-integral setting, which involved computing the zeta function on Euclidean $S^4$ and then analytically continuing to the Lorentzian time. They fixed the lapse to be unity. In our paper, we go beyond by considering all geometries that respect the $\mathbb{R}\times S^3$ symmetry, along with fluctuations over $S^3$ (in arbitrary parametrization). Our analysis essentially involves complex geometries via the $N_c$-integral. We evaluate the one-loop determinant at the complex saddle. We compute it exactly, whereas the analysis performed in \cite{Barvinsky:1992dz} was only asymptotic in Lorentzian time.

\subsection{One-loop expansion}
\label{PL_int} 

With the choices of contour specified, the integral in eq. (\ref{eq:grav_path_NC_form}) is highly oscillatory. To deal with such a conditionally convergent integral, we use the Picard-Lefschetz (PL) method (see \cite{Witten:2010cx, Lehners:2023yrj, Feldbrugge:2017kzv, Ailiga:2024mmt} for review on the PL method). In this method, one deforms the original integration contour $\mathcal{C}$ into a sum over steepest descent ($\mathcal{J}_\sigma$) thimbles passing through the saddles ($\sigma$) such that $\mathcal{C}=\sum_\sigma n_\sigma \mathcal{J}_\sigma$, where $n_\sigma={\rm Int} (\mathbb{D}, {\cal K}_\sg)$.
The integral over each thimble ($\mathcal{J}_\sigma$) is absolutely convergent. $\rm Int(. , .)$ counts the intersection between two curves and $n_\sigma=(0,\pm1)$. If $n_{\sigma}=0$, the saddle is irrelevant and hence, the deformed contour of integration doesn't pass through the saddle. Once the $n_\sigma$'s are determined, the integral in eq. (\ref{eq:grav_path_NC_form}) becomes
\beq
\label{eq:sumOthim}
\mathcal{Z}_\varepsilon[{\rm Bd}_f, {\rm Bd}_i]=\sum_\sg n_\sg \int_{{\cal J}_\sg} {\rm d}N_c
e^{i \mathcal{A}(N_c,\varepsilon)/\hbar}\, .
\eeq
To evaluate the integral using the saddle point, we perform the $\hbar$-expansion of the action and saddles:
\begin{equation}
\label{eq:ANc_1lp_form}
\begin{split}
&{\cal A}(N_c,\varepsilon) = {\cal A}_0(N_c) + i\hbar {\cal A}_1(N_c,\varepsilon) + \cdots \, ,\\
&N_\sg = N_\sg^{(q)} + \hbar \bl(
{\cal N}_\sg^{(q)} + {\cal N}_\sg^{(h,\varepsilon)}
\br) + \cdots \, ,
\end{split}
\end{equation}
where $N_\sg^{(q)}$ is the saddle point obtained from ${\cal A}_0(N_c)$, while $\bl(
{\cal N}_\sg^{(q)} + {\cal N}_\sg^{(h,\varepsilon)}
\br)$ corresponds to one-loop correction. Expanding ${\cal A}(N_c,\varepsilon)$ around the saddle point $N_\sg$, keeping terms up to $\mathcal{O}(\hbar)$, we get \cite{Ailiga:2024mmt}
\beq
\label{eq:LDordI}
\int_{{\cal J}_\sg} {\rm d}N_c\,
e^{i\mathcal{A}(N_c,\varepsilon)/\hbar} 
= \frac{e^{i\theta_\sg}e^{i\mathcal{A}^{\rm ghost}_h/\hbar}}{\sqrt{\bl\vert {\cal A}_0^{\prime\prime} (N_\sg^{(q)}) \br\rvert}}
\times 
\exp\Bl[ 
\frac{i}{\hbar}{\cal A}_0(N_\sg^{(q)})
- {\cal A}_1(N_\sg^{(q)},\varepsilon) + \cdots
\Br] \, ,
\eeq
Here, $\theta_\sigma$ is the direction of descent lines (without loop correction) at the corresponding saddle point ($N_\sigma$). It is given by $ \theta_\sigma =((2k-1)\pi-2\delta_{\sigma})/4$,
where $\delta_{\sg}$ is defined via $\mathcal{A}''(N_\sg)=|\mathcal{A}''(N_\sg)|e^{i\delta_\sg}$. In cases where the denominator in eq. (\ref{eq:LDordI}) vanishes, WKB approximation breaks down, and one needs to go to the cubic and higher orders to compute the wave function \cite{Ailiga:2024mmt}. The path integral in eq. (\ref{eq:sumOthim}) becomes the sum of relevant saddles, which are complex. In the following subsection, we define the boundary conditions ${\rm Bd_i}$ and ${\rm Bd_f}$ along with the background saddles ($N_\sg^{(q)}$).

\subsection{Complex saddles}
\label{subsubS:Nsq}
\begin{figure}[hbtp]
    \centering
\subfigure{\includegraphics[width=0.9\linewidth]{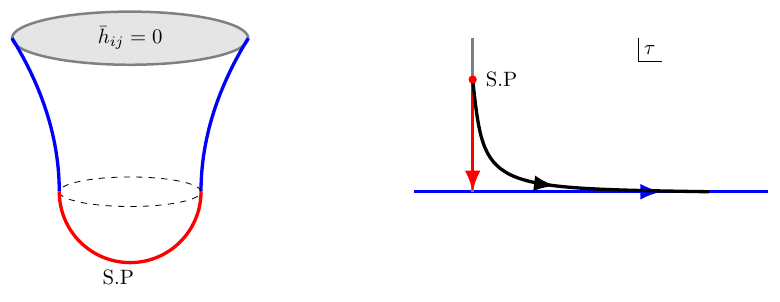}}
    \caption{Showing the background No-boundary universe where, on-shell $\bar{h}_{ij}=0$. The saddle geometry is depicted in the complex-$\tau$ plane, Euclidean (red) plus Lorentzian(blue). The equivalent geometry at the complex saddle is shown in black.}
    \label{fig:geometry_contour}
\end{figure}

To compute the wave function of the universe $\Psi[{\rm Bd_f}]$, we need to fix the boundary conditions ${\rm Bd_i}$ and ${\rm Bd_f}$. Considering the ``no-boundary" wave function \cite{Hartle:1983ai}, we fix the ${\rm Bd_i}$ to be either Neumann or Robin boundary conditions. These boundary choices are motivated by the on-shell fluctuation analysis showing the stable (Gaussian) behaviour \cite{DiTucci:2019bui, Narain:2021bff, Feldbrugge:2017mbc, Narain:2022msz, DiTucci:2019dji, Ailiga:2024wdx, Lehners:2023yrj, DiTucci:2018fdg, Feldbrugge:2017fcc, Ailiga:2024mmt, Matsui:2024bfn}. We fix the final boundary condition ${\rm Bd_f}$ to be either Dirichlet (fixed size) or fixed curvature/Hubble rate. The saddles are obtained by extremizing the on-shell action mentioned in eq. (\ref{eq:stot_onsh_rbc}) and (\ref{eq:stot_onsh_fixed_k}) for Robin-Dirichlet and Neumann-Fixed curvature ($K$) BC, respectively. The saddles are, for Robin-Dirichlet, 
\beq
\label{eq:NNB_sad}
N_{\pm}^{(\rm nb)}=
\frac{3}{\Lam}\Bl(-i\pm \sqrt{\frac{\Lam q_f}{3}-1}\Br),\hspace{4mm} N_{1,2}^{(\rm \cancel{\rm nb})}=-\frac{3}{\beta}
-\frac{3}{\Lambda}\Bl(- i  \pm\sqrt{\frac{\Lambda  q_f}{3}-1} \Br),
\eeq
where we choose $P_i=-3i$ and $\beta$ to be in the range: $\beta=-i\Lambda x/2$, $x>0$, \cite{Ailiga:2023wzl,Ailiga:2024mmt,Ailiga:2024wdx}. In this regime, both the saddles $N_{\pm}^{(\rm nb)}$ are relevant for $q_f>3/\Lam$ and $N_+^{\rm nb}$ is relevant for $q_f<3/\Lam$ according to the Picard-Lefschetz method. In the limit $\beta\rightarrow 0$, we get the saddles for the Neumann-Dirichlet case. For the Neumann-Fixed curvature BC case, the saddles are ($\Lambda=3H^2$)
\begin{equation}
\label{eq:fixed_k_saddle}
N_\pm^{(\rm nb)}=\frac{1}{H^2}\left(-i\pm\frac{H_1}{\sqrt{H^2-H_1^2}}\right).
\end{equation}
 In the Lorentzian regime $H_1<H$, both the saddles $N_\pm^{(\rm nb)}$ are relevant and contribute to the integral. In the Euclidean regime, $H_1$ is purely imaginary with $H_1=+i\gamma H$. For this choice, $ N_+^{(\rm nb)}$ is relevant \cite{DiTucci:2019bui}. We choose the $+$ sign, which is consistent with the fact that we fix the momentum $\pi_q=-3i$ with the ``$-$" sign for the (H-H) universe.
The no-boundary saddles mentioned in eqs. (\ref{eq:NNB_sad}) and (\ref{eq:fixed_k_saddle}) are the background saddles $N_\sg^{(q)}$, used in the one-loop computation in eq. (\ref{eq:LDordI}). The geometries at these saddles are complex; the imaginary part corresponds to the Euclidean section, and the real part to the Lorentzian evolution (see the sec. \ref{sec:comparison_de_sitter}). As shown in the figure. \ref{fig:geometry_contour}, the geometry interpolates between the south pole and the real de Sitter in the far future in the complex time ($\tau$)
 plane \cite{Lehners:2023yrj}.

\subsection{Semiclassical stability of fluctuation with fixed $K$}
\label{subsec:fxed_K_stability}

In this section, we consider $h_{ij}\neq 0$ on the final hypersurface and analyze the semiclassical behaviour of the perturbation. A similar study has been done in the past for a final fixed size of the universe ($q_f$), showing that the semiclassical fluctuations are stable \cite{DiTucci:2019bui, Narain:2021bff, Feldbrugge:2017mbc, Narain:2022msz, DiTucci:2019dji, Ailiga:2024wdx, Lehners:2023yrj, DiTucci:2018fdg, Feldbrugge:2017fcc, Ailiga:2024mmt, Matsui:2024bfn}; for fixed $K$, such a study has not been performed previously. As the extrinsic curvature ($K$) is related to the Hubble parameter ($H_1=K/3$), such a boundary condition is of great importance in cosmology. 
For metric fluctuation, we fix the boundary condition to be Dirichlet-Dirichlet. As explained earlier in sec. \ref{sec:fix-k}, such a boundary condition is allowed only for exponential parametrization. Indeed, as mentioned earlier such parametrization has been found useful in various cosmological studies.

The on-shell solution with the aforementioned boundary conditions is given by eq. (\ref{eq:hl_sol_gen}) with
\beq
\label{eq:hl_dpdq_NBC_nb}
d^{(l)}_{p} =  \frac{ h^{(l)}_1 \sqrt{\bar{q}(1)}}{\mathbb{P}_1^{-l-1}[\tau_1]} \, ,
\hspace{5mm}
d^{(l)}_{q} = 0 \,  ,
\eeq
where $h_l(t=1)= h^{(l)}_1$.
Utilizing the expression of $\mathbb{P}_1^{-l-1}(\tau_{\rm nb})$, the on-shell solution at the no-boundary saddles is given by
\begin{equation}
\label{eq:hLt_nbc_nb_form}
\begin{split}
\bar{h}_l^{\rm nb}(t) &=h^{(l)}_1 \sqrt{\frac{\bar{q}(1)}{\bar{q}_{\rm nb}(t)}}
\Bl(
\frac{1-\tau_{t}^{\rm nb}}{1+\tau_{t}^{\rm nb}}
\Br)^{(l+1)/2}
\Bl(
\frac{1+\tau_{1}^{\rm nb}}{1-\tau_{1}^{\rm nb}}
\Br)^{(l+1)/2}
\Bl(\frac{\tau_{t}^{\rm nb}+l+1}{\tau_{1}^{\rm nb}+l+1}
\Br) \, ,
\end{split}
\end{equation}
where, $\tau_t^{\rm nb}=1-i \Lam N_{\pm}^{\rm (nb)}t/3$ and $\bar{q}(t)$, mentioned in eq. (\ref{eq:qsol_RBC}), at the saddles becomes
\beq
\label{eq:q0ATnb_nbc}
\bar{q}^\pm_{\rm nb}(t) = t \bl\{
\bar{q}^\pm_{\rm nb}(1) t - 2(t-1) i N_{\pm}^{(\rm nb)}
\br\} \, .
\eeq
Note that $h_l^{\rm nb}(t)$ vanishes at the no-boundary saddles as $t^{l/2}$ for all $l\geq 2$.
Evaluating the on-shell action, mentioned in eq. (\ref{eq:hh_onSH_fix_K}), at the no-boundary saddles, we get for each $l$-mode 
\begin{equation} 
\label{eq:hh_onSH_NBC_nb}
\begin{split}
i S^{(l)}_{\rm grav} [\bar{q}, \bar{h}_l, N_{\pm}^{(\rm nb)}] 
&=-\frac{(h^{(l)}_1)^2 l (l+2)}{4 H^2 (l+1)-4 H_1 \left\{\pm i \sqrt{H^2-H_1^2}+H_1( l+1)\right\}}.
\end{split}
\end{equation}
To see the behaviour of fluctuation at the asymptotic future which corresponds to $H_1\rightarrow H$, we do the following expansion
\begin{equation}
    \label{eq:h1_equal_h_limit}
    \begin{split}
 i S^{(l)}_{\rm grav} [\bar{q}, \bar{h}_l, N_{\pm}^{(\rm nb)}] &= -\frac{(h_{1}^{(l)})^2 l (l+1) (l+2)}{4 H^2}\mp i\frac{(h_{1}^{(l)})^2 l (l+2)}{4 \sqrt{2} H^{3/2} \sqrt{H-H_1}}+\mathcal{O}(H-H_1).
 \end{split}
\end{equation}
Note that the semiclassical weight is exactly the same as obtained for fixed size; hence, it is independent of the boundary condition. The negative sign implies stability and corresponds to the Bunch-Davies vacuum. However, the leading phase differs from that in the fixed-size case and depends on the boundary condition. 

\subsection{KSW-allowability of saddles with fixed $K$}
\label{subsec:KSW_criterion}

In dealing with complex geometries, one often encounters configurations that are unphysical and, if included in the path integral, they lead to various pathologies \cite{Jonas:2022uqb, Lehners:2021mah, Hertog:2023vot}. It therefore becomes necessary to analyze whether such geometries are ``allowable", in the sense advocated in \cite{Kontsevich:2021dmb, Witten:2021nzp}, which we refer to as the ``KSW" criterion. In this section, we analyze this KSW-allowability of the no-boundary saddles when the curvature ($K$) is fixed on the final hypersurface. Similar analysis has been performed in \cite{Jonas:2022uqb, Lehners:2021mah, Ailiga:2025fny} when the final size is fixed, and it is found that the no-boundary saddle geometries eq. (\ref{eq:NNB_sad}) are allowed, whereas the non-no boundary geometries are disallowed. Such a criterion indirectly imposes a strong restriction on the allowed boundary conditions (see sec. \ref{sec:comparison_de_sitter} for an example of a boundary condition at odds with KSW) 
and it is important to check whether the fixing $K$ is compatible with it.

\begin{figure}[hbtp]
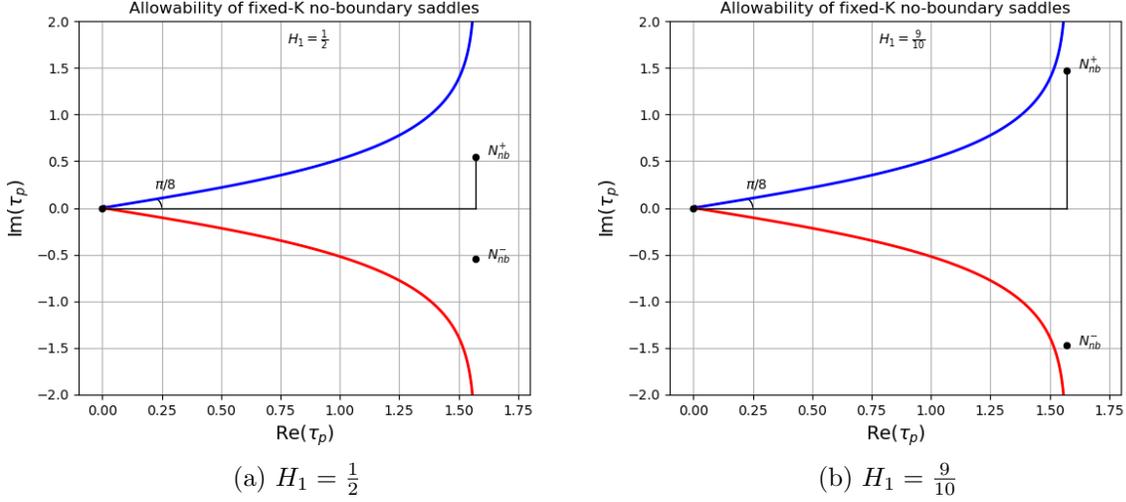

    \centering
\subfigure[$\,\,H_1=\frac{1}{2}$]{\includegraphics[width=0.49\linewidth]{filename2.pdf}}
\subfigure[$\,\,H_1=\frac{9}{10}$]{\includegraphics[width=0.49\linewidth]{filename3.pdf}}
    \caption{The plot showing the KSW allowability of no-boundary saddles for different values of $H_1<H$, with $H=1$. The extremal curves are plotted in blue and red lines with the angle $\pi/8$ near the origin. The saddles are allowed as there is an available time path connecting the origin to the final hypersurface (black dot) without crossing the extremal curve. As $H_1$ increases, the black dot moves closer to the extremal curve, but always stays below it.}
    \label{fig:extremal_curve_fix_K}
\end{figure}

Consider the complex metric $g_{\mu\nu}(x)$ on a $\mathfrak{D}\geq2$ dimensional manifold ($\mathcal{M}$). We define the Euclidean path integral of a $p$-form gauge field $A$ with the field strength $F=dA$, which is a $q=p+1$ form. According to the KSW allowability criterion, $g_{\mu\nu}$ is allowable, if it satisfies the condition \cite{Kontsevich:2021dmb,Witten:2021nzp}

\bea
\label{KSW}
\begin{split}
\mathcal{I}_q[A]& =\frac{1}{2q!}\int_M d^Dx\sqrt{det\,g}\,\,g^{\mu_1 \nu_1}\cdots g^{\mu_q \nu_q} F_{\mu_1 \mu_2...\mu_q} F_{\nu_1 \nu_2...\nu_q} \, ,\\
g_{\mu\nu} & \, \text{is KSW allowable iff} \, \,\, {\rm Re}\left(\sqrt{det\,g}\,\, g^{\mu_1 \nu_1}\cdots g^{\mu_q \nu_q}F_{\mu_1 \mu_2...\mu_q} F_{\nu_1 \nu_2...\nu_q}\right) > 0 \, ,
\end{split}
\eea
for all $q \in \{0, ... , D\}$, where $\mathcal{I}_q[A]$ is the Euclidean action. For the metrics which are diagonal in some real basis, i.e., $g_{\mu\nu}=\lambda_{\mu}(x)\delta_{\mu\nu}$, the criterion simplifies to 
\beq
\label{eq:KSW_criterion}
\Sigma \equiv \sum_{\mu=0}^{D-1}|\arg \, \lam_{\mu}(x)| < \pi \quad \forall x \in \mathcal{M}\, ,
\eeq
where $z$ is complex and $\arg (z) \in (-\pi,\pi]$. To analyze the allowability of the saddle geometries, we follow the method of ``extremal curve" \cite{Hertog:2023vot, Ailiga:2025fny}.
To start with, we define a Euclidean time coordinate $\tau_p$, such that $\tau_p(0)=0$ and $\tau_p(1)=\nu$. It is defined as 
\beq
\label{eq:Euc_phy_time}
{\rm d} \tau_{p} = i \frac{N_{c}}{\sqrt{q(t)}} {\rm d}t \, ,
\eeq
where $\bar{q}(t)$ is given in eq. (\ref{eq:qsol_RBC}).
With the new time-coordinate the metric becomes ($h_{ij}=0$)
\beq
\label{eq:frwmet_tau_p}
{\rm d}s^2 =  {\rm d} \tau_{p}^2 
+ q(\tau_{p}) \rho_{ij}dx^idx^j \, , \quad \because q(t(\tau_{p})) \equiv q(\tau_{p}) \, .
\eeq
At the no-boundary saddles (eq. (\ref{eq:fixed_k_saddle})), $\tau_p(t)$ and $q(\tau_p)$ take the following form
\begin{equation}
    \label{eq:time-path_saddle}
   \tau_p(t)= \sqrt{\frac{3}{\Lambda}}\left[\frac{\pi}{2} + i \sinh ^{-1}\left(\frac{\Lambda  N_{\rm nb} t}{3}+i\right)\right],\hspace{4mm}q(\tau_p)=\frac{3}{\Lambda }\sin ^2\left(\sqrt{\frac{\Lambda}{3}}\,\tau_p\right).
\end{equation}
Note that for small $\tau_p$, $q\sim \tau_p^2$ and we get Euclidean $S^4$.
We complexify the times $\tau_p$ keeping the angular part real, and consider the time contour $\tau_p(N_c,u)$, $u$ being a real parameter, such that the metric 
reads as
\beq
\label{eq:frwmet_changed_comp_t}
{\rm d}s^2 = \biggl(\frac{{\rm d}\tau_{p}}{{\rm d}u}\biggr)^2 {\rm d} u^2 
+ q(\tau_{p}(u)) \rho_{ij}dx^idx^j \, .
\eeq
The KSW function becomes
\beq
\label{eq:KSW_def}
\Sigma(N_c, \tau_{p}(u)) = \Sigma_{\rm temporal}(N_c, \tau_{p}(u))+ \Sigma_{\rm spatial}(N_c, \tau_{p}(u))\, ,
\eeq
where (``$\,'\,$" is derivative w.r.t $u$)
\begin{equation}
    \label{KSW_temp_spa_defs}
    \begin{split}
         \Sigma_{\rm temporal}(N_c,\tau_p(u)) = \left|\arg\left(\tau_{p}'(N_c, u)^2\right)\right|\, , \; \Sigma_{\rm spatial}(N_c,\tau_{p}) = 3\left|\arg q(N_c, \tau_{p})\right|\, .
    \end{split}
\end{equation}
We define an ``extremal curve" $\tau_{e}$ (in complex $\tau_p$ plane) that saturates the inequality in eq. (\ref{eq:KSW_criterion}): 
\beq
\label{eq:ext_curve_1}
\left|\arg(\tau_{e}'(N_c, u))^2\right| + 3 \left|\arg \bar{q}(N_c, \tau_{e}(u))\right| = \pi \, .
\eeq
Upon opening the modulus, the above equation is actually four different equations corresponding to four quadrants in the complex $\tau_p$-plane. Without losing any generality, one can consider any one of them. We consider
\begin{equation}
    \label{eq:extremal_curve}
    \arg(\tau_{e}'(N_c, u))^2 + 3 \arg \bar{q}(N_c,\tau_{e}(N_c,u)) = \pi
\end{equation}
Utilizing $\bar{q}(\tau_p)$ as given in eq. \ref{eq:time-path_saddle} and integrating the above equation, we get
\beq
\label{eq:red_ext_curve_sol}
\cos{(\sqrt{3\Lambda}\tau_x)} \cosh{(\sqrt{3\Lambda}\tau_y)} - 9 \cos{(\sqrt{\Lambda}\tau_x/\sqrt{3})}  \cosh{(\sqrt{\Lambda}\tau_y/\sqrt{3})} + 8 = 0 \, ,
\eeq
where we substituted $\tau_p=\tau_x+i\tau_y$. The above equation gives the equation for the extremal curve
with the initial conditions as $\tau_{y}(\tau_{x} = 0) = 0$ \footnote{The equation has multiple solutions; however, we consider those branches of the solution that make an angle $\pi/8$ with the horizontal axis \cite{Hertog:2023vot, Ailiga:2025fny}. It is consistent with the behaviour of $q(\tau_p)\sim \tau_p^2$, for small $\tau_p$.}. At $\tau_x=\sqrt{3}\pi/2\sqrt{\Lambda}$, the curve asymptotes to infinity in the imaginary direction.
Given this curve, we say the saddle is KSW allowable if the corresponding $\nu$ obeys \cite{Ailiga:2025fny, Hertog:2023vot}: 

\begin{equation}
\label{eq:ksw_allowability}
|\tau_{y}|> |\rm Im(\nu)|\hspace{4mm} \text{at}\hspace{4mm} \tau_x = \rm Re(\nu),
\end{equation}
where $\nu$ at the saddles (eq. (\ref{eq:fixed_k_saddle})) is given by the value of $\tau_p(t)$ at $t=1$
\begin{equation}
    \label{eq:nu_expression}
    \nu_\pm= 
    \sqrt{\frac{3}{\Lambda}}\biggl[\frac{\pi}{2} \pm i \sinh ^{-1}\biggl(\frac{H_1}{\sqrt{H^2-H_1^2}}\biggr)\biggr].
\end{equation}
The pictorial illustration of the extremal curve test is shown in figure \ref{fig:extremal_curve_fix_K}. From the figure, it is evident that both saddles are KSW-allowable, and there is an allowed path connecting the origin to the final hypersurface without crossing the extremal curve. In the Euclidean regime, $H_1=+iH\gamma$ and correspondingly $\nu$ is real. Hence, it is always KSW allowed.\\

\section{One-loop determinant at No-boundary saddles}
\label{NB_sad_cor} 
Having analyzed the semiclassical behaviour of the fluctuations, which is gaussian, it is crucial to know how the behaviour is modified once we incorporate the effects from one-loop fluctuations. In this section, we answer this by computing the one-loop determinant ($\mathcal{A}_h^\varepsilon$) mentioned in eq. (\ref{eq:QT_Nc_act_h_divST}) explicitly at the no-boundary saddles ($N_{\pm}^{\rm (nb)}$) given in eqs (\ref{eq:NNB_sad}) and (\ref{eq:fixed_k_saddle}) for different boundary choices. These are the saddles that contribute to the wave function of the universe ($\Psi[{\rm Bd_f}]$). Computation of this requires proper UV-renormalization and regularization. As we are dealing with Einstein-Hilbert gravity, one-loop divergences are expected. However, they can be renormalized by adding suitable counterterms \cite{tHooft:1974toh}. Below we extract these divergences and cancel them via appropriate counterterms. 
We consider the generalized zeta regularization, which allows us to focus only on physically meaningful UV divergences, cropping the rest.

\subsection{
Countertems and UV-renormalization}
\label{small_q0qf} 

%
At the no-boundary saddles $N_{\pm}^{\rm (nb)}$, the initial size of the universe vanishes: $\bar{q}(0)\equiv\bar{q}_0 \to 0$. At these saddles $\xi_l$ and $\tau_t$ in eq. (\ref{eq:hl_sol_gen}), for both boundary conditions, become 
\begin{equation}
\label{eq:xiL_tauT_nbc_nb1}
\xi_l^{\rm s}=1+l \, ,
\hspace{5mm}
\tau_{t}^{\rm nb}=1-i \Lam N_{\pm}^{\rm (nb)}t/3\, ,
\end{equation}
independent of $\varepsilon$, where $s$ superscript implies saddles. At $t=0$, $\tau_{0}^{\rm nb}=1$. At the no-boundary saddles $W_l^\varepsilon$ simplifies to $\mathbb{P}_1^{-l-1}(\tau_1^{\rm nb})\mathbb{Q}_1^{l+1}(\tau_0^{\rm nb})$,
where, $\mathbb{P}_1^{-l-1}(\tau_t^{\rm nb})$ can be expressed as 
(see appendix \ref{sec:asymp_mode_expansion})
\begin{equation}
\label{eq:legendre_polynom_simplified}
\mathbb{P}^{-l-1}_1(\tau_{t}^{\rm nb})=\frac{1}{\Gamma(l+3)}\Bl(\frac{\tau_{t}^{\rm nb}-1}{\tau_{t}^{\rm nb}+1}\Br)^{(l+1)/2}(\tau_{t}^{\rm nb}+l +1),\hspace{4mm}\forall \,\,\tau_{t}^{\rm nb}\,\, .
\end{equation} 
Clearly, when $\tau_0^{\rm nb}=1$, $\mathbb{P}^{-l-1}_1(\tau_{t}^{\rm nb})$ vanishes at $t=0$. However $\mathbb{Q}^{l+1}_1(\tau_{t}^{\rm nb})$ is finite at $t=1$.
To extract the UV-divergence near $t=0$, we write $\tau_t^{\rm nb}\sim 1-\Lambda \bar{q}(t)/6$.  From the asymptotic behaviour of $\mathbb{Q}_1^{l+1}(\tau_t^{\rm nb})$ near $t=0$, we get (see appendix \ref{sec:asymp_mode_expansion})
\beq
\label{eq:Wl_be_NoB}
W_l^\varepsilon(1,0) \mathbb{M}_\varepsilon(\xi_{l}^{s},N_{\rm nb})^{-1}\Br \rvert_{N_c \to N_{\pm}^{\rm (nb)}}
= \bar{q}_0^{-\xi_l^{\rm s}/2} \mathbb{W}_l(1,0) 
\,\, \xRightarrow[N_c \to N_{\pm}^{\rm (nb)}]{} \,\, 
\bar{q}_0^{-(l+1)/2} \mathbb{W}_l(1,0)
\, ,
\eeq
where $\mathbb{W}_l(1,0)$ is independent of $\varepsilon$ and is given by
\begin{equation}
    \label{eq:mathbb_W}
     \mathbb{W}_l(1,0)=\mathbb{P
    }_1^{-l-1}(\tau_1^{\rm nb})\mathbb{A}_l \, ,\hspace{5mm}\mathbb{A}_l=\frac{\Gamma(l+1)}{2} \left(\frac{12}{\Lambda}\right)^{(l+1)/2}.
\end{equation}
Hence, near the no-boundary saddles, the one-loop determinant ${\cal A}_{h}^\varepsilon(N_c)$, as mentioned in eq. (\ref{eq:Nc_act_total}), can be expressed as
\begin{equation}
    \label{eq:onel_loop_near_saddle}
    {\cal A}_{h}^\varepsilon(N_c) \br \rvert_{N_c \to N_{\pm}^{\rm (nb)}}=- \frac{2 i \hbar}{3} \ln \bar{q}(1) 
+ \frac{i \hbar}{2} \sum_{l=2}^\infty 
g_l  \ln \mathbb{W}_l(1,0) 
- 
\underbrace{
\frac{117i \hbar}{80} \ln 
\bar{q}_0 
}_{{\rm Log-div}}
+ {\cal O}(\hbar, N_c - N_{\pm}^{\rm (nb)},\varepsilon),
\end{equation}
where ${\cal O}(\hbar, N_c - N_{\pm}^{\rm (nb)},\varepsilon)$ depends on the parametrization and contributes away from the saddle.
Clearly, the one-loop contribution and the divergence at the no-boundary saddles are independent of parametrization. To obtain the coefficient of the log-divergence, we perform zeta regularization.

Together with the above divergence for vanishing initial size, there is another divergence in eq. (\ref{eq:onel_loop_near_saddle}) which appears 
either when $q_f$ is taken to zero or $H_1=i\gamma H$ with $\gamma$ taken to infinity. In the Euclidean regime, when $q_f<3/\Lam$ or $|H_1|>H$, only one of the two no-boundary saddles contributes to the path integral according to Picard-Lefschetz. It is $N_+^{\rm nb}$ for both fixed size and fixed $K$. At this particular saddle, we get
\begin{equation}
\label{eq:small_qf_div}
\mathbb{W}_l(1,0)\biggr|_{N_c\rightarrow N_+^{\rm nb}}\sim \begin{cases}
    q_f^{(l+1)/2}\hspace{10mm}\text{as}\hspace{3mm} q_f\rightarrow0\\
    |H_1|^{-(l+1)}\hspace{5mm}\text{as}\hspace{3mm} |H_1|\rightarrow\infty
\end{cases}
\end{equation}
for fixed size and fixed curvature, respectively. 
Combining the above asymptotic behaviour and the first term $\ln\bar{q}(1)$ in eq. (\ref{eq:onel_loop_near_saddle}) yields the total divergences. To renormalize the lapse action at the no-boundary saddles, one must add counterterms to cancel these UV divergences. After performing the $l$-summation via zeta regularization, the counterterm is given by 
\begin{equation}
\label{eq:Nc_act_total_nb_ct}
\eqnsizesmall
\begin{split}
\mathcal{A}_{\rm ct}(N_c)=
\left\{
{\renewcommand{\arraystretch}{1.5}
\begin{array}{l}
    \frac{117 i \hbar}{80} \ln \bar{q}_0
    - \frac{31 i \hbar}{240}\ln q_f \, \Theta\!\left(\frac{3}{\Lambda}-q_f\right)
    \hspace{14mm}\text{(fixed $q_f$)} \\[6pt]
    \frac{117 i \hbar}{80} \ln \bar{q}_0
    + \frac{31 i \hbar}{120}\ln |H_1| \, \Theta\!\left(|H_1|-H\right)
    \hspace{8mm}\text{(fixed $K$)}
\end{array}
}
\right.
\end{split}
\end{equation}
where $\Theta(x)$ is the Heaviside step function which gives one for $x>0$ and zero otherwise.
The counterterm has two terms: the first is independent of the boundary choices, while the second depends on them. However, both of them are independent of parametrization ($\varepsilon$), which meets the general expectation that the on-shell counterterm in gravity is parametrization independent; see sec. (\ref{eq:parameter_independen}).
Adding the counterterm, the UV-renormalized lapse action is given by
\bea
\label{eq:Nc_act_total_nb_fin}
&&
{\cal A}_{\rm ren}(N_c)
= {\cal A}_0(N_c) + i\hbar {\cal A}_1(N_c,\varepsilon) +\mathcal{A}_h^{\rm ghost}
+ {\cal A}_{\rm ct}(N_c) \, ,
\eea
where ${\cal A}_0(N_c) + i\hbar {\cal A}_1(N_c)$, $\mathcal{A}_h^{\rm ghost}$ and ${\cal A}_{\rm ct}(N_c)$ are mentioned in Eqs. (\ref{eq:Nc_act_total}), (\ref{eq:ghost_Action}) and (\ref{eq:Nc_act_total_nb_ct}) respectively. Once the lapse action is supplemented with the counterterm, it no longer suffers from UV divergences, leading to a finite effective action for the lapse ($N_c$). The infinite sum over $l$-modes is divergent and thus requires regularization and renormalization. We do it separately using the Hurwitz zeta function, as described in appendix \ref{sec:zeta_sum}
\footnote{In the other ways of evaluating the infinte sum, as taken in \cite{Ailiga:2024wdx}, one requires explicit counterterms in eq. (\ref{eq:Nc_act_total_nb_ct}).}.



\subsection{ 
Zeta regularization of $l$-mode summation
}
\label{sec:zeta_reg}
After cancelling the logarithmic divergences appearing in the one-loop effective action in eq. (\ref{eq:onel_loop_near_saddle}), we obtain a renormalized action that still contains an infinite sum over the angular momentum modes $l$. This summation is also divergent and needs a careful regularization procedure. We follow the method of zeta regularization which allows us to extract the `` physically meaningful " finite part cropping the divergent piece \cite{Elizalde:1994gf,Monin:2016bwf, Hawking:1976ja}. When evaluated on the no-boundary background given in eq. (\ref{eq:legendre_polynom_simplified}), the infinite sum takes the following form

\bea
\label{eq:log_exp_wl_nb_10}
    \sum_{l=2}^\infty g_l\ln \mathbb{W}_l(1,0)
    && =- \ln 2 \sum_{l=2}^\infty g_l +  \frac{1}{2}\ln\left[\frac{12 (1-\tau_1^{\rm nb})}{\Lam(1+\tau_1^{\rm nb})}\right]\sum_{l=2}^\infty g_l(l+1) 
    \notag \\
   && +\sum_{l=2}^\infty g_l\ln(\tau_{1}^{\rm nb}+l +1) - \sum_{l=2}^\infty g_l \ln (l+1)- \sum_{l=2}^\infty g_l \ln (l+2).  
\eea
In the above equation, the first and second series can be summed using ordinary zeta regularization, giving $\sum_{l=2}^\infty g_l = 8/3 $ and $\sum_{l=2}^\infty g_l(l+1)=191/60$ respectively. The other infinite summations appearing can be summed exactly using generalized zeta regularization, as shown in appendix \ref{sec:zeta_sum} (see eq. (\ref{eq:sum_2})). Performing the regularization and collecting all the terms, we get
\begin{eqnarray}
\label{eq:Log_wl_nb10_parts_fin}
\sum_{l=2}^{\infty} g_l\ln \mathbb{W}_l(1,0) \Biggr \rvert_{\zeta-{\rm reg}}
&&=-\frac{8}{3}\ln(2)+\frac{191}{120}\ln\biggl[\frac{12(1-\tau_1^{\rm nb})}{\Lambda(1+\tau_1^{\rm nb})}\biggr]+\log \left(2 \pi ^7 A^4\right)\notag\\
&&-2\zeta_H'(-2,\tau_1^{\rm nb}+3)+4 \tau_1^{\rm nb}\zeta_H'(-1,\tau_1^{\rm nb}+3)\notag\\
&& -2[(\tau_1^{\rm nb})^2-4] \zeta_H'(0,\tau_1^{\rm nb}+3)-\frac{\zeta (3)}{\pi ^2}-\frac{1}{3},
\end{eqnarray}
where $\zeta_H(s,a)$ is the Hurwitz-Zeta function and $(')$ denotes the differentiation w.r.t. $s$. $A$ is the Glaisher constant, $\zeta$ is the zeta function. As shown before in sec. \ref{sec:ghost_det}, the ghost action $\mathcal{A}_h^{\rm ghost}$ can be similarly regularized yielding a finite value. Performing the regularization for all the terms, we get the UV-renormalized and zeta regularized lapse action (both for fixed size and fixed $K$)
\bea
\label{eq:Nc_act_total_nb_fin_1}
&&
{\cal A}^{\zeta-{\rm reg}}_{\rm ren}(N_c,\varepsilon)
= {\cal A}_0(N_c) + i \hbar {\cal A}_1(N_c,\varepsilon)\biggr|_{\zeta-{\rm reg}} +\mathcal{A}_h^{\rm ghost}\biggr|_{\zeta-{\rm reg}}+{\cal A}_{\rm ct}(N_c)\notag\\
&&
=S^{(\bar{q})}_{\rm grav}(N_c)
+\frac{i\hbar}{2}\ln\Delta_q(N_c) 
- \frac{2 i \hbar}{3} \ln \bar{q}(1) 
- \frac{117i \hbar}{80} \ln 
\bar{q}_0+\frac{i\hbar}{2} \sum_{l=2}^{\infty} g_l\ln \mathbb{W}_l(1,0) \Biggr \rvert_{\zeta-{\rm reg}}
\notag \\
&&
+\mathcal{A}_h^{\rm ghost}\Biggr \rvert_{\zeta-{\rm reg}}+ {\cal A}_{\rm ct}(N_c)+{\cal O}(\hbar, N_c - N_{\pm}^{\rm (nb)},\varepsilon).
\eea
The renormalized zeta regularized lapse action ${\cal A}^{\zeta-{\rm reg}}_{\rm ren}(N_c,\varepsilon)$ is utilized in sec. (\ref{sec:wave_fun_PL_method}) to compute the wavefunction of the universe $\Psi[{\rm Bd}_f]$ (where, ${\rm Bd}_f$ could be either fixed size or, fixed $K$) using the WKB and Picard-Lefschetz method.


\subsection{Constraints and parametrization independence}
\label{eq:parameter_independen}
In this section, we discuss the parametrization ($\varepsilon$) independence observed in the lapse action ${\cal A}^{\zeta-{\rm reg}}_{\rm ren}(N_c)$, in eq. (\ref{eq:Nc_act_total_nb_fin_1}) when evaluated at the no-boundary saddles. This establishes that different parametrizations of the metric fluctuation do not affect the wave function. To start with, we note that constraints of the gravitational theory in the ADM formulation are the Hamiltonian and Diffeomorphism constraints, which are given by 
\begin{equation}
    \label{eq:constraints}
    \frac{\delta S_{\rm grav}[g_\mn]}{\delta N_c}=0,\hspace{4mm}\text{and}\hspace{4mm}\bar{D}^i(K_{ij}-K\gamma_{ij})=0,
\end{equation}
respectively,
where $S_{\rm grav}[g_\mn]$ is mentioned in eq. (\ref{eq:EHact_exp}). The expressions of $\gamma_{ij}$ and $K_{ij}$ can be read off from eqs. (\ref{eq:frwmet}) and (\ref{eq:chris_gam}), respectively. In the transverse-traceless gauge condition, the diffeomorphism constraint is satisfied. On the other hand, the Hamiltonian constraint is given by
\begin{equation}
\label{eq:parameter_independent}
\frac{\delta S_{\rm grav}[g_\mn]}{\delta N_c}=0\hspace{3mm} \Rightarrow \,\, \hspace{3mm}\frac{\dot{\bar{q}}^2}{8N_c^2}+\frac{\bar{q}\ddot{\bar{q}}}{2N_c^2}+\frac{1-\Lambda\bar{q}}{2}+h_{ij} \,\,\text{corrections}=0.
\end{equation}
The leading-order terms in the above eq. (\ref{eq:parameter_independent}) vanish at the no-boundary saddles (eqs. (\ref{eq:NNB_sad}) and (\ref{eq:fixed_k_saddle})). The remaining correction terms will vanish when the $h_{ij}$ correction to the saddles is accounted. As we put $\bar{h}_{ij}=0$, the remaining terms trivially vanish, and saddles receive no correction due to $h_{ij}$. Now, in the one-loop operator ($\mathcal{D}_h$), mentioned in eq. (\ref{eq:determin}), parametrization ($\varepsilon$) dependent term is multiplied by the leading-order Hamiltonian constraint (eq. \ref{eq:parameter_independent}), which vanishes at the saddles. As a consequence, the counterterms and the one-loop effective action evaluated at the saddles are independent of parametrization. This aligns with the general expectation, which states that the on-shell (satisfying Einstein's equation) effective action should be independent of parametrization. As a consequence of this, when we compute the wave function $\Psi[{\rm Bd}_f]$ up to one loop, it becomes independent of parametrization.


\section{Computing wave function via Picard-Lefschetz} 
\label{sec:wave_fun_PL_method} 
With all the necessary ingredients in place, we now compute the wave function $\Psi[{\rm Bd}_f]$, defined in Eq. (\ref{eq:wave_function}), up to one-loop order for two choices of final boundary conditions: fixed curvature ($K$) and fixed size ($q_f$). Since we set the on-shell fluctuation to zero ( $\bar{h}_{ij}=0$), the wave function receives one-loop contributions arising from virtual graviton fluctuations. We wish to understand the one-loop behaviour as a function of either $q_f$ or $K$ in the wave function. In conventional flat-spacetime quantum field theory, one would naively expect loop corrections to remain perturbatively suppressed relative to the leading, typically tree-level, contribution. In de Sitter spacetime, however, this expectation can be challenged, as quantum corrections may exhibit secular growth in the deep infrared \cite{Akhmedov:2013vka, Akhmedov:2019cfd, Akhmedov:2024npw, Miao:2024shs,Kamenshchik:2024ybm}. Such growth may signal a breakdown of fixed-order perturbation theory and might not be a property of the physical wave function of the universe. Although there is currently no consensus on its physical interpretation or resolution, several approaches have been proposed. These include resumming the secularly enhanced contributions \cite{Cespedes:2023aal, Baumgart:2019clc, Honda:2023unh,Palma:2025oux}, introducing and carefully treating an infrared cutoff \cite{Huenupi:2024ksc}, and employing suitable regularization schemes \cite{Xue:2011hm}. In this work, by wave function, we mean the perturbative form of the wave function as defined in eq. (\ref{eq:wave_function}).

After carrying out the path integrals over the background, metric fluctuations, and ghost fields, the wave function reduces to the following one-dimensional integral 
\beq
\label{eq:grav_path_NC_form_fin}
\Psi_{\zeta-{\rm reg}}[{\rm Bd}_f]
= \int_{\mathcal{C}} {\rm d} N_c \,\,
\exp \bl\{i {\cal A}_{\rm ren}^{\zeta-{\rm reg}}(N_c,\varepsilon)/\hbar \br\} \, ,
\eeq
where ${\cal A}^{\zeta-{\rm reg}}_{\rm ren}(N_c)$ is finite zeta-regularized lapse action defined in Eq. (\ref{eq:Nc_act_total_nb_fin_1}). The above integral can be evaluated using the Picard-Lefschetz and WKB methods. To start with, we perform the $\hbar$-decomposition of the action and saddles similar to eq. (\ref{eq:ANc_1lp_form}). In leading order, the $h_{ij}$ correction to the saddles is not relevant. Similarly, the correction to the thimbles/ $\theta_\sigma$ wouldn't matter in the one-loop approximation. Hence, using eqs. (\ref{eq:sumOthim}) and (\ref{eq:LDordI}), we get the wave function
\beq
\label{eq:LDordI_fin}
\Psi_{\zeta-{\rm reg}}[{\rm Bd}_f]
= 
\sum_\sg \frac{n_\sigma e^{i\theta_\sg}}{\sqrt{\bl\vert {\cal A}_{0}^{\prime\prime} (N_\sg^{(q)}) \br\rvert}}
\times 
\exp\Bl[ 
\frac{i}{\hbar}{\cal A}_{\rm ren}^{\zeta-{\rm reg}}(N_\sg^{(q)})
+ \cdots
\Br] \, ,
\eeq
where $\sigma$ is the relevant saddles depending on the choice of contour ($\mathcal{C}$) and $ {\rm Bd}_f$ is the final boundary condition, which we take either fixed size or fixed $K$. $N_\sg^{(q)}$ is the background saddle for these boundary conditions. As explained earlier, the dependence on $\varepsilon$ disappears in the wave function. Eq. (\ref{eq:LDordI_fin}) is written for generic boundary choices, which in special cases get appropriately evaluated. 

\begin{figure}[hbtp]
\centering
\subfigure[\,\,Lorentzian Regime]{\includegraphics[width=0.42\linewidth]{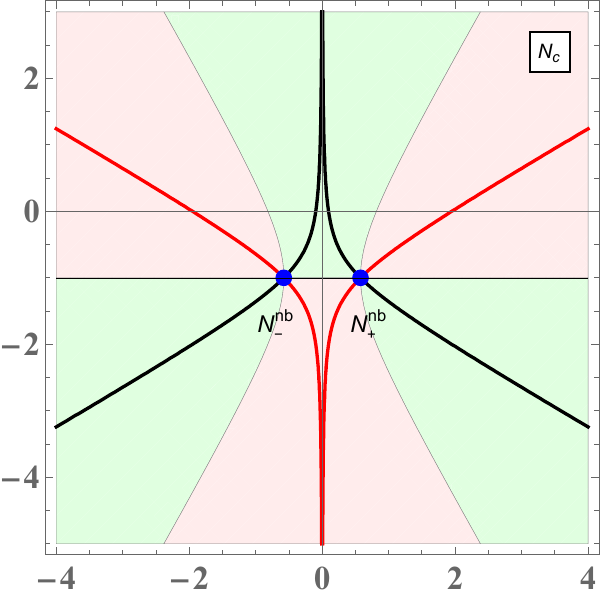}}
\,\,\,\,\,\,
\subfigure[\,\,Euclidean Regime]{\includegraphics[width=0.4\linewidth]{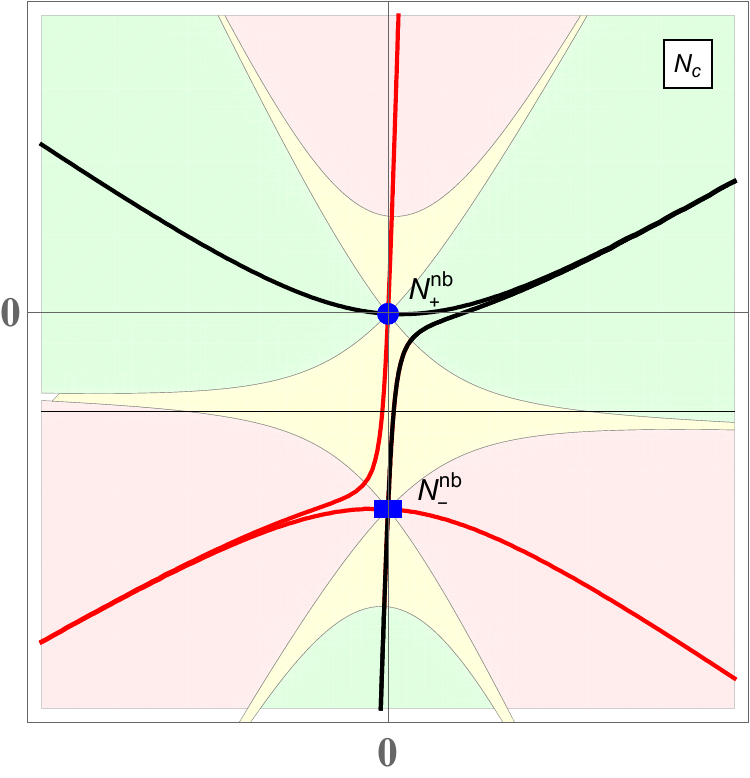}}
\caption{The PL plot shows the relevant saddles (Blue dots) and irrelevant saddle (Blue square) in two different regimes. a) $H_1<H$, Lorentzian phase $H_1=1/2,H=1$, b) $H_1=+i\gamma H,\gamma>0$, Euclidean phase, $\gamma=6,H=1$. As explained in the main text, we complexify Newton's constant $G=|G|e^{i\theta},\theta=\pi/10$ to resolve the Stokes degeneracy that appears between the saddles. Steepest descent lines are plotted in black, while the ascent lines are plotted in red.}
\label{fig:relevant saddle}
\end{figure}

In the case of No-boundary Universe, as mentioned in sec. (\ref{subsubS:Nsq}), in the Lorentzian regime when either $H_1<H$ or, $q_f>3/\Lam$, both the no-boundary saddles contribute to the wave function with $n_\sigma=1$, see figure \ref{fig:relevant saddle}. In the Euclidean regime, only the $N_+^{\rm nb}$ saddle contributes. Hence, in the WKB approximation, the wave function is given by
\begin{equation}
\label{eq:Gtrans_qf>3/lam_gen}
\begin{aligned}[t]
\Psi^{\rm nb}_{\zeta\text{-reg}}[{\rm Bd}_f] \approx
\begin{cases}
\displaystyle
\sum_{\sg=\pm} e^{i\theta_\sg }
\exp[i\mathcal{A}_{\rm ren}^{\zeta\text{-reg}}(N^{\rm nb}_\sg)/\hbar]
\left|{\cal A}_{0}^{\prime\prime}(N_{\sg}^{\rm nb})\right|^{-1/2}
& {\rm (Lorentzian\,\, Regime)} \\[6pt]
\displaystyle
e^{i\theta_+ }
\exp[i\mathcal{A}_{\rm ren}^{\zeta\text{-reg}}(N^{\rm nb}_+)/\hbar]
\left|{\cal A}_{0}^{\prime\prime}(N_{+}^{\rm nb})\right|^{-1/2}
& {\rm (Euclidean\,\, Regime)}
\end{cases}
\end{aligned}
\end{equation}
where the zeta-regularized one-loop action is given by
\bea
\label{eq:Nc_act_total_nb_fin_at_saddle}
{\cal A}^{\zeta-{\rm reg}}_{\rm ren}(N_\pm^{\rm nb})
= {\cal A}_0(N_\pm^{\rm nb})+\frac{i \hbar}{2} \ln \D_q(N_\pm^{\rm nb}) - \frac{2 i \hbar}{3} \ln \bar{q}(1)
+ \frac{i \hbar}{2} \sum_{l=2}^\infty 
g_l  \ln \mathbb{W}_l(1,0)+\mathcal{A}_h^{\rm ghost}.
\eea
$\mathcal{A}_0(N_c)$ is the on-shell action mentioned in eqs. (\ref{eq:stot_onsh_rbc}) and (\ref{eq:stot_onsh_fixed_k}) for fixed $q_f$ and fixed $K$ respectively. The factor $\D_q(N_\pm^{\rm nb})$ is discussed in sec. (\ref{SubS:PI_qt}). The remaining terms in eq. (\ref{eq:Nc_act_total_nb_fin_at_saddle}) are coming from the one-loop fluctuation determinant. The ghost contribution ($\mathcal{A}_h^{\rm ghost}$) is computed in sec. \ref{sec:ghost_det}. In the following subsections, we will compute $\Psi_{\zeta-{\rm reg}}[{\rm Bd}_f]$ explicitly in different regimes utilizing eq. (\ref{eq:Gtrans_qf>3/lam_gen}). 

\subsection{Wave function with fixed $K$}
\label{s_sec:lorentzian_regime_fix_K}


\subsubsection{Lorentzian Regime: $\bold{H}_1(=H_1/H)<1$}
\label{ss_sec:lorentzian_regime_fix_K}

\begin{figure}
\centering
\subfigure{\includegraphics[width=0.55\linewidth]{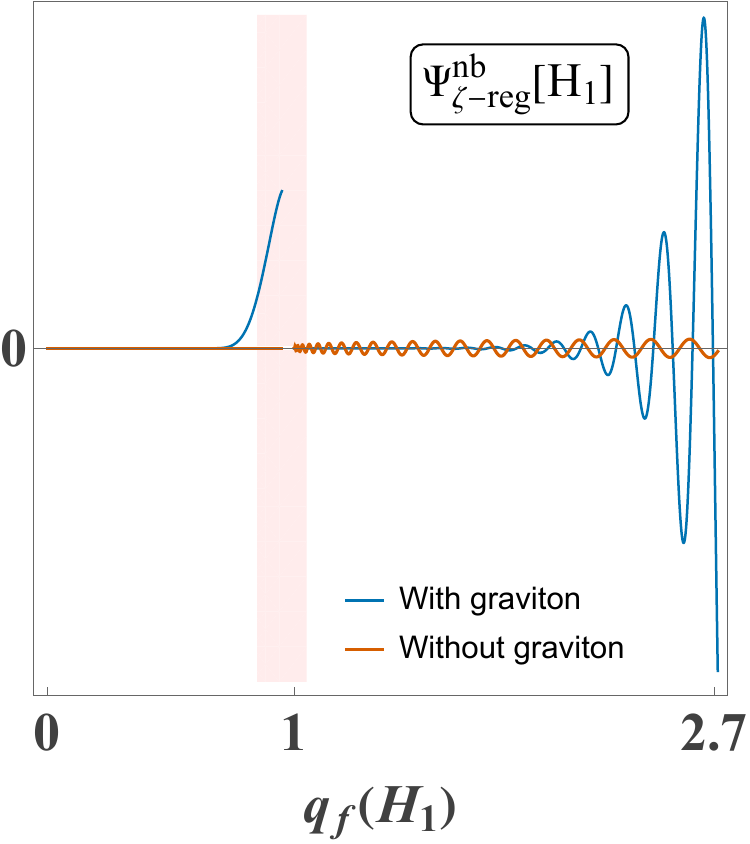}}
\caption{The plot showing $\Psi^{\rm nb}_{\zeta-{\rm reg}}[H_1]$ as a function of final size of the universe ($q_f(H_1)=1/(H^2-H_1^2)$). The blue curve shows the growing nature of the wave function originating from the one-loop gravity fluctuation. The orange curve is the background. The pink-shaded region is where the WKB approximation breaks down; the left region is the Euclidean, whereas the right is the Lorentzian regime. The vertical axis is scaled appropriately, and we set $H=1$ for illustration purposes.}
\label{fig:wave_function_fix_K_plot}
\end{figure}

In this section, we compute the wave function when the extrinsic curvature ($K$) is fixed at the final hypersurface. This boundary condition corresponds to fixing the Hubble radius ($H_1=K/3$), which is more physically sensible compared to the fixed size \cite{DiTucci:2019bui, Abdalla:2026mxn}. We consider the no-boundary condition on the initial hypersurface. Considering the closed slicing of the de-Sitter space, we require $H_1<H$. This corresponds to the Lorentzian regime, where the relevant saddles are $N_{\pm}^{\rm nb}$ (eq. (\ref{eq:fixed_k_saddle})). As illustrated in Fig.\ref{fig:relevant saddle}, the steepest-ascent contours associated with both saddle points intersect the original integration contour, implying that both saddles contribute to the path integral. The zeta-regularized sum, as mentioned in eq. (\ref{eq:Log_wl_nb10_parts_fin}), at these saddles evaluates to
\begin{eqnarray}
\label{eq:Log_parts_fin_zeta_regu_fixed_k}
\sum_{l=2}^{\infty} g_l\ln \mathbb{W}_l(1,0)&& \Biggr \rvert_{\zeta-{\rm reg}}
=-\frac{8}{3}\ln(2)+\frac{191}{120}\ln\biggl[\frac{12(\sqrt{1-\bold{H}_1^2}\pm i\,\bold{H}_1)}{\Lambda(\sqrt{1-\bold{H}_1^2}\mp i\,\bold{H}_1)}\biggr]+\log \left(2 \pi ^7 A^4\right)\notag\\
&&-2\zeta_H'(-2,3\mp i\,\bold{H}_1/\sqrt{1-\bold{H}_1^2})\mp 4i \frac{\bold{H}_1}{\sqrt{1-\bold{H}_1^2}}\zeta_H'(-1,3\mp i\,\bold{H}_1/\sqrt{1-\bold{H}_1^2})\notag\\
&& +\frac{2(4-3\bold{H}_1^2)}{1-\bold{H}_1^2}\,\zeta_H'(0,3\mp i\,\bold{H}_1/\sqrt{1-\bold{H}_1^2})-\frac{\zeta (3)}{\pi ^2}-\frac{1}{3},
\end{eqnarray}
where we define dimensionless $\bold{H}_1=H_1/H$. The terms in the above equation constitute the leading one-loop contribution. These terms arise from the $h_{ij}$ fluctuation, and, as pointed out earlier, the one-loop terms from the background are sub-leading. Using the asymptotic expansions of the derivatives of the Hurwitz-Zeta function given in appendix \ref{sec:zeta_sum}, one can obtain the asymptotic behaviour $H_1\rightarrow H$. This limit corresponds to the IR limit. Expressing the one-loop action as $\mathcal{A}_{\rm ren}^{\zeta-{\rm reg}}(N_c)={\rm Re}\{ \mathcal{A}_{\rm ren}^{\zeta-{\rm reg}}(N_c)\}+i\,{\rm Im}\{\mathcal{A}_{\rm ren}^{\zeta-{\rm reg}}(N_c)\}$, the leading and subleading terms in the real and imaginary parts of the action are given by

\begin{equation}
\label{eq:Afin_ReIm_fixed_k}
   \begin{split}
        &{\rm Re}\{ \mathcal{A}_{\rm ren}^{\zeta-{\rm reg}}(N^{\rm nb}_\pm)\} \Br\rvert_{1-\rm loop}\approx  \mp \hbar\biggl[\frac{1}{36\sqrt{2}(1-\bold{H}_1)^{3/2}}\{11+\ln8+3\ln (1-\bold{H}_1)\}\\
        &\hspace{40mm}+\frac{29}{16\sqrt{2}(1-\bold{H}_1)^{1/2}}\ln (1-\bold{H}_1)+\cdots\biggr] \, ,\\
        &{\rm Im}\{\mathcal{A}_{\rm ren}^{\zeta-{\rm reg}}(N^{\rm nb}_\pm)\}\Br\rvert_{1-\rm loop} \approx  \hbar\biggl[-\frac{\pi }{12 \sqrt{2} (1-\bold{H}_1)^{3/2}}-\frac{29\pi}{16\sqrt{2}(1- \bold{H}_1)^{1/2}}+ \cdots\biggr]\, .
   \end{split} 
\end{equation}
The dots are the subsubleading terms in the expansion. To the best of our knowledge, the above asymptotic expansion is new and has not been reported earlier. Putting all the factors and summing over the two relevant saddles following eq. (\ref{eq:Gtrans_qf>3/lam_gen}), the  wave function for fixed curvature $(K)$ becomes
\bea
\label{eq:H_1<H_limit}
&&
\Psi^{\rm nb}_{\zeta-{\rm reg}}[H_1<H]\biggr \rvert_{H_1\to H}\notag\\
&&
\sim \frac{1}{(1-\bold{H}_1^2)^{1/4}}\exp\left(\frac{V_3}{4\pi G\hbar H^2}+\frac{\pi  }{12 \sqrt{2} (1-\bold{H}_1)^{3/2}}+\frac{29\pi}{16\sqrt{2}(1- \bold{H}_1)^{1/2}}+ \cdots+\mathcal{O}(\hbar)\right)
\notag \\
&&
\times \cos\biggl[\frac{V_3}{4\pi\sqrt{2} GH^2\hbar}\frac{1}{\sqrt{1-\bold{H}_1}}+\frac{1}{36\sqrt{2}(1-\bold{H}_1)^{3/2}}\{11+\ln8+3\ln(1-\bold{H}_1)\}\notag\\
&&+\frac{29}{16\sqrt{2}(1-\bold{H}_1)^{1/2}}\ln(1-\bold{H}_1)+\cdots+\cdots+\mathcal{O}(\hbar)-\frac{\pi}{4}\biggr] \, ,
\eea
and $-\pi/4$ corresponds to the direction of the thimbles near the saddle in the asymptotic limit, see fig. \ref{fig:relevant saddle}. The wave function is real, as it is expected from the asymptotic behaviour in eq. (\ref{eq:Afin_ReIm_fixed_k}); however it holds true for a finite $H_1$ as well, see  appendix \ref{sec:symm}. The leading correction to the exponential factor grows as the universe expands. It is divergent in the IR limit $\bold{H}_1\rightarrow 1$, see figure \ref{fig:wave_function_fix_K_plot}. Precisely, it scales with volume of the spatial hypersurface $\bar{q}(1)^{3/2}\sim(1-\bold{H}_1)^{-3/2}$. The expression of the wave function in eq. (\ref{eq:H_1<H_limit}) is new and not known earlier in the literature. Such behaviour can be interpreted as a signature of infrared divergences, known to appear in de Sitter QFT calculations. Such enhancement is often referred to as secular growth in literature \cite{Akhmedov:2013vka, Akhmedov:2019cfd, Akhmedov:2024npw,Miao:2024shs}. 
Another point we would like to emphasize is that the one-loop correction to the phase dominates over the semiclassical phase. This is different from the fixed-size case, found earlier in \cite{Ailiga:2024wdx}, see also sec.\ref{PL_Nc_sad_qfLG}. As noted in \cite{DiTucci:2019bui}, the semiclassical phase oscillates with $\sqrt{\bar{q}(1)}\sim 1/\sqrt{1-\boldsymbol{H}_1}$, whereas we find that the one-loop-corrected phase oscillates with the spatial volume, $\bar{q}(1)^{3/2}$.\\

\subsubsection{Euclidean Regime: $\bold{H}_1=i\gamma,\gamma>0$}
\label{ss_sec:euclidean_regime_fix_K}
In this section, we compute the wave function in the Euclidean regime where $\bold{H}_1=+i\gamma$, where $\gamma>0$ and is purely imaginary. As explained in sec. \ref{subsubS:Nsq}, the sign of $\gamma$ reflects the Hartle-Hawking nature of the wave function, where the momentum conjugate to the scale factor is negative imaginary. The zeta regularized sum, as mentioned in eq. (\ref{eq:Log_wl_nb10_parts_fin}), at the $N_\pm^{\rm nb}$ saddles is given by

\begin{eqnarray}
\label{eq:Log_parts_fin_zeta_regu_fixed_k_appendix}
\sum_{l=2}^{\infty} g_l\ln \mathbb{W}_l(1,0)&& \Biggr \rvert_{\zeta-{\rm reg}}
=-\frac{8}{3}\ln(2)+\frac{191}{120}\ln\biggl[\frac{12(\sqrt{1+\gamma^2}\mp\gamma)}{\Lambda(\sqrt{1+\gamma^2}\pm\gamma)}\biggr]+\log \left(2 \pi ^7 A^4\right)\notag\\
&&-2\zeta_H'(-2,3\pm\gamma/\sqrt{1+\gamma^2})\pm 4 \frac{\gamma}{\sqrt{1+\gamma^2}}\zeta_H'(-1,3\pm\gamma/\sqrt{1+\gamma^2})\notag\\
&& 
+\frac{2(4+3\gamma^2)}{1+\gamma^2}\,\zeta_H'(0,3\pm\gamma/\sqrt{1+\gamma^2})-\frac{\zeta (3)}{\pi ^2}-\frac{1}{3},
\end{eqnarray}
Note that the above expression is completely real as it is evaluated at the Euclidean saddles.
As explained earlier, one needs to add a counterterm to subtract the divergence in the $\gamma\rightarrow\infty$ limit, which appears at the saddle $N_+^{\rm nb}$, see eq. (\ref{eq:Nc_act_total_nb_ct}). In this regime, the saddles are purely imaginary and lie on the Stokes ray (where the steepest descent thimble from one saddle overlaps with the steepest ascent thimble from another saddle). This is the degenerate situation where the Picard-Lefschetz method breaks down. To determine the relevance of the saddles, we complexify Newton's constant ($G$). Methods for resolving such degeneracies in the Lefschetz thimbles have been discussed in detail in \cite{Ailiga:2025fny}. In figure \ref{fig:relevant saddle}, we consider such rotation $G=|G|e^{i\theta},0<|\theta|\ll 1$. Resolving the degeneracy, we get the relevant saddle to be $N_+^{\rm nb}$ for either sign of $\theta$. Once the relevant saddle is determined, one can compute the wave function in this regime utilizing the eqs. (\ref{eq:Gtrans_qf>3/lam_gen}) and (\ref{eq:Log_parts_fin_zeta_regu_fixed_k_appendix}). We plot the wave function in fig. \ref{fig:wave_function_fix_K_plot}.


\subsection{Wave function with Fixed size: $q_f>3/\Lam$}
\label{PL_Nc_sad_qfLG} 

In this section, we will compute the wave function for a fixed final size of the universe ($q_f$). We set the initial boundary condition to a Robin condition with parameter $0\leq x\leq 1$ to ensure that the no-boundary saddles are relevant (see sec. \ref{subsubS:Nsq} for details). A similar analysis was attempted in \cite{Ailiga:2024wdx}, where a different regularization method was used. Here, we revisit this issue using Hurwitz-zeta regularization, a well-established technique in quantum field theory in curved spacetime \cite{Hawking:1976ja, Elizalde:1994gf}. This regularization allows us to systematically isolate the physically meaningful infrared-divergent contributions, thereby motivating us to a careful reassessment of the problem. Here, we only analyze the asymptotic behaviour at large $q_f$, which is the IR limit; see appendix \ref{sec:wave_function_other_regime} for details of the exact analysis for finite $q_f$. This also helps us to understand the structural dependence of IR behaviour on the final boundary condition.

As before, we split the one-loop action $\mathcal{A}_{\rm ren}^{\zeta-{\rm reg}}$ into real and imaginary parts (see eqs. \ref{eq:Log_wl_nb10_parts_fin} and \ref{eq:Nc_act_total_nb_fin_1}). Using the asymptotic expansions of the derivatives of the Hurwitz-Zeta function given in appendix \ref{sec:zeta_sum}, one can obtain the asymptotic behaviour $q_f\rightarrow\infty$. This limit corresponds to the IR limit. The leading and subleading terms in the real and imaginary parts of the action are 
\begin{equation}
\label{eq:Afin_ReIm_LG_qf}
   \begin{split}
       &{\rm Re}\{ \mathcal{A}_{\rm ren}^{\zeta-{\rm reg}}(N^{\rm nb}_\pm)\} \Br\rvert_{1-\rm loop}\approx \mp \hbar\left[\frac{1}{18}(11-3\ln \bold{q_f})\bold{q_f}^{3/2}-\frac{7}{4}\bold{q_f}^{1/2}\ln \bold{q_f}+\cdots\right] \, ,\\
    &{\rm Im}\{ \mathcal{A}_{\rm ren}^{\zeta-{\rm reg}}(N^{\rm nb}_\pm)\}\Br\rvert_{1-\rm loop} \approx  \hbar\left[-\frac{\pi}{6}\bold{q_f}^{3/2}-\frac{7\pi}{4}\bold{q_f}^{1/2}+ \cdots\right]\, ,
   \end{split} 
\end{equation}
where, we have defined dimensionless $\bold{q_f}=\Lam q_f/3$.
Note that while the imaginary part at the two no-boundary saddles is the same, the real part differs by an overall minus sign. Putting all the factors and summing over the two complex saddles following eq. (\ref{eq:Gtrans_qf>3/lam_gen}), the wave function for a fixed final size of the universe becomes

\bea
\label{eq:Gtr_qf>3/lam_LG}
&&
\Psi^{\rm nb}_{\zeta-{\rm reg}}[q_f>3/\Lambda]\Biggr \rvert_{q_f\to \infty}\sim \,\frac{1}{\mathcal{F}_\infty}\times \exp\left(\frac{3V_3}{4\pi G\hbar\Lambda}+\frac{\pi}{6}\bold{q_f}^{3/2}+\frac{7\pi}{4}\bold{q_f}^{1/2}+ \cdots+\mathcal{O}(\hbar)\right)
\notag \\
&&
\times \cos\biggl[\frac{3V_3}{4\pi G\hbar\Lambda}\bold{q_f}^{3/2}+\frac{1}{18}(11-3\ln \bold{q_f})\bold{q_f}^{3/2}-\frac{7}{4}\bold{q_f}^{1/2}\ln \bold{q_f}+\cdots+\mathcal{O}(\hbar)-\bar{\theta}_s^+\biggr] \, ,
\eea
where $\mathcal{F}_\infty$ comes from the denominator in eq. \ref{eq:Gtrans_qf>3/lam_gen}:
\begin{equation}
\label{eq:calF_form_LGqf}
    \begin{split}
    \mathcal{F}_\infty &\sim \bold{q_f}^{1/4}\hspace{5mm}\text{for $x=0$}\, ,\\
     &\sim \text{const}\hspace{5mm}\text{for $0<x<1$} \, ,
    \end{split}
\end{equation}
and $\bar{\theta}_s^+=\pi/4$ for $x=0$, and $\bar{\theta}_s^+=0$ for $0<x<1$ in the large $\bold{q_f}$ limit. The real and imaginary part of subsequent subleading terms in the expression falls as $\bold{q_f}$. Note that, as in the fixed-$K$ case, in the IR limit one also observes a similar divergence, though the coefficient is different. It scales with the volume of the spatial hypersurface, as in the fixed-$K$ case. The correction to the phase is of the same order as the semiclassical phase and similar to what was obtained earlier for fixed $K$. The leading exponential behaviour, $\exp(+\pi \mathbf{q}_f^{3/2}/6)$, is exactly the same to that observed earlier in \cite{Ailiga:2024wdx}. Although the leading term is identical to that found in \cite{Ailiga:2024wdx}, the subleading term differs. This discrepancy can be attributed to the different regularization prescription employed in \cite{Ailiga:2024wdx}. A key advantage of the zeta-function regularization employed here is that it allows us to extract the physically reliable IR divergent terms. Finally, for both boundary conditions, i.e., fixed size and fixed $K$, the leading and subleading terms exhibit similar divergences. We therefore conclude that the IR issues are qualitatively independent of the final boundary condition. Furthermore, evaluating the mode sum using alternative regularization methods confirms that the leading IR behaviour is independent of the choice of regularization scheme (see appendix \ref{sec:wave_function_other_regime} for details). 

\section{de-Sitter wave function and IR behaviour}
\label{sec:comparison_de_sitter}

So far, we have focused on studying the no-boundary wave-function of the Universe mentioned in eq. (\ref{eq:wave_function}). It is seen that the dominant and {\it relevant} saddles contributing to the path-integral correspond to complex geometries, which we refer to as {\it no-boundary saddles} (see figure \ref{fig:geometry_contour}). The wave-function, when computed to one-loop for these boundary choices, shows that the one-loop contribution from TT-fluctuations witnesses an IR growth, overpowering the contribution from the background (where the on-shell metric-fluctuations are not included). It is important to investigate whether this behaviour persists if pure-Lorentzian de Sitter (dS) is considered instead. To put it more clearly, we wish to study the IR behaviour of the wave function of dS and compare it with that of the no-boundary wave function.  This comparison allows us to identify which features of the IR behaviour are sensitive to the choice of boundary conditions and which remain universal.

\begin{figure}[hbtp]
\centering
\includegraphics[width=0.8\linewidth]{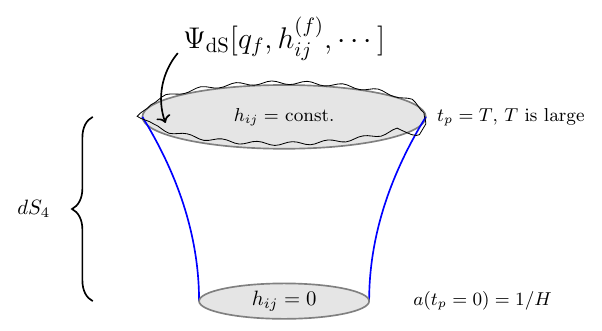}
\caption{Saddle geometry 
describing the expanding phase in real de-Sitter ($dS_4$) from the throat at $t=0$ to $t=T$, with $T$ taken large. We choose the initial surface fluctuation to vanish ($h_{ij}=0$). $q_f(H_1),h_{ij}^{(f)}$ are the final parameters.}
\label{fig:de-sitter}
\end{figure}
The quantity we are interested in can be written in an analogous manner as in eq. (\ref{eq:wave_function}), 
\beq
\label{eq:ds_wave_function}
\Psi_{\rm dS}[{\rm Bd}_f]\equiv\mathcal{Z}[{\rm Bd}_f, {\rm Bd}_i]=\int_{\rm Bd_i}^{\rm Bd_f}{\cal D} g_\mn \, e^{i\mathsf{S}[g_\mn]/\hbar}\, ,
\eeq
where ${\rm Bd}_i$ refers to the classical boundary condition \cite{Feldbrugge:2017kzv}.
The de Sitter metric with graviton fluctuation, in exponential parametrization, is 
\begin{equation}
\label{eq:ds_stan_metric}
ds^2=-N_p^2 dt_p^2+a(t_p)^2\rho_{ik}(e^{h})^k_jdx^idx^j.
\end{equation}
When $h=0$, $N_p=1$, $a(t_p)=\cosh(H t_p)/H$ describes the expanding de-Sitter phase, starting from the throat, where $a(t_p=0)=1/H$, and extending to the asymptotic future at $t_p=\infty$, see figure \ref{fig:de-sitter}. On performing a change of variable: ($N_p(t_p)$): $N_p(t_p) {\rm d} t_p = (N(t)/a(t)) {\rm d} t$ and $q(t) = a^2(t)$ as before (where $t\in[0,1]$), it is noticed that to achieve pure de Sitter geometry, one need to impose boundary conditions: $\pi_q=0$ at the initial hypersurface (${\rm Bd}_i$) and a fixed extrinsic curvature at the final hypersurface (${\rm Bd}_f$), see eq. (\ref{eq:fixing_K}) \cite{Abdalla:2026mxn}. This corresponds to setting $P_i=0$ in eq. (\ref{eq:qsol_RBC}) and eq. (\ref{eq:stot_onsh_fixed_k}) which are the e.o.m and the on-shell action, respectively. 
\bea
\label{eq:ds_path_qform}
\Psi_{\rm dS}[{\rm Bd}_f]
=&& \int_{\mathcal{C}} {\rm d} N_c \,[\mathbb{G}]
\int_{P_i=0}^{ {\rm Bd}_f} {\cal D} q(t) {\cal D} h_{ij}(t, {\bf x})
\exp \left(\frac{i}{\hbar} S_{\rm dS}[q, h_{ij}, N_c] \right) 
\notag \\
=&&
\int_{\mathcal{C}} {\rm d} N_c \,\,
\exp \bl\{i {\cal A}_{\rm dS}(N_c)/\hbar \br\} \, ,
\eea
where $S_{\rm dS}[q, h_{ij}, N_c]$ is the gravity action with the gauge-fixing condition imposed and is mentioned in 
eq. (\ref{eq:EHact_exp}), $[\mathbb{G}]$ is the Faddeev-Popov ghost determinant and ${\cal A}_{\rm dS}(N_c)$ is the quantum corrected lapse action which includes contribution from $q$ and $h_{ij}$ path-integrals. 

For these boundary choices, the on-shell solutions $\bar{q}(t)$ and $\bar{h}_l(t)$ can be obtained from eqs (\ref{eq:qsol_RBC}) and (\ref{eq:hl_dpdq_RBC}) respectively. The on-shell gravity action for such boundary choices can be obtained from eq. (\ref{eq:EHact_exp_onshell}) and (\ref{eq:hh_onSH_fix_K}) by plugging $P_i=0$. This leads to the following lapse-$N_c$ action for each mode ($l$) of fluctuation 

\begin{equation}
\label{eq:lapse-action_de-sitter}
\begin{split}
&S_{\rm dS}^{(l)}[N_c,H_1,h_1^{(l)}]=S_{\rm dS}^{\rm (\bar{q})}[N_c,H_1] +S_{\rm dS}^{(\rm{\bar{h}})}[N_c,H_1,h_1^{(l)}]\\
\text{where},\hspace{10mm} & S_{\rm dS}^{\rm (\bar{q})}[N_c,H_1] =\frac{V_3}{8\pi G}\biggl[N_c^3 \left(H^4-\frac{H^6}{H_1^2}\right)+3 N_c\biggr]\\
&S_{\rm dS}^{(\rm{\bar{h}})}[N_c,H_1,h_1^{(l)}]=-\frac{(h_{1}^{(l)})^2N_c^3H^6}{64\pi GH_1^4}
\Bl[H_1^2
-H^2\bl\{\ln{W}_l^{\varepsilon=1}(t,0)\br\}'\big|_{t=1}\Br], 
\end{split}
\end{equation}
where ${W}_l^{\varepsilon}(t,0)$ is given in eq. (\ref{eq:W_l10_func}). The complete action for fluctuation is obtained by summing over all the $l$-modes with the degeneracy $g_l=2(l+3)(l-1)$. 

Following the background field formalism as described previously, the path-integral breaks into two parts as mentioned in eq. (\ref{eq:grav_path_qhform_1}), but this time with the initial condition ${\rm Bd}_i \equiv P_i=0$. The path-integral over $q(t)$ can be computed following subsection \ref{SubS:PI_qt}, but with $P_i=0$. From eq. (\ref{eq:lapse-action_de-sitter}), one can immediately work out the saddles of the theory for the case of $(h_{1}^{(l)}=0)$. These are given by
\begin{equation}
\label{eq:de-sitter}
N_\pm^{\rm dS}=\pm \frac{H_1}{H^2\sqrt{H^2-H_1^2}} \, .
\end{equation}
Note that this is the same quantity that appears in the real part of the no-boundary saddles as mentioned in eq. (\ref{eq:fixed_k_saddle}). The on-shell action for $(h_{1}^{(l)}=0)$ at these saddles evaluates to $\pm 2H_1/H^2\sqrt{H^2-H_1^2}$. 
\begin{figure}[hbtp]
\centering
    \includegraphics[width=0.5\linewidth]{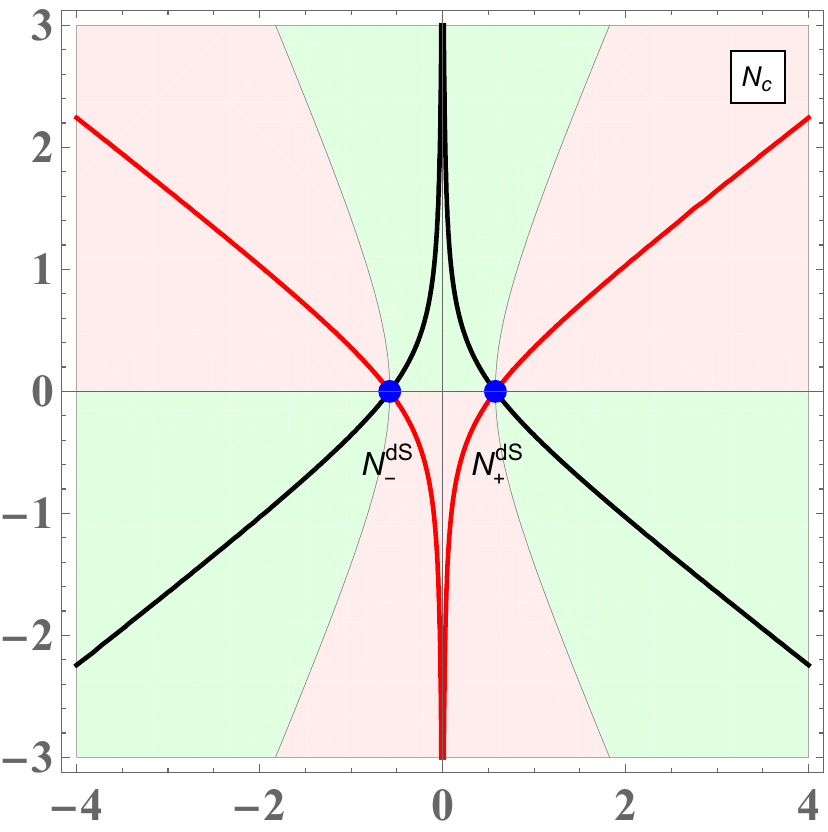}
    \caption{Picard-Lefschetz analysis for the classical boundary condition describing the real de-Sitter. We set $H=1$ and $H_1=1/2$ for plotting. The color convention is the same as the other figures. Both the saddles are relevant as they lie on the integration contour $N_c\in (-\infty,\infty)$}
\label{fig:pl-de-siiter}
\end{figure}
For the case of zero fluctuations, the lapse-$N_c$ integral mentioned in eq. (\ref{eq:ds_path_qform}), can be computed via Picard-Lefschetz methods. The saddles, flow-line structure, allowed/disallowed region are depicted in figure \ref{fig:pl-de-siiter}. Both saddles are seen to be {\it relevant} and to contribute to the path integral with the appropriate phase ($\theta_\sigma$). The dS wavefunction without the contribution from the fluctuations-$h_{ij}$ is given by, at the saddle point approximation

\begin{equation}
    \label{eq:psi_ds_zeroth order}
    \Psi_{\rm dS}[H_1]=\frac{\sqrt{H_1}}{\sqrt{6}H(H^2-H_1^2)^{1/4}}\cos\biggl(\frac{V_3}{8\pi G\hbar}\frac{2H_1}{H^2\sqrt{H^2-H_1^2}}-\frac{\pi}{4}\biggr).
\end{equation}
The quantity $\Psi_{\rm dS}$ is a pure phase as the saddle geometries are real de-Sitter. The prefactor multiplied to the phase arises from the WKB approximation. We now proceed to compute the contribution from fluctuations, which arise not only from on-shell terms but also from one-loop corrections. In the choice where the perturbations vanish at the two hypersurfaces, the on-shell contribution from the fluctuations vanish. However, the off-shell contributions coming from one-loop effects survive. In what follows, we focus on precisely these terms.

\subsection{dS Fluctuations: one-loop}
\label{subs:Fl_det_realdS_sad}

Following the procedure described in the subsection \ref{SubS:PI_hij}, the path-integral over the fluctuations in the present case of dS leads to the determinant of a differential operator which can be computed using the Gel'fand-Yaglom method described in subsection \ref{sec:GY_method}. For the dS case, this differential operator is given by (see near eq. (\ref{eq:determin}) with $\varepsilon=1$)
\begin{equation}
\label{eq:pure-de-sitter-action}
\begin{split}
&\mathcal{D}_{h}^l=-\frac{d}{dt}\left(\frac{\bar{q}^2}{N_c}\frac{d}{dt}\right)-N_c l(l+2),\hspace{4mm}\text{with}\hspace{4mm} \bar{q}(t)=H^2N_c^2 \left(H^2+(t^2-1)H_1^2\right)/H_1^2.
\end{split}
\end{equation}
As explained in sec. (\ref{sec:GY_method}), for the Dirichlet type of boundary condition, the determinant of this operator according to the Gel'fand-Yaglom method is $ \det(\mathcal{D}_{h}^l)=u_2(1)$, where $u_2(t)$ satisfies the differential equation $\mathcal{D}_{h}^l u_2(t)=0$, with the boundary condition: $u_2(0)=0, \dot{u}_2(0)=N_c/\bar{q}(0)^2$ (see eq. \ref{eq:INC_cond_u1u2}). The set of independent solutions are $\mathbb{P}^{-\xi_l}_1\{i H_1 (H^2 - H_1^2)^{-1/2} t\}/\sqrt{\bar{q}(t)}$ and $\mathbb{Q}^{\xi_l}_1\{iH_1 (H^2 - H_1^2)^{-1/2} t\}/\sqrt{\bar{q}(t)}$, as discussed in sec. \ref{sec:GY_method}, where $\xi_l=\{1+H_1^2 l (l+2)/H^4\left(H^2-H_1^2\right) N_c^2\}^{1/2}$ and it simplifies to $l+1$ at both these saddles. Note that the argument of $\mathbb{P}$ and $\mathbb{Q}$ is independent of $N_c$. Solving for $u_2(t)$ with the above boundary conditions, we get
the determinant for each $l$-mode
\bea
\label{eq:det_simplify}
\D^{(l)}_h(N_c) && =\det\mathcal{D}_{h}^l \biggr \rvert_{N_c} =\frac{iH_1^3e^{-i\pi\xi_l}}{H^5N_c^3(H^2-H_1^2)} 
\notag \\
&&
\times \biggl[\mathbb{P}_1^{-\xi_l}\biggl(\frac{i H_1}{\sqrt{H^2 - H_1^2}}\biggr)\mathbb{Q}_1^{\xi_l}(0)
-\mathbb{Q}_1^{\xi_l}\biggl(\frac{i H_1}{\sqrt{H^2 - H_1^2}}\biggr)\mathbb{P}_1^{-\xi_l}(0)\biggr] \, .
\eea

This determinant is for generic $N_c$, implying it holds true for all complex metrics respecting the $\mathbb{R}\times S^3$ symmetry. For points in the complex-$N_c$ plane lying on the real axis (which correspond to Lorentzian geometries), the above determinant becomes real. This also includes the real-dS saddles mentioned in eq. (\ref{eq:de-sitter}). The above determinant is given for a particular $l$-mode. To compute the one-loop corrected lapse-$N_c$ action, these need to be exponentiated (similar to the eq. \ref{eq:h_PI_mode_degen}), leading to an infinite summation over the $l$-modes, each coming with a degeneracy factor $g_l$. The one-loop contribution to the lapse-$N_c$ action is given by
\begin{equation}
\label{eq:ds_lapse_action}
\mathcal{A}_h^{\rm dS}(N_c)=\frac{i\hbar}{2}\sum_{l=2}^\infty g_l\ln \, \D^{(l)}_h(N_c)  \, ,
\end{equation}
where $\D^{(l)}_h(N_c)$ is mentioned in eq. (\ref{eq:det_simplify}). 

It is expected that this will be divergent (as Einstein-Hilbert gravity is nonrenormalizable in 4D): (1) due to the infinite summation over the $l$-modes, and (2) contact divergences arising when the final boundary is close to the initial boundary. The former, in Hurwitz-Zeta regularization, doesn't explicitly show up, as the regularization procedure naturally crops it (see appendix \ref{sec:zeta_sum} for details). The later however, cannot be avoided and leads to UV-divergence. The contact divergences can be extracted by analysing the short-distance behaviour of the determinant. 

\begin{figure}
    \centering
    \includegraphics[width=0.5\linewidth]{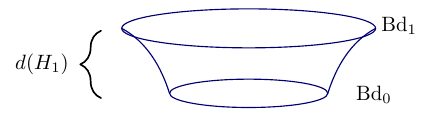}
    \caption{The Divergence shows up as the final hypersurface (${\rm Bd}_1$) approaches/contacts the initial hypersurface (${\rm Bd}_0$). The proper distance between the hypersurfaces $d(H_1)=H^{-1}\tanh^{-1} (H_1/H) \approx H_1/H^2$ approaches zero as the hypersurfaces come close together.}
    \label{fig:contact_divergence}
\end{figure}
The proper distance between two points $(t_1,r,\theta,\phi)$ and $(t_2,r,\theta,\phi)$ with the same spatial coordinate is defined via the integral $\int_{t_1}^{t_2}N_cdt/\sqrt{\bar{q}}$. If these two points are lying on the initial and final hypersurfaces, respectively, then the proper distance between them is given by
\begin{equation}
\label{eq:proper_distance}
d(H_1)=\int_0^1\frac{N_cdt}{\sqrt{\bar{q}(t)}}=\frac{1}{H}\tanh^{-1} (H_1/H) \approx H_1/H^2 \,\, ({\rm for} \,\, H_1 \ll H) \, ,
\end{equation}
This is depicted in figure \ref{fig:contact_divergence}. In the limit of the final hyper-surface approaching the initial one, leading to ``contact'', $d(H_1) \to 0$. This implies that the short distance limit corresponds to $H_1\to0$. In this limit, the determinant stated in eq. (\ref{eq:det_simplify}), for generic $N_c$, has the following expression 
\begin{equation}
\label{eq:det_small_H1}
\D^{(l)}_h(N_c) \biggr\rvert_{H_1\to 0}=\frac{H_1^4}{H^8 N_c^3}\biggl(1+\frac{4H_1^2}{3H^2}\biggr) 
+\mathcal{O}(H_1^8)
\end{equation}
In the path-integral, this determinant leads to a divergence, as it appears in the path-integral in the form $\{ \Delta_h^l(N_c) \}^{-g_l/2}$, where $g_l$ is the degeneracy factor. Clearly, for $H_1\ll H$, $\{\Delta_h^l(N_c) \}^{-g_l/2} \sim \bl(H_1\br)^{-2g_l}$, which is divergent for $H_1\to0$. 

The one-loop contribution to the lapse-$N_c$ action, mentioned in eq. (\ref{eq:ds_lapse_action}), accordingly contains a divergent contribution arising in limit $H_1\to0$. These need to be cropped by addition of suitable counter-term, leading to a renormalized lapse action, which is then utilized further-on in the Picard-Lefschetz analysis of the lapse-$N_c$ integral, needed for the computation of the wave-function. The one loop action in eq. (\ref{eq:ds_lapse_action}) in the small $H_1$ limit becomes the following
\begin{equation}
    \label{eq:log-div-l-sum}
\mathcal{A}_h^{\rm dS}(N_c)\biggr|_{H_1\rightarrow 0}=\frac{i\hbar}{2}\ln(H_1^4/H^8N_c^3)\sum_{l=2}^\infty g_l +\mathcal{O}(H_1^2),
\end{equation}
where the first term is the divergent which needs to be renormalized by adding a counterterm. Performing the sum over $l$-modes, we get the counterterm to be $\mathcal{A}_{\rm ct}^{\rm dS}(N_c)$, leading to the renormalized lapse action $\mathcal{A}_h^{\rm dS,\, ren}(N_c)$
\begin{equation}
\label{eq:renormalized_action}
\mathcal{A}_{\rm ct}^{\rm dS}(N_c)=-\frac{4i\hbar}{3}\ln(H_1^4/H^8N_c^3) 
\hspace{5mm}\Rightarrow\hspace{5mm}
\mathcal{A}_h^{\rm dS,\, ren}(N_c)=\mathcal{A}_h^{\rm dS}(N_c)+\mathcal{A}_{\rm ct}^{\rm dS}(N_c)
\, .
\end{equation}
We are interested in computing the wave function $\Psi_{\rm dS}$, as defined in eq. (\ref{eq:ds_path_qform}), including one-loop effects from fluctuations. This is to be computed by usage of Picard-Lefschetz methods by following the procedure described in sec. \ref{PL_int}, which leads to dS wavefunction following eq. (\ref{eq:LDordI_fin}), similar to the no-boundary case mentioned in eq. (\ref{eq:Gtrans_qf>3/lam_gen}). The dS-wavefunction $\Psi_{\rm dS}$ is given by
\begin{equation}
    \label{eq:wave_fun_ds}
    \displaystyle
   \Psi_{\rm dS}[{\rm{Bd_f}}]= \sum_{\sg=\pm} e^{i\theta_\sg }
\exp[i\mathcal{A}^{\rm dS, \,\,ren}(N^{\rm dS}_\sg)/\hbar]
\left|{\cal A}_{0}^{\prime\prime}(N_{\sg}^{\rm dS})\right|^{-1/2} \, ,
\end{equation}
where ${\cal A}_{0}(N_c)$ is background on-shell action mentioned in eq. (\ref{eq:lapse-action_de-sitter}) and $\mathcal{A}^{\rm dS, \,\,ren}(N^{\rm dS}_\sg)$ is the entire renormalized lapse action at the saddles including the background and fluctuation.
Interestingly, at the saddles, this quantity admits a simpler form as at the saddles $\xi_l =l+1$ (an integer). This allows $\mathbb{P}_1^{-\xi_l}(z)$ and $\mathbb{Q}_1^{\xi_l}(z)$ to be written as an algebraic function of $z$. The renormalized lapse action in eq.  (\ref{eq:renormalized_action}) at the saddles reduces to
\begin{equation}
\label{eq:det_simplify_real}
\begin{split}
&\mathcal{A}_h^{\rm dS,\, ren}(N_+^{\rm dS})=\frac{i\hbar}{4}\sum_{l=2}^\infty g_l\biggl[\ln\{(l+1)^2(H^2-H_1^2)+H_1^2\}+2\ln\sin\Theta(l,H_1)+\\
&\hspace{30mm}2\{\ln (H/H_1)-\ln(l)-\
\ln(l+2)\}\biggr], \\
&\text{where},\,\, \Theta(l,H_1)=(l+1)\tan^{-1}\biggl[\frac{H_1}{\sqrt{H^2-H_1^2}}\biggr]-\tan^{-1}\biggl[\frac{H_1}{(l+1)\sqrt{H^2-H_1^2}}\biggr].
\end{split}
\end{equation}
The above one-loop action is finite in the $H_1\rightarrow 0$ limit (UV-limit), as the divergence gets canceled by the counter terms. For the real negative saddle, $\Theta$ becomes $-\Theta$ and hence, $\Delta_h^l(N_+^{\rm dS})=-\Delta_h^l(N_-^{\rm dS})$, which has lead to emergence of $\sin$-function in the one-loop corrected lapse action. 

This is where the trouble arises, as $\sin \Theta(l, H_1)$ can become zero for some values of $l$ and/or $H_1$. One can explicitly find $l$'s that lead to vanishing $\sin \Theta(l, H_1)$ for a fixed $H_1$. When this happens, it leads to a divergence in the lapse action! This is a serious situation and resembles the ``pinching''-effect in thermal QFTs. One can get rid of such ``pinch''-singularities by either resummation methods or working in Schwinger-Keldysh formalism \cite{Millington:2012pf,Dadic:1998yd}. In this paper, we don't focus on ``pinch''-singularities as it is beyond the scope of this work. Partially, it may also be due to the Lorentzian background (real-dS saddles) on which we are computing the functional determinant. We realize these problems don't appear if we complexify the cosmological constant $\Lam$ by a tiny amount. In a sense, this leads to disturbing the real poles of the ``propagator'' by $i\ep$, as $\Lam$ leads to a mass-term in the $h_{ij}$-Hessian. We investigate this in more detail in next subsection to see if this overcomes these spurious divergences and eventually analyse the IR behaviour of the dS-wavefunction.

\subsection{Complexified de-Sitter and IR behaviour}
\label{subs:complex-de-sitter}

In the previous subsection, we noticed that classical boundary choice ($\pi_q=0$ at the initial hyper-surface and fixed extrinsic curvature on the final hyper-surface) leads to real saddles. Furthermore, it is seen that Picard-Lefschetz methods alone are not sufficient to compute the one-loop wavefunction, as the functional determinant becomes ill-defined. One cannot proceed with the lapse-$N_c$ integral via Picard-Lefschetz until one is able to overcome the spurious singularities in a systematic meaningful manner. 

\begin{figure}[hbtp]
\centering
\includegraphics[width=0.5\linewidth]{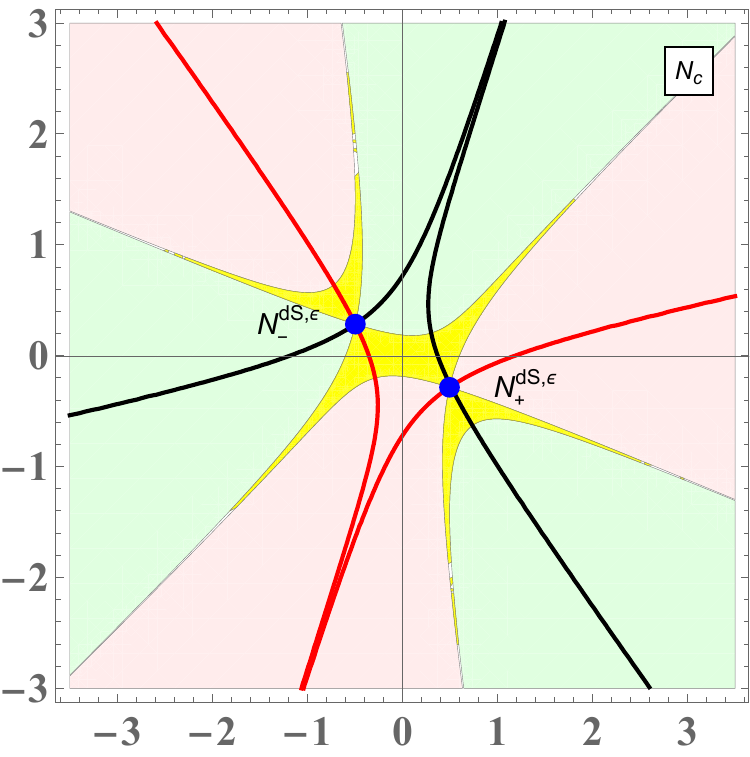}
\caption{PL plot with $\Lambda$ complex $\bar{\ep}=\pi/10$, $H_1=1/2$. Both complex saddles are relevant because the steepest ascent (red curve) from each intersects the integration contour. The original integration contour along the real line extends slightly into the disallowed region and thus needs to be rotated slightly to ensure the integral is well-defined \cite{Ailiga:2025fny}. The deformations to the saddles are magnified for illustration.}
\label{fig:ep_def_PL_plot}
\end{figure}

It is realised that this can be overcome by a slight complexification of the real saddles. If the real saddles are $\ep$-deformed in an ad hoc manner, there is no guarantee that the $\ep$-deformed real saddles will also be saddles of the system (which implies that $\xi_l$ won't be an integer at these saddles). To ensure that $\ep$-deformation is implemented in such a manner so that the $\ep$-deformed $N^{\rm dS}_\pm$ are also saddles, we deform $\Lam(=3H^2)$ as
\bea
\label{eq:Lam-defo}
&& \Lam \to \Lam e^{i\bar{\ep}}
\hspace{5mm}
\Rightarrow
\hspace{5mm}
N^{{\rm dS},\ep}_\pm = N^{\rm dS}_\pm \mp i \ep\, ,
\hspace{5mm}
\ep>0 \, ,
\\
{\rm where}
\hspace{5mm}
&&
\epsilon= \bar{\ep} \bl\{H_1\left(3 H^2-2 H_1^2\right) \left(H^2-H_1^2\right)^{-3/2}/(2H^2)\br\} \, ,
\notag 
\eea
is the leading correction to the saddles. Such $\ep$-modification of $\Lam$ affects the ``mass'' term in the $h_{ij}$-Hessian, in a sense {\it pushing the poles off the real-axis}.
One of the real-dS saddle gets pushed above the real line, while the other is pushed below. The sign of $\ep$ is chosen so as to provide necessary convergence in the path-integral for each of the $l$-modes. 
It is worth asking if $N^{{\rm dS},\ep}_\pm$ are {\it relevant}-saddles in the computation of the lapse-$N_c$ integral. As the steepest ascent flow lines emanating from them intersect the original integration contour, these $\ep$-deformed saddles remain {\it relevant} in the path-integral (see figure \ref{fig:ep_def_PL_plot}). Although the flow-line structure and the allowed/disallowed region are slightly modified by the $\ep$-deformation. It should also be noted that the original integration contour now partially lies in the disallowed region, while the Lefschetz thimbles remain in the allowed region. 

Once the $\ep$-deformation is carefully achieved, we proceed further with the computation of the one-loop functional determinant. The differential operator stated in eq. (\ref{eq:pure-de-sitter-action}) gets $\ep$-modification 
\footnote{As we are working with exponential parametrization, there is no additional $\Lam$-dependent ``mass" term in the $\mathbb{D}$, which will come if one works with linear parametrization.}
\bea
\label{eq:diff_op_change}
&&
\mathcal{D}_{h}^l\rightarrow \mathcal{D}_{h}^{l,\ep} = \mathcal{D}_{h}^l+i\epsilon \,\mathbb{D}\, ,
\hspace{10mm}
\mathbb{D}=-\frac{d}{dt}\left(A(t)\frac{d}{dt}\right) 
\, ,
\notag \\
{\rm where}
&& 
A(t)=[2 H^4 N_c^3 \left\{H^2+H_1^2 \left(t^2-1\right)\right\} \left\{2 H^2+H_1^2 \left(t^2-1\right)\right\}]/H_1^4 \, , 
\eea
The small-$\ep$ deformation of the differential operator brought out due to the complex deformation of $\Lam$ offers a way to compute the functional determinant in Lorentzian dS where spurious singularities are bypassed, as will be seen below. Note, this is the correction at all points in the complex-$N_c$ plane. If we now consider this operator at the corrected saddles, then this operator becomes the following
\bea
\label{eq:diff_op_change_at_saddle}
&&
\mathcal{D}_{h}^l\rightarrow \mathcal{D}_{h}^{l,\ep} = \mathcal{D}_{h}^l+i\epsilon \,\mathbb{D}_\pm \, ,
\hspace{10mm}
\mathbb{D}_\pm=\frac{d}{dt}\left(A_\pm(t)\frac{d}{dt}\right)\pm l(l+2) 
\, ,
\notag \\
{\rm where}
&& 
A_\pm(t)=\pm \frac{z t^2+1}{z \bl\{z +3\br\}}
\bl\{3 z^2 t^2+ z \left(5 t^2-1\right)+1\br\} \, ,
\eea
and $z = H^4 (N_\pm^{\rm dS})^2$. Interestingly, at the corrected saddles $N^{{\rm dS},\ep}_\pm$, $\xi_l = l+1$ (it receives higher order ($\ep^2$) correction).

The one-loop determinant in the case of small $\ep$-deformation can be computed either by directly applying the Gelfand-Yanglom method to the operator stated in eq. (\ref{eq:diff_op_change}) or by realizing that the differential operator $\Delta_h^l$ and $u_2(1)$ are analytic in $\Lam$ for $\Lam>3H_1^2$ (the Lorentzian regime we are working with) for generic $N_c$. This later knowledge can be exploited to easily obtain the $\det\mathcal{D}_{h}^{l,\ep} =\Delta_h^{l,\ep}$ by using eq. (\ref{eq:det_simplify}) and replacing $\Lam \to \Lam e^{i\bar{\ep}}$. The complete one-loop corrected action can be obtained from eq. (\ref{eq:ds_lapse_action}) by summing over all the $l$-modes. As before, the one-loop action will be divergent in the limit $H_1\to 0$. The leading divergence will be the same, and is obtained by replacing $H^2\to H^2e^{i\bar{\ep}}$ in eq. (\ref{eq:det_small_H1}). This leads to the same counterterm as given in eq. (\ref{eq:renormalized_action}).

Moreover, in the computation of $\Psi_{\rm dS}$ using Picard-Lefschetz and saddle-point methods (see eq. \ref{eq:wave_fun_ds}), we need the determinant only at the saddles. The $\ep$-corrected one-loop determinant (for each $l$-mode) at the $\ep$-deformed saddle simplifies to 
\bea
\label{eq:det_com_sad_plus}
&&
\Delta_h^{l,\ep}\biggr\rvert_{N_+^{{\rm dS},\ep}}
=-\Delta_h^{l,\ep}\biggr|_{N_-^{{{\rm dS},\ep}}}
= \frac{H}{l(l+2)}\sqrt{(l+1)^2(H^2-H_1^2)+H_1^2}
\notag \\
&&
\times 
\biggl[
\sin\Theta(l,H_1)
+ \frac{i \ep}{H_1}\biggl\{ 
\sin\Theta(l,H_1) A(z,H)
- H_1\cos\Theta(l,H_1) B(z,H)
\biggr\} + {\cal O}(\ep^2)
\biggr] \, ,
 \\
&& A(z,H)=\frac{H^3\left(l^2 (z+2)+2 l (z+2)+2 (z+1)\right)}{\sqrt{z+1} (z+3) [(l+1)^2+z]},\hspace{3mm} B(z,H)= \frac{ H^2 l(l+1)(l+2)  }{(z+3)[(l+1)^2+z]}\, .
\notag
\eea
Note that $\Lambda$ deformation introduces an $i\ep$ correction to the determinant.
The renormalized lapse action at the deformed saddles is given by
\bea
\label{eq:ren_action_eo_deform}
&&\mathcal{A}_{h,\ep}^{\rm dS,\, ren}(N_+^{{\rm dS},\ep})=\frac{i\hbar}{4}\sum_{l=2}^\infty g_l\biggl[\ln\{(l+1)^2(H^2-H_1^2)+H_1^2\}+2\{\ln (H/H_1)-\ln(l)-\ln(l+2)\}\notag\\
&&+2\ln\biggl[\sin\Theta(l,H_1)
+ \frac{i \ep}{H_1}\biggl\{ 
\sin\Theta(l,H_1) A(z,H)
- H_1\cos\Theta(l,H_1) B(z,H)
\biggr\}+\mathcal{O}(\ep)^2\biggr]\biggr].
\eea
The above one-loop action reduces to eq. (\ref{eq:det_simplify_real}) when $\ep=0$. However, for non-zero $\ep$, the terms in the second line, the quantity inside the logarithm is complex and always non-vanishing. The ``pinch '' singularities which appear for vanishing $\sin\Theta$ are regulated by $i\ep$ modification. The one-loop renormalized lapse action is well-defined for all values of $l$ and $H_1$. The $l$- summation appearing in the first line can be done exactly, while the sum in the second line is quite non-trivial and can only be analyzed in the limit $H_1\to H$. This corresponds to the IR limit. Unlike the no-boundary case, where the series can be summed using the Hurwitz-Zeta function, in the present case, it is not possible. The wave function with the $\ep$ deformation can be obtained from eq. (\ref{eq:wave_fun_ds}). It is given by
\begin{equation}
\label{eq:wave_function_ep_def}
\begin{split}
\Psi_{\rm dS}^{\ep}[H_1]=&\frac{\sqrt{H_1}\exp\biggl[i\mathcal{A}_{h,\ep}^{\rm dS,\, ren}(N_+^{{\rm dS},\ep})/\hbar\biggr]}{\sqrt{6}H(H^2-H_1^2)^{1/4}}\biggl[\exp\biggl\{-\frac{i\pi}{4}+\frac{i}{\hbar}\frac{2H_1}{H^2\sqrt{H^2-H_1^2}}\biggr\}\\
&+\exp\biggl\{-\frac{13 i \pi}{12}-\frac{i}{\hbar}\frac{2H_1}{H^2\sqrt{H^2-H_1^2}}\biggr\}\biggr],
\end{split}
\end{equation}
where we sum over two saddles $N_\pm^{{\rm dS},\ep}$ as both are relevant and contribute to the wave function; see figure \ref{fig:ep_def_PL_plot}. The quantity $\sqrt{H_1}/\sqrt{6}H(H^2-H_1^2)^{1/4}$ is the WKB prefactor $[{\cal A}_{0}^{\prime\prime}(N_{\sg}^{\rm dS})]^{-1/2}$, see eq. (\ref{eq:wave_fun_ds}). The phase $-\pi/4$ arises from $\theta_\sigma$ for $N_+^{{\rm dS},\ep}$ (where $\ep$ correction is ignored) and $-13i\pi/12$ originates from two sources one is the $\theta_\sigma$ another is the extra overall minus sign present in determinant at $N_-^{{\rm dS},\ep}$ saddle, see eq. (\ref{eq:det_com_sad_plus}). $\Psi_{\rm dS}^{\ep}$ is complex because the saddles are complex and, unlike the no-boundary saddles, are not complex conjugates of each other, see appendix \ref{sec:symm} for details. Working in the leading order in $\epsilon$, the renormalized action $\mathcal{A}_{h,\ep}^{\rm dS,\, ren}(N_+^{{\rm dS},\ep})$ takes the following simplified form 
\begin{equation}
    \label{eq:asymptotic_limit_de-sitter}
    \begin{split}
    &\mathcal{A}_{h,\ep}^{\rm dS,\, ren}(N_+^{{\rm dS},\ep}) \\
    \sim &\frac{i\hbar}{2}\sum_{l=2}^\infty(l-1)(l+3)\biggl\{\ln\biggl(l+1+i\frac{H_1}{\sqrt{H^2-H_1^2}}\biggr)+\ln\biggl(l+1-i\frac{H_1}{\sqrt{H^2-H_1^2}}\biggr)\\
    &+\ln(H^2-H_1^2)+2\{\ln (H/H_1)-\ln(l)-\ln(l+2)\}+2i\Theta_l+ \underbrace{2\ln\left[1-e^{-2i\Theta_l-2\ep B}\right]}_{\mathcal{R}_\ep(l,H_1)}\biggr\}.
    \end{split}
\end{equation}
In the above expression, we retain terms linear in $\ep$ and exponentiate it for later convenience. All the summations except the remainder term $\mathcal{R}_\ep(l,H_1)$ term can be computed exactly using the Hurwitz-zeta regularization as explained in appendix \ref{sec:zeta_sum}. It is interesting to note that the structure of the logarithms in the first line above is the same as we noticed earlier for the no-boundary saddle; see eq. (\ref{eq:log_exp_wl_nb_10}). The quantity $(\tau_1^{\rm nb}+l+1)$ is exactly the same to $(l+1\pm iH_1/\sqrt{H^2-H_1^2})$ at the two saddles. This arises because the no-boundary geometry asymptotes to de Sitter in the late time; see figure (\ref{fig:geometry_contour}). Likewise, the term $i\Theta_l$ yields exactly the same phase as obtained for the no-boundary universe. Hence, except for the summation of the remainder terms $\mathcal{R}_\ep$, in the IR limit, one obtains exactly the same behaviour. Any deviation from the no-boundary is thus encoded in this remainder sum. Note that when $\epsilon>0$, $\mathcal{R}_\ep(l,H_1)$ is convergent for large $l$ and hence the sum doesn't require any regularization. This explains the direction of rotation ($\ep$) as considered in eq. (\ref{eq:Lam-defo}). This sum can only be computed in the IR limit ($H_1\rightarrow H$), yielding the leading correction to the coefficients; see appendix \ref{sec:summation_reminder}. In this limit, the $l$-modes that affect the coefficients of the leading and subleading terms are of order $l\sim \mathcal{O}(1/\sqrt{H-H_1})$. Hence, one requires uniform expansion by taking $l\rightarrow\infty$ and $H_1\rightarrow H$ simultaneously and extracting the leading terms. This we explicitly discussed in appendix  \ref{sec:summation_reminder}. Explicitly, the leading IR contribution from the remainder sum is given by (where we define dimensionless $\bold{H}_1=H_1/H$)
\begin{equation}
    \label{eq:reminder_sum_main}
    \begin{split}
    &\sum_{l=2}^\infty (l-1)(l+3)\ln\left[1-e^{-2i\Theta_l-2\ep B}\right]\biggr|_{H_1\rightarrow H}\\
    &=\underbrace{\frac{I_2}{4\sqrt{2}}}_{\delta_{3/2}}\frac{1}{(1-\bold{H}_1)^{3/2}}+\underbrace{\frac{-3 I_2-88 I_0}{48 \sqrt{2}}}_{\delta_{1/2}}\frac{1}{(1-\bold{H}_1)^{1/2}}+\mathcal{O}(\ln(1-\bold{H}_1)),\\
    \text{where}, \quad & I_0\approx -2.51429 - 1.71889 i,\,\, I_2\approx 0.286291-0.460498 i.
    \end{split}
\end{equation}
First, note that the coefficients in the above equation are complex; hence, it will affect both the exponential magnitude and phase of the wave function. After performing the summation, we set $\ep=0$ at the end, and it doesn't introduce additional divergences. The precise role of introducing the regulator by slightly complexifying the cosmological constant($\Lambda$) is to sensibly define the above summation so that one can compute it, and in the end, one takes $\ep=0$ consistently. Second, note that the above sum only introduces terms with scales of $1/(1-\bold{H}_1)^{3/2} $ and $ 1/(1-\bold{H}_1)^{1/2}$, which are already present in the determinant. This precise scaling of terms in de Sitter space matches that of the no-boundary universe. This matching of the scaling of the leading terms between the de Sitter and the no-boundary Universe is expected as the latter approaches de Sitter asymptotically at late times. However, this asymptotic agreement of the geometries does not always ensure that the corresponding numerical coefficients in front of the terms coincide. In the present case, they are different. Explicitly, the difference in the one-loop between the de Sitter and no-boundary saddle is 
\begin{equation}
\label{eq:log_sum_ren_act_ds}
\begin{split}
&\mathcal{A}^{\rm dS,\, ren}(N_+^{{\rm dS}}) -\mathcal{A}_{\rm ren}^{\rm \zeta-reg}(N_+^{\rm nb})\biggr|_{1-{\rm loop}}\overset{\bold{H}_1\rightarrow 1}{\Longrightarrow} i\hbar\biggl[\frac{\delta_{3/2}}{(1-\bold{H}_1)^{3/2}}+\frac{\delta_{1/2}}{(1-\bold{H}_1)^{1/2}}+\mathcal{O}(\ln(1-\bold{H}_1))\biggr], 
\end{split}
\end{equation}
where $\mathcal{A}_{\rm ren}^{\rm \zeta-reg}(N_+^{\rm nb})$ is for the no-boundary saddle as in eq. (\ref{eq:Afin_ReIm_fixed_k}) and $\mathcal{A}^{\rm dS,\, ren}(N_+^{{\rm dS}})$ is the similar one-loop action at the corresponding de-Sitter saddle. Plugging eq. (\ref{eq:asymptotic_limit_de-sitter}) and (\ref{eq:reminder_sum_main}) in the expression for the wave function in eq. (\ref{eq:wave_function_ep_def}), we get, in the IR limit (restoring $8\pi G$)
\begin{equation}
    \label{eq:de-sitter_ep_def_wave_fun}
    \begin{split}
     &\Psi_{\rm dS}[H_1\rightarrow H]\sim \frac{1}{(1-\bold{H}_1^2)^{1/4}}\,\,\,
     \exp\biggl[\left(\frac{\pi}{12\sqrt{2}}-\delta_{3/2}\right)\frac{1}{(1-\bold{H}_1)^{3/2}}+\left(\frac{29\pi}{16\sqrt{2}}-\delta_{1/2}\right)\frac{1}{\sqrt{1-\bold{H}_1}}\\
     &-i\biggl\{\frac{1}{36\sqrt{2}(1-\bold{H}_1)^{3/2}}\{11+\ln8+3\ln(1-\bold{H}_1)\}+\frac{29}{16\sqrt{2}(1-\bold{H}_1)^{1/2}}\ln(1-\bold{H}_1)\biggr\}\cdots\biggr]\\
     &\biggl[\exp\biggl\{\frac{i}{\hbar}\frac{V_3}{4\pi\sqrt{2} GH^2}\frac{1}{\sqrt{1-\bold{H}_1}}-\frac{i\pi}{4}\biggr\}
+\exp\biggl\{-\frac{i}{\hbar}\frac{V_3}{4\pi\sqrt{2} GH^2}\frac{1}{\sqrt{1-\bold{H}_1}}-\frac{13 i\pi}{12}\biggr\}\biggr].
\end{split}
\end{equation}
The above expression is the final form of the de Sitter wave function in the infrared regime, including the one-loop corrections arising from graviton fluctuations. The coefficients of the leading and subleading terms in the exponential magnitude are different from those in the no-boundary universe. The difference is encoded in ``$\delta_{3/2}$" and ``$\delta_{1/2}$". In fact, we further found that the coefficients are smaller than those for the no-boundary universe as ${\rm Re}\,\delta_{3/2}>0,{\rm Re}\,\delta_{1/2}>0$. The leading one-loop contribution scales ($\sim\exp(a\times\,\,\text{spatial volume}),a>0$) which is the same as found earlier in the no-boundary wave function in eq. (\ref{eq:H_1<H_limit}).  The difference in the IR coefficients needs proper explanation, which we elaborate in sec. \ref{sec:uv_dependence}. For completeness, we now verify the KSW-allowability of the $\ep$-deformed complex-dS saddles and also justify the complexification (which is also consistent with the convergence of the mode sum) in the light of the KSW criterion, which puts restrictions on the allowable complex metrics in the path integral.
\subsection{KSW allowability of complex de-Sitter saddles}
\label{sec:allow_compl_sad}

\begin{figure}[hbtp]
    \centering
    \subfigure[$H_1=0.5,\ep=0.02$]{\includegraphics[width=0.44\linewidth]{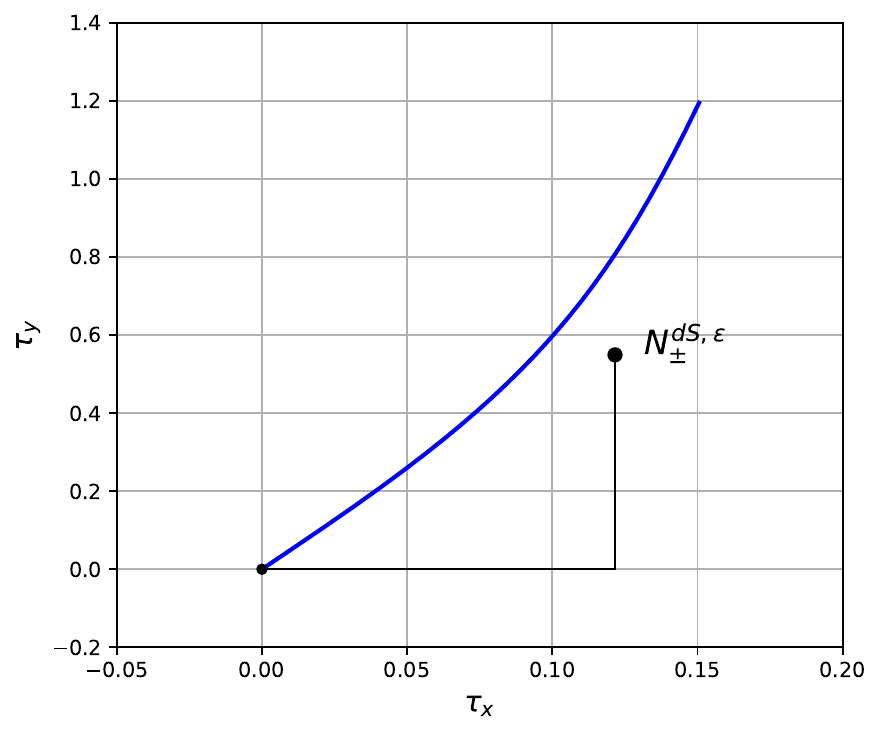}}
    \subfigure[$H_1=0.8,\ep=0.01$]{\includegraphics[width=0.44\linewidth]{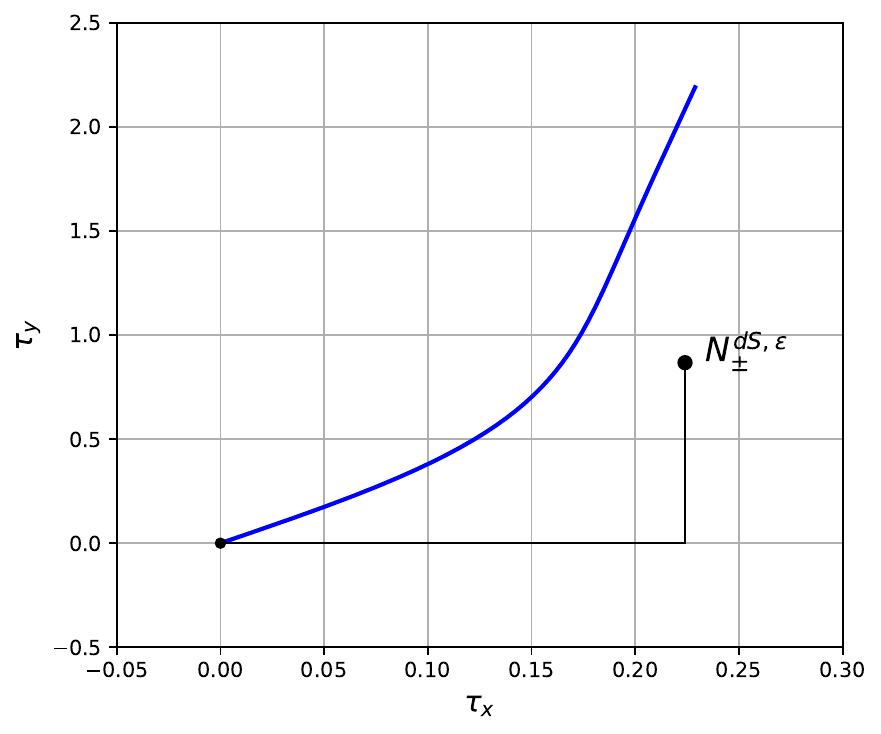}}
    \caption{Plots showing the KSW allowability of the complex de-Sitter saddles for various parameters; we set $H=1$. As the black dot (${\rm Re}\,(\nu),{\rm Im}\,(\nu)$) is always below the extremal curve (blue line), the complex saddles are KSW allowed. The initial angle depends on the parameters. Clearly, the parameters satisfy the bound, as mentioned in the text.}
    \label{fig:KSW_allobility}
\end{figure}

To overcome the difficulties encountered in computing the wave function for Lorentzian de Sitter, we slightly complexified the saddles in such a way that they remain saddles of the path integral. This is achieved by allowing $\Lambda$ to be slightly complex. As explained in sec. (\ref{subs:complex-de-sitter}) and also near eq. (\ref{eq:reminder_sum_main}), such complexification ensures convergence of the $l$-mode summation and helps us to extract the correct IR coefficients. Such a complexification is also quite natural upon realising that the real geometries are, in fact, not allowable by the KSW criterion; they lie on the boundary of the allowable domain \cite{Kontsevich:2021dmb, Witten:2021nzp}. One requires a slight complexification to push the real saddles into the allowable domain. Such a criterion has been shown to be useful for discarding complex geometries that would otherwise appear pathological in the path integral in cosmology; see, for example, \cite{Jonas:2022uqb, Lehners:2021mah, Ailiga:2025fny, Ailiga:2026wju, Hertog:2023vot}. In this section, we analyze the compatibility of such complexification with the KSW criterion. In particular, we are interested in whether the complex de Sitter saddles given in Eq. (\ref{eq:Lam-defo}) are KSW-allowable.  
As explained in sec. \ref{subsec:KSW_criterion}, we proceed by defining the euclidean time coordinate $\tau_p(t)$: ${\rm d} \tau_{p} = i N_c/\sqrt{\bar{q}(t)}$, where at the $\bar{q}(t)$ is given in eq. (\ref{eq:pure-de-sitter-action}). Performing the integration, we get with the initial condition $\tau_p(0)=0$:
\begin{equation}
    \label{eq:tau_p}
    \tau_p(t)=\frac{i}{H}\text{arctanh} \left(\frac{t H_1}{\sqrt{H^2+(t^2-1)H_1^2}}\right).
\end{equation}
Note that $\tau_p(t)$ is independent of $N_c$ and the same for all saddles. In $\tau_p$, coordinate $\bar{q}(t)$ takes becomes $\bar{q}(\tau_p)=N_c^2 H^2\left(H^2-H_1^2\right) \cos ^2(H \tau_p )/H_1^2$.
The equation of the extremal curve ($\tau_e$) which follows from eq. (\ref{eq:ext_curve_1}) and eq. (\ref{eq:extremal_curve}) can be expressed as
\begin{equation}
    \label{eq:extremal_curve_de-siiter}
    {\rm Re}\int_0^{\tau_p} \bar{q} (\tau_p')^{3/2} d\tau_p'\biggr|_{\tau_p=\tau_x+i\tau_y}=0.
\end{equation}
The above integration can be performed exactly. Substituting the deformation as given in eq. (\ref{eq:Lam-defo}) and equating the real part to zero, one obtains the extremal curve for a given saddle in the complex $\tau_p$-plane. The saddles will be KSW allowed if $|\tau_y|>|{\rm Im}(\nu)|$ at $\tau_x = \rm Re(\nu)$, where $\nu=\tau_p(t=1)$. For the classical boundary condition, $\nu$ is the same at both saddles.

At $t=0$, $q(=0)=H^2N^2(H^2-H_1^2)/H_1^2$ which is complex with the $\Lambda$-deformation.
Evaluating the spatial part of the extremal condition in eq. (\ref{eq:ext_curve_1}), we find that the allowed deformation ($\bar{\epsilon}$) should be such that $3|\arg(q(0))|<\pi$. Evaluating this at the deformed saddles as given in eq. (\ref{eq:Lam-defo}), we get the bound $|\bar{\ep}|<\sqrt{3}(H^2-H_1^2)/(3H^2-2H_1^2)$. As $H_1$ approaches $H$ in the asymptotic future, the allowed value of $\bar{\ep}$ becomes smaller. This is not much of an obstruction, as one requires only an infinitesimal deformation to complexify the geometries.  From the figure \ref{fig:KSW_allobility}, we conclude that the deformed saddles are always allowed as they lie below the extremal curve. This establishes that the complexification is consistent with the KSW criterion for the allowable complex metrics.

\section{UV dependence in IR} 
\label{sec:uv_dependence}
\begin{figure}[hbtp]
\centering
\includegraphics[width=0.72\linewidth]{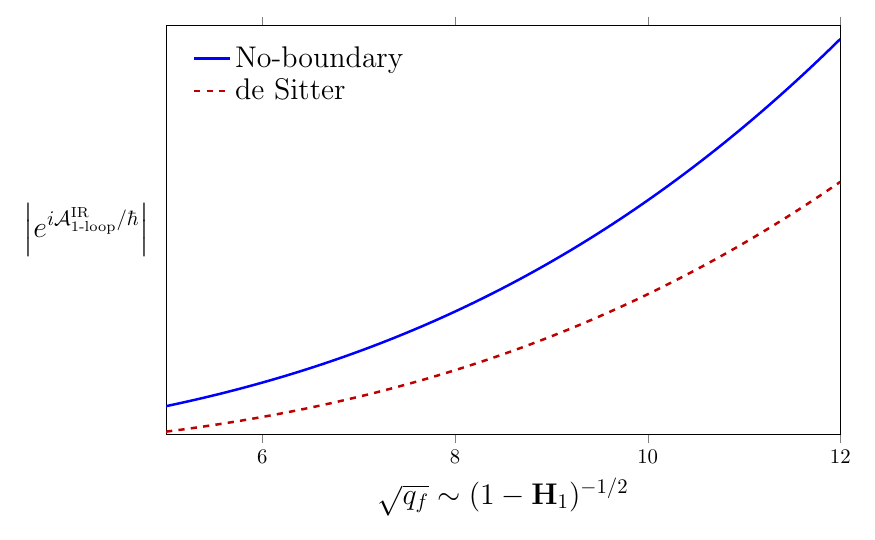}
\caption{The plot shows the absolute value of the one-loop determinant ($|e^{iA_{1-{\rm loop}}^{\rm IR}/\hbar}|$) at the saddle in the infrared where the size of the universe $\sqrt{q_f}\sim(1-\bold{H}_1)^{-1/2}$ becomes large. The blue/red curve depicts the no-boundary/de-sitter saddles. Both curves are monotonically increasing, indicating qualitatively similar IR divergence. However, the two curves differ, indicating the sensitivity of the IR coefficients to UV fluctuation conditions, as explained in the text.}
\label{fig:uv_dependence_in_IR}
\end{figure}
In this section, we explain the difference observed in the IR behaviour, specifically in the leading IR coefficients between the Lorentzian de-Sitter and no-boundary universe. From the expression for the wave function in eqs. (\ref{eq:H_1<H_limit}) and (\ref{eq:de-sitter_ep_def_wave_fun}) for No-boundary and de-Sitter, respectively, we found that although the leading/subleading behaviour in the deep IR regime is qualitatively the same between them, the precise numerical coefficients are found to be different. Both geometries have the same future asymptotes; see fig. (\ref{fig:geometry_contour}). However, this agreement doesn't guarantee that the leading/subleading coefficients will be the same. It only fixes the qualitative scaling behaviour, which is of order $(1-\bold{H}_1)^{-3/2}$ and $(1-\bold{H}_1)^{-1/2}$. The numerical coefficients depend on the details of the state of fluctuation on the initial hypersurface. Let us consider the absolute value of the one-loop correction evaluated at the saddle. In the IR limit, one gets at $N_{\pm}^{\rm nb}$ and $N_{\pm}^{\rm dS}$ saddles, respectively
\begin{equation}
\label{eq:absolute_one_loop_IR}
\begin{split}
&\biggl|e^{i\mathcal{A}^{\rm IR}_{1-{\rm loop}}/\hbar}\biggr|_{N_{\pm}^{\rm nb}}\sim \exp\biggl[\frac{\pi}{12\sqrt{2}}\frac{1}{(1-\bold{H}_1)^{3/2}}+\frac{29\pi}{16\sqrt{2}}\frac{1}{\sqrt{1-\bold{H}_1}}+\cdots\biggr]\\
& \biggl|e^{i\mathcal{A}^{\rm IR}_{1-{\rm loop}}/\hbar}\biggr|_{N_{\pm}^{\rm dS}}\sim \exp\biggl[\biggl(\frac{\pi}{12\sqrt{2}}-\Re(\delta_{3/2})\biggr)\frac{1}{(1-\bold{H}_1)^{3/2}}+\biggl(\frac{29\pi}{16\sqrt{2}}-\Re(\delta_{1/2})\biggr)\frac{1}{\sqrt{1-\bold{H}_1}}+\cdots\biggr],\\
&\text{where},\quad\Re(\delta_{3/2})\approx0.0506096,\quad\Re(\delta_{1/2})\approx 3.24678.
\end{split}
\end{equation}
The real parts of $\delta_{3/2}$ and $\delta_{1/2}$ modify the numerical coefficients in front of the leading/subleading term but never change the sign.
In the fig. \ref{fig:uv_dependence_in_IR}, we have plotted the above one-loop contributions as a function of the final size of the universe ($\sqrt{q_f}\sim(1-\bold{H}_1)^{-1/2}$).
To understand the difference in the coefficients, we realize that the no-boundary cap with the $h_{ij}=0$ condition on the initial(UV) surface prepares the state of fluctuations on the de-Sitter throat, which is different from the Dirichlet. Specifically, it prepares a state which imposes the following complex Robin condition on each $l$-mode of fluctuation at the de Sitter throat
\begin{equation}
    \label{Eq:robin_condition}
  \text{NB cap}\qquad\Longrightarrow\qquad  \dot h_l(0)=iN_cH^2\frac{l(l+2)}{l+1}h_l(0).
\end{equation}
The above boundary condition follows directly from eq. (\ref{eq:hLt_nbc_nb_form}).
Since we chose the Dirichlet condition on the initial dS hypersurface (at the throat) ($h_l=0$), which is completely different from what the no-boundary state prepares (eq. (\ref{Eq:robin_condition})), the precise numerical coefficients of the leading IR terms turn out to be different. This strongly suggests that the leading and subleading coefficients in the IR are sensitive to the state of fluctuation on the initial hypersurface. We have checked that, indeed, when the fluctuation condition on the initial dS-hypersurface is taken to be eq. (\ref{Eq:robin_condition}), the one-loop determinant yields identical IR coefficients. Thus, we find that the IR divergence behaviour is quite generic in one-loop computations in universes with a positive cosmological constant and is present for a variety of initial conditions. However, the difference in the coefficients suggests that the IR behaviour is sensitive to the details of the UV fluctuation condition imposed on the initial hypersurface. This dependence of IR behaviour on the UV boundary condition is an important conclusion of our work and hasn't been discussed previously. This completes our comparison of the IR behaviour of the complex Hartle–Hawking Universe with that of de Sitter, thereby addressing the question posed in the Introduction.

\section{Conclusions and outlook}
\label{sec:conclusion}
In this paper, we study the path integral over complex geometries with a positive cosmological constant at the one-loop order. Our primary goal is to understand the IR behaviour of complex saddles and investigate how different boundary conditions might affect it. We focus on two types of complex saddles: the no-boundary saddle and the complex de Sitter saddle. The one-loop IR behaviour of no-boundary saddles was previously studied in \cite{Ailiga:2024wdx} with linear Dirichlet boundary conditions. Here, we first extend the analysis by incorporating a nonlinear boundary condition that fixes extrinsic curvature ($K$). 
We found that in both cases the leading/subleading terms in the one loop exhibit similar IR behaviour, independent of the metric parametrization (see eqs. (\ref{eq:H_1<H_limit}), (\ref{eq:Gtr_qf>3/lam_LG})); thereby establishing the robustness of the result. However, we notice that for fixed $K$, the leading one-loop correction to the phase exhibits a potentially different scaling from the background; see eq. (\ref{eq:H_1<H_limit}).
Then we proceed to study the IR behaviour of complex de-Sitter saddles and compare it with that of no-boundary saddles to assess how the choice of initial/UV boundary condition might influence the IR behaviour. We found that although the leading/subleading IR-divergent terms scale exactly the same way as in the no-boundary case, the precise numerical coefficients in front of them differ between the two scenarios. The difference in the coefficient can be understood from the difference in the state of fluctuation on the initial dS-hypersurface. Hence, our study finds evidence that the leading IR coefficients are sensitive to the details of the UV fluctuation condition; see eqs. (\ref{eq:log_sum_ren_act_ds}), (\ref{eq:de-sitter_ep_def_wave_fun}).  

Path integral requires a set of boundary conditions for both the background and the metric fluctuation field. As noticed in \cite{Brizuela:2023vmb, Ailiga:2024wdx}, these boundary conditions are interlinked via the general covariance. Given a boundary condition on the background, it automatically determines the allowed boundary condition for fluctuations. Focusing on the (H-H) no-boundary condition for the background, we derive the allowed boundary choices for the fluctuation. At the initial hypersurface, we impose a general linear Robin boundary condition, while at the final hypersurface, we fix either the size (Dirichlet condition) or the extrinsic curvature/Hubble radius (conformal condition) of the hypersurface, consistent with the no-boundary proposal. While the other boundary conditions are linear, fixing curvature is non-linear. The allowed boundary choices for metric fluctuations, in general, can be complicated non-linear combinations of generalized coordinates and momenta; for example, as seen in the case of linear split \cite{Ailiga:2024wdx}. However, we observe that when the fluctuation is parametrized as an exponential, all the allowed boundary conditions become ``linear"; see eq. (\ref{eq:allowed_bc_exp_para}). This is a key advantage of using the exponential parametrization. We show this by introducing a parameter $\varepsilon$ into the metric ansatz and ensuring that it satisfies well-posed variation; the linear split corresponds to $\varepsilon=0$, while the exponential parametrization corresponds to $\varepsilon=1$. We find that boundary conditions become linear only for the special value of $\varepsilon=1$.

Having figured out the appropriate boundary conditions for both the background ($q(t)$) and the fluctuation ($h_{ij}$), we next study the one-loop path integral. To be consistent with the No-boundary condition, we take the fluctuation to be vanishing (Dirichlet) on the initial hypersurface. We first perform gauge fixing and compute the ghost determinant. We then use the background field formalism to compute the remaining path integral over $q(t)$ and $h_{ij}$. Considering the on-shell fluctuation to vanish on the final hypersurface, we get rid of non-diagonal terms in the quadratic fluctuation. The path integral over the background can be computed exactly for the linear boundary conditions. The fluctuation path integral is computed up to one loop using the Gelfand-Yaglom method. This method uses the generalized zeta function to compute the determinant of any second-order Sturm-Liouville operator. We observe that for exponential parametrization, the differential operator takes the same form as that of a massless minimally coupled scalar field.

Before computing the one-loop corrections, we analyze the semiclassical stability of these saddles when the on-shell fluctuations are non-vanishing at the final hypersurface (Dirichlet conditions). While for fixed size, the stability is known to be stable and Gaussian \cite{DiTucci:2019bui, Narain:2021bff, Feldbrugge:2017mbc, Narain:2022msz, DiTucci:2019dji, Ailiga:2024wdx, Lehners:2023yrj, DiTucci:2018fdg, Feldbrugge:2017fcc, Ailiga:2024mmt, Matsui:2024bfn}; for fixed $K$, such a study has not been done previously. For fixed $K$, we find that the fluctuation remains Gaussian, although its phase differs from that obtained under fixed-size boundary conditions; see eq. (\ref{eq:h1_equal_h_limit}). We also note that when fixing $K$, such a Dirichlet condition on the fluctuation is allowed only for the exponential parametrization ($\varepsilon=1$). This further highlights the utility of the exponential parametrization in cosmology, as it naturally accommodates physically motivated boundary conditions. Then we also show that the fixed-$K$ boundary condition is compatible with the KSW criterion, since complex saddles are always allowed.

Now we proceed to evaluate the one-loop fluctuation determinant at the complex no-boundary saddles. At these saddles, there is UV divergence, as expected, since we are dealing with Einstein-Hilbert gravity. We extract these divergences and renormalize them by adding an appropriate counterterm. We find the counterterms for both boundary conditions: fixed-size and fixed-curvature cases. We observe that these counterterms contain a part that is independent of the boundary conditions and another that depends on them; see eq. (\ref{eq:Nc_act_total_nb_ct}). However, these counterterms are independent of parametrization ($\varepsilon$). The resulting renormalized lapse action contains an infinite summation over the tensor modes of fluctuations. We regularize this infinite sum using Hurwitz zeta regularization, a widely used technique in QFTs in curved spacetime \cite{Monin:2016bwf, Elizalde:1994gf, Hawking:1976ja} and obtain the finite part. This method allows us to extract the physically meaningful finite part by cropping the divergent piece. Finally, we utilize the renormalized and $\zeta$-regularized one-loop lapse action to compute the Hartle-Hawking wave function utilizing the Picard-Lefschetz method and WKB approximation. We evaluate the wave function in both the Lorentzian and Euclidean regimes, and for both fixed final size ($q_f$) and fixed extrinsic curvature ($K$). In the Lorentzian regime, two anti-linear no-boundary saddles contribute to the path integral, yielding a real wave function. In the Euclidean regime, the saddle becomes purely imaginary, and the wave function remains real.

We analyze the asymptotic behaviour of the wave function in the deep Lorentzian regime (IR). These correspond to either $q_f\rightarrow\infty$ (fixed size) or $H_1\rightarrow H$ (fixed curvature $K=3H_1$). We observe that in these limits the contribution from the one-loop determinant grows as the universe expands. The growth is proportional to the exponential of the volume of the spatial hypersurface and independent of the parametrization. This conclusion agrees with that reported earlier in \cite{Ailiga:2024wdx}, where a different regularization method was used to regularize the infinite sum. Here we employ $\zeta$-regularization throughout, which is physically more reliable. It allows us to extract the physically meaningful IR divergent terms. While \cite{Ailiga:2024wdx} considers the Dirichlet case, here we extend to the fixed-curvature ($K$) case; see eqs. (\ref{eq:H_1<H_limit}) and (\ref{eq:Gtr_qf>3/lam_LG}). We notice that for fixed $K$, the leading one-loop correction to the phase exhibits a potentially different scaling from the background; see eq. (\ref{eq:H_1<H_limit}). This establishes the robustness of the IR behaviour across different boundary conditions. Furthermore, evaluating the mode sum using alternative regularization method confirms that the leading behaviour is independent of the choice of regularization scheme.

We next study the behaviour of the wavefunction in Lorentzian de Sitter and compare its UV and IR behaviour with that of the no-boundary wavefunction. We wish to determine which features of one-loop IR behaviour remain the same and which are sensitive to the choice of initial boundary condition. The dS-wavefunction is computed by studying the path-integral over scale-factor $q(t)$, fluctuation $h_{ij}$, and an ordinary integral over lapse-$N_c$. The path integral over fluctuation is done to one-loop, while the integral over the lapse is performed in the complex-$N_c$ plane via Picard-Lefschetz methods. Two real saddles (which are {\it relevant} and dominant) contribute to the path integral. The one-loop-corrected lapse action suffers from contact divergence, which is subtracted by counterterms. Besides this contact divergence, we notice a host of ``spurious'' singularities which seem analogous to {\it pinching}. We notice that a tiny complex deformation of cosmological constant ($\Lam$) manages to bypass these {\it pinch}-singularities, as the real dS saddles undergo tiny complex deformation. The lapse-integral performed along the thimbles of these deformed saddles bypasses the ``spurious'' singularities. 

With the above $i\epsilon$-prescription, the mode sum can be computed only in the IR limit ($H_1/H=\bold{H}_1\rightarrow 1$), due to the complicated structure of the determinant. We extract the leading/subleading IR divergent terms. We note that the leading IR behaviour (scaling of the terms which are $((1-\bold{H_1})^{-3/2},(1-\bold{H_1})^{-1/2}$) agrees qualitatively with the no-boundary wave function. However, the precise numerical coefficients in front of the leading terms differ between the two cases: no-boundary and de-Sitter; see eqs. (\ref{eq:log_sum_ren_act_ds}) and (\ref{eq:de-sitter_ep_def_wave_fun}). The same scaling of the leading/subleading terms in de Sitter with the no-boundary Universe is ensured as the latter approaches de Sitter asymptotically at late times. However, this asymptotic agreement does not always imply that the corresponding numerical coefficients are identical.
The difference in the coefficients suggests that the IR behaviour is sensitive to the details of initial (UV) state of fluctuation condition. The no-boundary cap with the $h_{ij}=0$ condition prepares the state of fluctuations on the throat, which differs from the Dirichlet condition we chose in the de-Sitter. Although the geometry/fluctuations approach the same final state in both cases, their distinct initial states leave an imprint on the coefficients of the leading/subleading IR terms. This is a new result in our paper and, to the best of our knowledge, has not been discussed previously. It should be mentioned that to compute the dS-wavefunction, we utilized a version of $i\ep$-prescription along with Picard-Lefschetz methods, as the later was not sufficient to find the convergent contours. This prescription ensures convergence of the mode sum. Finally, we also show that the resulting complex $\ep$-deformed de-Sitter saddles are KSW-allowed.

While the manuscript manages to address one interesting problem, a comparison of the IR-behaviour of the no-boundary and pure dS-wavefunction, the study is carried out for certain choices of boundary conditions: fixed size or extrinsic curvature at the final hypersurface and Robin boundary condition at the initial hypersurface. Whether the qualitative features remain the same for more general boundary choices is unclear, but we expect this to hold. It will be interesting to study the IR behaviour under a generic $\alpha$-vacuum boundary condition and see how it gets modified. We plan to address this in future studies. Finally, it will be worth investigating the one-loop IR properties of other instantons \cite{Betzios:2024oli, Rubakov:1988wx}, which include past asymptotic EAdS boundaries and wormhole geometries, but, when analytically continued to Lorentzian signature, lead to an expanding universe.

\section{Acknowledgment}

We would like to thank Justin David for the useful discussions involving infinite sums. We also thank Manishankar Ailiga for useful discussions. GN would like to thank the organizers of the conference ``Frontiers of Gravity 2026'' held at IISER Mohali, where some of discussions regarding ongoing work took place. We are thankful to the ongoing {\it Gauge-Gravity by the Ghats} seminar series, for providing a platform for various useful interactions. We also thank ChatGPT and Google Gemini for assistance with the literature survey and language editing, which are independently verified. GN also acknowledges the startup support from IISc, leading to workstation purchase, where some of the simulations have been performed.

\section{Appendix}
\subsection{Quadratic expansion of various curvature quantities}
\label{sec:curvature_expansion}
In this appendix, we write the expansions of the various quantities up to quadratic order in the fluctuations. The inverse metric ($\g^{ij}$), determinant and Christoffel connection have the following expansions
\bea
\label{eq:exp1}
&&
\g^{ij} = a^{-2} \left[
\rho^{ij} - h^{ij} +\frac{2-\varepsilon}{2} h^{ik} h_k{}^j + \cdots
\right]\hspace{20mm}\varepsilon=\{0,1\}\notag\\
&&
\sqrt{\g} = a^{3} \sqrt{\rho} \left[
1 + \frac{h}{2}+\frac{\varepsilon-1}{4}h_{ij}h^{ij}+ \frac{h^2}{8} +\cdots
\right] \, ,
\notag \\
        &&
\Theta_i{}^k{}_j = \bar{\Theta}_i{}^k{}_j 
+ 
\frac{1}{2}\left(\bar{D}_i h^k{}_j + \bar{D}_j h^k {}_i - \bar{D}^k h_{ij} \right)
-\frac{1}{2} h^{km} \left(\bar{D}_i h_{mj} + \bar{D}_j h_{mi} - \bar{D}_m h_{ij}\right)\, 
\notag \\
&&+\frac{\varepsilon}{4}\rho^{km}\biggl[\bar{D}_i (h_{mp}h^{p}{}_j) + \bar{D}_j (h_{mp}h^p{}_i) - \bar{D}_m (h_{ip}h^{p}{}_j)\biggr] + \cdots=\bar{\Theta}_i{}^k{}_j+\Theta^{(1)}_i{}^k{}_j+\Theta^{(2)}_i{}^k{}_j+\cdots\,
\eea
where, $h = \rho^{ij} h_{ij}$, and the indices of $h_{ij}$ are raised/ lowered 
using $\rho_{ij}$, covariant derivative ($\bar{D}_i$) and connection ($\bar{\Theta}_i{}^k{}_j$) are defined on $\rho_{ij}$. The term proportional to $\varepsilon$ in $\Theta_i{}^k{}_j$ is a total derivative. The extrinsic curvature on hypersurfaces ($K_{ij}$) and its trace ($K$) are defined by
\begin{equation}
\label{eq:chris_gam}
K_{ij}=\frac{\gamma_{ij}'}{2N_p},\hspace{5mm} K=\gamma^{ij}K_{ij}=\frac{(\ln\sqrt{\gamma})'}{N_p}.
\end{equation}
Using the definitions mentioned in eq. (\ref{eq:3Rgamma}), (\ref{eq:exp1}) and (\ref{eq:chris_gam}), we get 
\bea
\label{eq:K_exp}
&&
K = \frac{3a^\prime}{a N_p} 
+ \frac{1}{2N_p} \left[
h^\prime +(\varepsilon-1) h^{im} h^\prime_{im}+ \cdots
\right] \, , 
\notag \\
&&
K_{ij} K^{ij} = \frac{1}{4N_p^2} \left[
\frac{12 a^{\prime 2}}{a^2} + \frac{4 a^\prime h^\prime}{a}
+ \frac{4(\varepsilon-1) a^\prime h^{im} h^\prime_{im}}{a}
+ h^\prime_{im} h^{\prime im} \cdots
\right] \, ,
\notag \\
&&
{\cal R} = a^{-2} \biggl[
\bar{\cal R} 
+ \left(\bar{D}_i \bar{D}_j h^{ij} - \bar{\Box} h - \bar{\cal R}_{ij} h^{ij}\right)
+ \rho^{ij} \left(\bar{D}_k \Theta^{(2)}_i {}^k{}_j
- \bar{D}_i \Theta^{(2)}_k {}^k{}_j \right) 
\notag \\
&&
+ \bar{\cal R}_{ijmn} h^{im} h^{jn} 
- h^{ij} \bar{D}_i \bar{D}_k h^k{}_j 
+ \frac{1}{2} h^{ij} \bar{\Box} h_{ij} 
+ \frac{1}{2} h^{ij} \bar{D}_i \bar{D}_j h 
+ \frac{1}{2} \bar{D}^i h \bar{D}^j h_{ij}  
\notag \\
&& - \frac{1}{4} \bar{D}^i h \bar{D}_i h 
+ \frac{1}{4} \bar{D}^k h^{ij} \bar{D}_k h_{ij} 
- \frac{1}{2}\bar{D}^k h^{ij} \bar{D}_i h_{jk} -\frac{\varepsilon}{2}\bar{\cal R}_{ij}h^{ik}h_{k}{}^j+\cdots
\biggr]
\, ,
\eea
where $\bar{\cal R}_{ijmn}$, $\bar{\cal R}_{ij}$, $\bar{\cal R}$ 
are the Riemann tensor, Ricci tensor and Ricci scalar, respectively, defined on the metric 
$\rho_{ij}$. It is worth noticing that whenever $\varepsilon=1$ (exponential parametrization), the cross term $h_{im}h_{im}^\prime$ is absent in the above expressions.  This fact simplifies the calculations. Also, in the gauge that describes traceless graviton fluctuation $h'=0$, extrinsic curvature $K$ doesn't change due to the fluctuation. It is, in fact, true in all orders:
\begin{equation}
  \gamma=\det \gamma_{ij}=\det (a^2\rho_{ik}(e^h)^k_j)=\det (a^2\rho_{ik})\det[ (e^h)^k_j]=\det (a^2\rho_{ik})e^{{\rm Tr}(h)}=\det (a^2\rho_{ik}).  \notag
\end{equation}
 
\subsection{
 Diffeomorphisms, gauge fixing and ghosts}
\label{sec:gauge-fixing_ghost} 

In this appendix, we discuss the gauge fixing condition and compute the ghost determinant. We use the background field formalism, in which fields are decomposed into a background and a fluctuation \cite{DeWitt:1980jv, Abbott:1980hw}. Consider the quantum (perturbative) gauge transformations that leave the background fixed and change only the quantum fluctuations. 
The gauge-fixing generates an additional ghost determinant, which can be computed via the Faddeev-Popov method \cite{Faddeev:1967fc}.

Consider the diffeomorphisms generated by the vector field $V^{\rho}(t_p, {\bf x})$ which leads to the following transformation of metric
\beq 
\label{eq:gaugetrgamma}
\de_{\rm D} g_\mn = {\cal L}_V g_\mn = V^\rho \partial_\rho
g_\mn + g_{\mu\rho} \partial_\nu V^\rho +
g_{\nu\rho}\partial_\mu V^\rho \, ,
\eeq
where ${\cal L}_V g_\mn$ is the Lie derivative of
$g_\mn$ along $V^\rho$. The corresponding ADM variables $N_p,\g_{ij}$ changes as
\bea
\label{eq:gaugeTR}
&&
\de_{\rm D}(N_p) 
= \pt_0(N_p V^0) + V^k \pt_k N_p \, ,
\notag \\
&&
\de_{\rm D} (\g_{ij}) = V^0 \pt_0 \g_{ij}
+ V^k \pt_k \g_{ij} + \g_{ik} \pt_j V^k + \g_{jk} \pt_i V^k \, .
\eea
The gauge transformations might generate cross-terms (shift vector) in the metric even if 
the ansatz eq. (\ref{eq:frwmet}) does not have any. To ensure that no cross-term is generated, we restrict ourselves to a subset of full diffeomorphisms generated by $V^\rho (t_p, {\bf x})= \{V^0 (t_p), V^i ({\bf x})\}$\footnote{
The transformation of the shift vector/cross-term is $\de_{\rm D} N_i = \pt_0(N_i V^0) + V^j \pt_j N_i 
+ N_j \pt_i V^j + (-N^2 + N_k N^k) \pt_i V^0 + \g_{ij} \pt_0 V^j$. For diffs $V^\rho (t_p, {\bf x})= \{V^0 (t_p), V^i ({\bf x})\}$, $\de_{\rm D} N_i=0$, if one starts with $N_i=0$ and hence doesn't generate any new terms.}. Such diffeomorphisms always preserve the metric's diagonal form. In the proper-time gauge where $N_p'=0$, it leads to a trivial (constant) ghost contribution \cite{Halliwell:1988wc}.
Finally, one is left only with the transformation of the 
spatial metric $\g_{ij}$. Plugging $\g_{ij} = a^2(t_p)(\rho_{ij}
+ h_{ij}+\frac{\varepsilon}{2}h_{ik}h^{k}{}_j+\cdots)$ in eq. (\ref{eq:gaugeTR}), we get,
\beq
\label{eq:hij_trans}
\begin{split}
\de_{\rm D} h_{ij}+&\frac{\varepsilon}{2}\delta_D h_{ik}h^{k}{}_j+\frac{\varepsilon}{2}h_{ik}\delta_D h^{k}{}_j+\cdots\\
& =\frac{2a^\prime}{a} V^0 \rho_{ij}
+ \bar{D}_i V_j + \bar{D}_j V_i
+ \frac{2a^\prime}{a} V^0 h_{ij} + V^0 \pt_0 h_{ij}+ V^k \bar{D}_k h_{ij}+ h_{ik} \bar{D}_j V^k 
\\
&+ h_{jk} \bar{D}_i V^k+\varepsilon\biggl[\frac{a'}{a}V^0h_{ik}h^k{}_{j}+\frac{V^0}{2}\partial_0 h_{ik}h^{k}{}_j+\frac{V^0}{2}h_{ik}\partial_0h^k{}_j+\frac{V^k}{2}\bar{D}_kh_{im}h^m{}_j\\
&+\frac{V^k}{2}h_{im}\bar{D}_kh^m{}_j+\frac{1}{2}h_{jm}h^m{}_k\bar{D}_iV^k+\frac{1}{2}h_{im}h^m{}_k\bar{D}_jV^k\biggr]+\cdots\, .
\end{split}
\eeq
Both sides of the above equation are infinite series that can be solved order by order. The transformation for the fluctuation up to the linear order becomes
\begin{equation}
    \label{eq:gauge_trans_hij}
    \begin{split}
    \delta_D h_{mn}&=\frac{2a'}{a}V^0\rho_{mn}+(\bar{D}_mV_n+\bar{D}_nV_m)+(1-\varepsilon)\frac{2a'}{a}V^0h_{mn}+V^0\partial_0h_{mn}+V^k\bar{D}_kh_{mn}\\
&+h_{mk}\bar{D}_nV^k+h_{nk}\bar{D}_mV^k-\frac{\varepsilon}{2}h^i_m(\bar{D}_nV_i+\bar{D}_iV_n) -\frac{\varepsilon}{2}h^i_n(\bar{D}_mV_i+\bar{D}_iV_m)+\mathcal{O}(h^2).
\end{split}
\end{equation}
In the above expansion, terms that are independent of $h_{ij}$ are the leading ones that contribute to the one-loop. Under the transformation eq. (\ref{eq:gauge_trans_hij}), the action ($S_{\rm grav}$) eq. (\ref{eq:EHact_ADM}) remains invariant. To break it, we choose 
transverse and traceless gauge $\bar{D}^m h_{mi} = h^i_i=0$.
In this gauge,  $h_{ij}$ has two d.o.fs and describes physical gravitons where longitudinal modes are absent. The gauge fixing condition $\bar{D}^m h_{mi}=0$
will lead to a Faddeev-Popov determinant ($\det \D_{nm}$). Up to linear order, it is
\bea
\label{eq:ghost_det}
&&
\D_{nm} = \rho_{mn} \nb^2 + \bar{D}_m \bar{D}_n
+ \bar{D}^i \bar{D}_m h_{in}
+\frac{2-\varepsilon}{2}\biggl[\bar{D}_m h_{in} \bar{D}^i+ h_{im} \bar{D}^i \bar{D}_n + \bar{D}^i h_{mn} \bar{D}_i\,\notag\\
&& \hspace{10mm} \,+ h_{mn} \nb^2\biggr]
-\frac{\varepsilon}{2}\rho_{mn}h_{ip}\bar{D}^p\bar{D}^i-\frac{\varepsilon}{2}h_{in}\bar{D}_m\bar{D}^i+\mathcal{O}(h^2)
\, ,
\eea
where $\nb^2 = \bar{D}_i \bar{D}^i$ is the Laplacian on $\rho_{ij}$. 
At one loop, terms independent of the $h_{ij}$ contribute.
It is interesting to note that the scale factor ($a$) doesn't appear in the above determinant. Hence, for the restricted set of diffeomorphisms we consider, the ghost determinant is defined on the comoving hypersurfaces on which the ghost fields live. The gauge condition $h_i{}^i=0$ leads to trivial ghosts as it has no differential operator \cite{Lin:2017ool}.

\subsection{Mode functions, their asymptotic forms and near singularity behaviour}
\label{sec:def_mode_fun}
In this appendix, we briefly collect the definitions, their asymptotic forms, and near-singularity behaviour used in the main text. We start with the Legendre-P function, which is given by (for $\mu$ not a positive integer) \cite{dlmf, Abram-steg}
\begin{equation}
    \label{eq:legendre_p}
    \mathbb{P}_\lam^{\mu}(z)=\frac{1}{\Gamma(1-\mu)}\Bl(\frac{z+1}{z-1}\Br)^{\mu/2}{}_2F_1(-\lambda,\lambda+1;1-\mu;\frac{1-z}{2}),\hspace{4mm}\text{for}\,\, |1-z|<2.
\end{equation}
\label{sec:asymp_mode_expansion}
where, ${}_2F_1(a,b;c,z)$ is the hypergeometric function, defined for $|z|<1$,
\begin{equation}
    \label{eq:hypergeometric_function}
{}_2F_1(a,b;c;z)=\sum_{n=0}^\infty\frac{(a)_n(b)_n}{(c)_n}\frac{z^n}{n!}=1+\frac{ab}{c}\frac{z}{1!}+\frac{a(a+1)b(b+1)}{c(c+1)}\frac{z^2}{2!}+\cdots
\end{equation}
and $(a)_n$ is Pochhammer symbol defined as $(a)_n=\Gamma(a+n)/\Gamma(a),\,\,a\neq 0,-1,-2,\cdots$. For $|z|>1$, it can be analytically continued. In the case when $a$ is a negative integer, then the infinite series in eq. (\ref{eq:hypergeometric_function})
truncates. Since ${}_2F_1(a,b;c;z)$ becomes a polynomial/holomorphic function, it can be easily analytically continued to $|z|>1$. For tensor perturbation, we have $\lambda=1$ for all $N_c$. Hence, we get
\begin{equation}
    \label{eq:legendre_simplify}
    \mathbb{P}_1^{-\xi_l}(\tau_t)=\frac{1}{\Gamma(2+\xi_l)}\biggl(\frac{\tau_t-1}{\tau_t+1}\biggr)^{\xi_l/2}(\tau_t+\xi_l),\hspace{4mm} \forall\,\, \tau_t
\end{equation}
where $\xi_l$ and $\tau_t$ are given in eq. (\ref{eq:hl_sol_gen}).
At the no-boundary saddles $\xi_l=l+1,\tau_t=1-i\Lambda N_ct/3$. In the asymptotic limit when $|\tau_t|$ is large, we get
\begin{equation}
    \mathbb{P}_1^{-\xi_l}(\tau_t)\sim\mathcal{O}(\tau_t),\hspace{4mm}\text{for} |\tau_t|\gg1.
\end{equation}
On the other hand, the Legendre-Q function is given by \cite{dlmf,Abram-steg}
\begin{equation}
\label{eq:legendre-Q}
\mathbb{Q}_\lam^\mu(z)=\frac{\sqrt{\pi}\Gamma[\lam+\mu+1]e^{i\mu\pi}(z^2-1)^{\mu/2}}{2^{\lam+1}\Gamma(\lam+3/2)z^{\lam+\mu+1}}{}_2F_1\biggl(\frac{\lam+\mu+1}{2},\frac{\lam+\mu+2}{2};\lam+\frac{3}{2};\frac{1}{z^2}\biggr),\hspace{3mm}\text{for} |z|>1
\end{equation}
where at the saddles $\mu=l+1$, $\lambda=1$ and $z=\tau_t^{\rm nb}$. Note that for $|z|\gg 1$, $\mathbb{Q}_\lam^\mu(z)\sim\mathcal{O}(z^{-\lam-1})$ and hence it vanishes. This behaviour is different from the other mode function, which grows in the asymptotic limit. 

\noindent For $|z|<1$, one needs to analytically continue using the relation (see \cite{Abram-steg}, eq. 15.3.7),
\begin{equation}
\label{eq:hypergeometric_relation}
\begin{split}
    &_2F_1(a,b;c;z)=\frac{\Gamma(c)\Gamma(b-a)}{\Gamma(b)\Gamma(c-a)}(-z)^{-a}{}_2F_1(a,1-c+a;1-b+a;\frac{1}{z})\\
    &+\frac{\Gamma(c)\Gamma(a-b)}{\Gamma(a)\Gamma(c-b)}(-z)^{-b}{}_2F_1(b,1-c+b;1-a+b;\frac{1}{z}),\hspace{3mm}|\arg(-z)|<\pi.
\end{split}
\end{equation}
Near $z=1$, $\mathbb{Q}_\lam^\mu(z)$ admits the following asymptotic expansion
\begin{equation}
    \label{eq:legendre-Q_apen}
\mathbb{Q}_\lam^\mu(z)\sim (1-z)^{-\mu/2}\frac{2^{\mu/2}}{2}\Gamma(\mu)\cos(\pi\mu)+\cdots,\hspace{4mm}\mu\neq \frac{1}{2},\frac{3}{2},\cdots
\end{equation}
When $\mu$ is half integer, it vanishes as $(1-z)^{\mu/2}$ \cite{dlmf}
\begin{equation}
    \label{eq:legendre-Q-half-integer-mu}
    \mathbb{Q}_{\lambda}^{\mu}(z)
    \sim
    (-1)^{\mu+\frac{1}{2}}
    \frac{
        \pi\,\Gamma(\lambda+\mu+1)
    }{
        2\,\Gamma(\mu+1)\Gamma(\lambda-\mu+1)
    }
    \left(\frac{1-z}{2}\right)^{\mu/2},
    \qquad
    \mu=\frac{1}{2},\frac{3}{2},\frac{5}{2},\ldots\,.
\end{equation}

\subsection{Generalized zeta regularization of infinite series}
\label{sec:zeta_sum}

In this appendix, we will compute the zeta-regularized finite sum of the infinite series appearing in the gravity and ghost parts of the effective action. It has the following form, for real $x$ and $y$,

\begin{equation}
    \label{eq:Ser_sum}
    \begin{split}
    S_1(a) = \sum_{l=1}^\infty (l+x)(l+y) \log (l+a) \, 
    \end{split}
\end{equation}
where $Re[a]>-1$. The above sum is divergent, and the full UV divergence structure can be obtained using the integral representation of $\ln(y)$: $\log(y) = \int_{0}^{\infty} \frac{ds}{s}\, (e^{-s} - e^{-sy})$, where one notices logarithmic and power law divergences at $s=0$ \cite{Vassilevich:2003xt,Ailiga:2024wdx}.
Below, we will describe a method that gives the renormalised value of the infinite sum, stripping off these divergences. Using the identity $\ln(a)=-\frac{\partial}{\partial s}(a^{-s})|_{s=0^+}$,
the renormalized sum for eq. (\ref{eq:Ser_sum}), can be written as
\begin{equation}
    \label{eq:sum_identity}
     S_1^{\rm ren}(a) = -\sum_{l=1}^\infty (l+x)(l+y) \frac{\partial}{\partial s}[(l+a)^{-s}]\biggr|_{s=0^+}.
\end{equation}
Interchanging the summation and differentiation,
we get
\begin{equation}
    \label{eq:interchange}
    S_1^{\rm ren}(a)=-\frac{\partial}{\partial s}\sum_{l=1}^\infty\frac{(l+x)(l+y)}{(l+a)^s}\biggr|_{s=0^+}.
\end{equation}
To evaluate it, we rewrite each term in the summation as
\begin{equation}
    \label{eq:sum_identity_1}
      \frac{(l+x)(l+y)}{(l+a)^s}=\frac{1}{(l+a)^{s-2}}+\frac{x+y-2a}{(l+a)^{s-1}}+\frac{a^2+xy-a(x+y)}{(l+a)^{s}}.
\end{equation}
Each of the terms in the above equation can be summed using the Hurwitz-zeta function $\zeta_H(s,a)$:
\begin{equation}
 \label{eq:hurwitz_zeta_function}
    \zeta_H(s,a)=\sum_{l=0}^\infty\frac{1}{(l+a)^s}.
\end{equation}
After summing the series and evaluating the derivative at $s=0^+$, we get the following renormalized part
\begin{equation}
    \label{eq:sum_finite_1}
    S_1^{\rm ren}(a)= -\zeta_H'(-2,a+1)+(2 a-x-y)\zeta_H'(-1,a+1)-(a-x) (a-y) \zeta_H'(0,a+1),
\end{equation}
where, prime ($'$) represents derivative w.r.t the first argument, i.e., $(\partial/\partial s)\zeta_H(s,a)$. For $Re[a]>-1$, the zeta functions appearing in the finite sum are analytic.

In the case when the sum starts from $l=2$, i.e.,
\begin{equation}
\label{eq:sum_2}
S_2(a)=\sum_{l=2}^\infty (l+x)(l+y) \log (l+a) \,,
\end{equation}
where $Re[a]>-2$, the zeta regularized and renormalized sum can be computed in a similar way, giving
\begin{equation}
\label{eq:sum_finite_2}
S_2^{\rm ren}(a)=-\zeta_H'(-2,a+2)+(2 a-x-y)\zeta_H'(-1,a+2)-(a-x) (a-y) \zeta_H'(0,a+2).
\end{equation}
For $Re[a]>-2$, the zeta functions appearing in the finite sum are analytic. It is useful to evaluate the sum when the second argument ($a$) is large. To get it, we
utilize the asymptotic expansions of the $\zeta_H'(m,q)$ for $m=0,-1,-2$, which are given by \cite{Elizalde,dlmf}
\begin{equation}
\label{eq:asymptotic_expansion}
\begin{split}
&\zeta_H'(0,q)=q\ln(q)-q-\frac{1}{2}\ln(q)+\sum_{k=0}^\infty\frac{B_{2k+2}\,\,q^{-(2k+1)}}{(2k+2)(2k+1)},\\
&\zeta_H'(-1,q)=\frac{q^2}{4}(2\ln(q)-1)+\frac{\ln(q)}{12}(1-6q)+\frac{1}{12}-\sum_{k=1}^\infty\frac{B_{2k+2}\,\,q^{-2k}}{(2k+2)(2k+1)(2k)}\\
&\zeta_H'(-2,q)=\frac{q^3}{9}(3\ln(q)-1)+\frac{q\ln(q)}{6}(1-3q)+\frac{q}{12}+2\sum_{k=1}^\infty\frac{B_{2k+2}\,\,q^{-(2k-1)}}{(2k+2)(2k+1)(2k)(2k-1)}
\end{split}
\end{equation}
where, the Bernoulli numbers $B_{2k}$ are given by
\begin{equation}
\label{eq:bernouli_number}
\begin{split}
B_{2k}=\frac{(-1)^{k+1}2(2k)!\zeta(2k)}{(2\pi)^{2k}},\,\,\hspace{4mm} B_{2k+1}=0,\hspace{4mm}k=1,2,3,\cdots.
\end{split}
\end{equation}
Utilizing the above asymptotic expansions, we get the following finite parts for the infinite sums in eqs. (\ref{eq:sum_finite_1}) and (\ref{eq:sum_finite_2}), respectively
\begin{equation}
\label{eq:finite_sum_combined}
\begin{split}
     S_1^{\rm ren}(a)=&\frac{1}{18} a^3 (11-6 \log (a))+\frac{1}{4} a^2 (2 \log (a)-3) (x+y)-a x y (\log (a)-1)\\
    &-\frac{1}{12} \log (a) (6 x y+x+y)+\cdots\\
    S_2^{\rm ren}(a)=&S_1^{\rm ren}(a)-(x+1)(y+1)\ln(1+a)
\end{split}
\end{equation}
where $\cdots$ terms include the infinite sums over $B_{2k+2}$ which vanish at $a\rightarrow\infty$. Also note that the leading terms in both the series are independent of $x$ and $y$.

\subsection{Gel'fand-Yaglom method for one-dimensional path integral}
\label{sec:GY_append} 
To make the paper self-contained, in this appendix, we briefly review the Gel'fand-Yaglom method of computing the functional integral in one dimensional path integral \cite{Kirsten:2003py, Kirsten:2004qv}. For any such path integral the quadratic fluctuation operator can be written as
\begin{equation}
\label{eq:STL_opp}
\mathcal{D}=-\frac{d}{dt}\left(P(t)\frac{d}{dt}\right)+Q(t) \, ,
\end{equation}
where, $t\in [0,1]$. It is the most general second-order Sturm-Liouville operator and the corresponding eigenvalue equation is
\begin{equation}
\label{eq:eigenEQ}
\mathcal{D}u_{(\lambda)}(t)=\lambda u_{(\lambda)}(t) \, .
\end{equation}
We define $v_{(\lambda)}\equiv P(t)\dot{u}_{(\lambda)}$ (``dot'' being $t$ derivative) and go to the first order formalism. Denote two independent solutions of eq. (\ref{eq:eigenEQ}) as $u_{(\lam)}^1$ and $u_{(\lam)}^2$. $u_{(\lam)}$(and also $v_{(\lam)}$) is the linear combination of these which are
by the generalized ``linear" boundary condition
\begin{equation}
\label{eq:mat_MN}
M\left(\begin{matrix}
u_{(\lambda)}(0)\\
v_{(\lambda)}(0)
\end{matrix}
\right)
+N\left(\begin{matrix}
u_{(\lambda)}(1)\\
v_{(\lambda)}(1)
\end{matrix}\right)
=\left(\begin{matrix}
0\\
0
\end{matrix}\right) \, ,
\end{equation}
where $M$ and $N$ are $2\times 2$ matrix. Define the matrix $E_{(\lam)}(t)$ such that,
\begin{equation}
    \label{eq:E_matrix}
    E_{(\lam)}(t)=\biggl(\begin{array}{cc}
        u_{(\lam)}^1 &  u_{(\lam)}^2\\
        v_{(\lam)}^1 & v_{(\lam)}^2
    \end{array}\biggr);\hspace{5mm}\biggl(\begin{array}{c}
         u_{(\lam)}(t)  \\
        v_{(\lam)}  (t)
    \end{array}\biggr)=E_{(\lam)}(t)\biggl(\begin{array}{c}
         u_{(\lam)}(0)  \\
        v_{(\lam)}  (0)
    \end{array}\biggr).
\end{equation}
Note that $E_{(\lam)}(t)$ is ``wronskian" matrix and $\det E_{(\lam)}(t)$ is independent of $t$; we set it to unity. This fixes $E_{(\lam)}(0)=I_2$, with $I_2$ being $2\times 2$ identity matrix. 
Now, from eq. (\ref{eq:mat_MN}) and (\ref{eq:E_matrix}), we get
\begin{equation}
    \label{eq:eigenvalue_eqn}
    [M+NE_{(\lam)}(1)]\biggl(\begin{array}{c}
         u_{(\lam)}(0)  \\
        v_{(\lam)}(0)
    \end{array}\biggr)=0
\end{equation}
implying that $\det [M+NE_{(\lam)}(1)]=0$ when $\lam$ is the eigenvalue. Hence the function $[M+NE_{(\lam)}(1)]$ has zeros at $\lam$ which are the eigenvalues of $\mathcal{D}$. Hence, the corresponding zeta function for the operator $\mathcal{D}$ is 
\begin{equation}
    \label{eq:zeta_function}
    \zeta_\mathcal{D}(s)=\frac{1}{2\pi i}\int_\g d\lam\,\, \lam^{-s}\frac{d}{d\lam}\ln\det[M+NE_{(\lam)}(1)],
\end{equation}
where $\g$ is the contour (counterclockwise) enclosing all the eigenvalues \cite{Kirsten:2003py, Kirsten:2004qv}. One can easily evaluate the contour integral along the cut ($\theta$ is the angle for the cut), giving 
\begin{equation}
    \label{eq:contour_integral}
    \zeta_\mathcal{D}(s)=\frac{e^{is(\pi-\theta)}\sin(\pi s)}{\pi}\int_0^\infty d\lam \,\lam^{-s}\frac{d}{d\lam}\ln\det[M+NE_{(\lam e^{i\theta})}(1)].
\end{equation}
Integrating by parts and then evaluating the derivative w.r.t $s$ at $s=0$, we get
\begin{equation}
    \label{eq:det_gel_fand}
    \det\mathcal{D}=\exp(-\zeta_\mathcal{D}'(0))=\det[M+NE_{(0)}(1)].
\end{equation}
Interestingly, to compute the determinant, one doesn't need the eigenvalues, since we set $\lam=0$. This implies that one needs only to solve the homogeneous equation $\mathcal{D}u=0$. The metrics $M$ and $N$ depend on the boundary conditions and are given by ( Dirichlet-Dirichlet case)
\begin{equation}
    \label{eq:M_and_N matrices_appendix}
    M=\left(\begin{matrix}
    1&0\\
    0&0
\end{matrix}\right),\hspace{5mm}N= \left(\begin{matrix}
    0&0\\
    1&0
\end{matrix}\right) \, ,
\end{equation}
which satisfies the condition $u_{(\lam)}(0)=0$ and $u_{(\lam)}(1)=0$. Finally, it is important to note that this method only works for linear boundary conditions (eq. (\ref{eq:mat_MN})) and cannot be applied to the non-linear ones.

\subsection{Contribution from the remainder sum in the IR}
\label{sec:summation_reminder}
In this appendix, we extract the contribution from the summation of the remainder term $\mathcal{R}_\ep(l,H_1)$ (see eq. (\ref{eq:asymptotic_limit_de-sitter})) in the IR limit, which is given by
\begin{equation}
    \label{eq:reminder}
   \mathcal{S}_\ep= \sum_{l=2}^\infty (l-1)(l+3)\ln\biggl[1-e^{-2i\Theta_l-2\ep B}\biggr],
\end{equation}
where $\Theta_l(H_1)=(l+1)\tan^{-1}\left[H_1/\sqrt{H^2-H_1^2}\right]-\tan^{-1}\left[H_1/(l+1)\sqrt{H^2-H_1^2}\right]$ as given in eq. (\ref{eq:det_simplify_real}) and $B$ is given in eq. (\ref{eq:det_com_sad_plus}).

To evaluate it, we write $x=l/a$, where $a=H_1/\sqrt{H^2-H_1^2}$, and do the series expansion at large $a$ (IR) while keeping $ x$ fixed. It is a type of uniform expansion which simultaneously takes large $l$-modes and large $a$ \cite{Kirsten:2004qv, Kirsten:2003py, Dunne:2006ct}. This essentially includes the modes near $a$,i.e, $l\sim\mathcal{O}(a)$ in the IR.  Doing the expansion of $\Theta_l$, we get the following 
\begin{equation}
    \label{eq:expansion_D}
   \Theta_l(H_1)=\frac{\ell\pi}{2}-D_\ell(x,a),\quad D_l(x,a)=F(x)+\frac{F'(x)}a
       +\frac1{a^2}\left[\frac{F''(x)}2-\frac{x}{3}\right]
       +O(a^{-3}),
\end{equation}
where, $F(x)=x-\tan^{-1}(x)$ and prime ($'$) denotes derivative w.r.to $x$. Splitting the sum in eq. (\ref{eq:reminder}) into even and odd, approximating the sum as an integral, we get
\begin{equation}
    \label{eq:remider_expansion}
\mathcal{S}_\ep(a)=\frac{1}{2}\int_0^\infty dx (a^3 x^2+2 a^2 x-3a)\ln\left[1-e^{4iD_l(x,a)-4\ep B}\right].
\end{equation}
Changing the sum to an integral is accurate to leading order in $a^3$ and $a$. However, it is can change the terms of order $\mathcal{O}(\ln(a)),\mathcal{O}(1)$ number. The above sum is convergent for $\ep>0$ and requires no regularization. We first perform the sum and then take the limit as $\ep$ goes to zero. The limit is smooth and doesn't introduce any further divergences. Hence, we get 
\begin{equation}
    \label{eq:sum_ep_zero_limit}
   \mathcal{S}(a)= \lim_{\ep\rightarrow 0}\mathcal{S}_\ep(a)=\frac{1}{2}\int_{\mathcal{C}_\phi} dx  (a^3 x^2+2 a^2 x-3a)\ln\left[1-e^{4iD_l(x,a)}\right],\quad \mathcal{C}_\phi=e^{i\phi}(0,\infty),\phi>0.
\end{equation}
It is easy to check that the integral is independent of the rotation angle $0<\phi<\pi/2$. The direction of the contour is anti-clockwise and arises from the direction of rotation of the cosmological constant ($\Lambda$). This essentially avoids the pole lying on the real $x$-axis, which was the reason behind rotating the cosmological constant. Defining $\mathcal{L}_a(x)$ as the logarithm inside the integral, we get in the same expansion
\begin{equation}
    \label{eq:expansio_L}
    \begin{split}
    \mathcal{L}_a=\ln\left[1-e^{4iD_l(x,a)}\right],\qquad &\mathcal{L}_a(x)=L+\frac{L'}a
                   +\frac1{a^2}\left(\frac{L''}{2}+B\right)
                   +O(a^{-3}),\\
                   \text{where}, \qquad&L=\ln\left[1-e^{4iF}\right],\quad B=\frac{4\ii x}{3}\frac{e^{4iF}}{1-e^{4iF}}.
                   \end{split}
\end{equation}
Substituting the above expansion of $\mathcal{L}_a(x)$ in eq. (\ref{eq:remider_expansion}), we get
collecting the order of $a^3$ and $a$
\begin{equation}
    \begin{split}
         \mathcal{S}(a)={}&\frac{a^3}{2}\int_{\mathcal{C}_\phi} x^2L\,\dd x
 +\frac{a^2}{2}\int_{\mathcal{C}_\phi}(2xL+x^2L')\,\dd x\\
 &+\frac{a}{2}\int_{\mathcal{C}_\phi}\left(-3L+2xL'+\frac{x^2}{2}L''+x^2B\right)\dd x
 +\cdots.\label{eq:collect}
    \end{split}
\end{equation}
Note that the coefficient in front of $a^2$ is a purely boundary term and vanishes along the ray
\begin{equation}
    \label{Eq:surtace_term}
    x^2L\biggl|_{0}^{\infty e^{i\phi}}=0
\end{equation}
Evaluating the remaining integrals (in a similar way, the surface terms generated by parts vanish), we get
\begin{equation}
    \label{eq:correction}
    \mathcal{S}(a)=\frac{a^3I_2}{2}
       +\frac{a}{2}\left(I_2-\frac{11}{3}I_0\right)+\cdots,
\end{equation}
where, the $I_2$ and $I_0$ are given by the following integrals
\begin{equation}
    \label{eq:integrals}
    I_0=\int_{\mathcal C_\phi}L(x)\,\dd x,\qquad
 I_2=\int_{\mathcal C_\phi}x^2L(x)\,\dd x.
\end{equation}
The above integrals can be easily evaluated in Mathematica, yielding
\begin{equation}
    \label{Eq:numerical_evauation}
    I_0\approx -2.51429 - 1.71889 i,\,\,\qquad I_2\approx0.286291\, -0.460498 i.
\end{equation}
Hence, the contribution from the summation of the remainder term, we get ($\bold{H}_1=H_1/H$)
\begin{equation}
    \label{eq:reminder_expre_appen}
    \mathcal{S}(a)=\frac{I_2}{4\sqrt{2}}\frac{1}{(1-\bold{H}_1)^{3/2}}-\frac{(88 I_0 + 3 I_2)}{48 \sqrt{2}}\frac{1}{(1-\bold{H}_1)^{1/2}}+\cdots,
\end{equation}
where $\cdots$ are logarithmic and $\mathcal{O}(1)$ constant.

\subsection{Antilinearity and Real wave function}
\label{sec:symm}
In this appendix, we briefly comment on the antilinearity seen in the lapse action \cite{Ailiga:2025fny}, namely (see eq. (\ref{eq:Nc_act_total_nb_fin_1}))
\begin{equation}
\label{eq:anti_lilear_checking}
{\cal A}_{\rm ren}^{\zeta-{\rm reg}}(-N_c^*,\varepsilon)=-{\cal A}_{\rm ren}^{\zeta-{\rm reg}}(N_c,\varepsilon)^*.
\end{equation}
Let's consider the case of Neumann-fixed $K$ boundary condition. To start with the we note that the counterterm action mentioned in eq. (\ref{eq:Nc_act_total_nb_ct}) satisfies $\mathcal{A}_{\rm ct}(-N_c^*)=-\mathcal{A}_{\rm ct}(N_c)^*$, as $\bar{q}(t)$ mentioned in eq. (\ref{eq:qsol_RBC}) satisfies $\bar{q}(N_c,t)^*=\bar{q}(-N_c^*,t)$ for $t\in[0,1]$. To see if the anti-linearity is respected at the one-loop fluctuation determinant, we note that $\xi_l$ and $\tau_t$ in eq. (\ref{eq:hl_sol_gen})
\begin{equation}
    \label{eq:xi_tau_gen_Nc}
    \begin{split}
    &\xi_l(N_c)=i\left\{\frac{H^2 (1-2 \varepsilon)}{H^2-H_1^2}-\frac{H_1^2 \{9 l (l+2)+\Lambda N_c (1-2 \varepsilon ) (\Lambda  N_c+6 i)+9\}}{\left(H^2-H_1^2\right) (\Lambda  N_c+3 i)^2}\right\}^{1/2},\\
    &\tau_t(N_c)=\frac{H_1 (\Lambda  N_c t+3 i)}{\sqrt{H_1^2-H^2} (\Lambda  N_c+3 i)},
    \end{split}
\end{equation}
satisfies $\xi_l(-N_c^*)=-\xi_l(N_c)^*$ and $\tau_t(-N_c^*)=-\tau_t(N_c)^*$. It implies the fundamental solutions $\mathbb{P}$ and $\mathbb{Q}$ satisfy the same property. This ensures that the antilinearity is satisfied for the on-shell fluctuation and one-loop corrections. In \cite{Ailiga:2025fny}, we showed that the background on-shell and one-loop prefactors also satisfy anti-linearity, which we wouldn't repeat here. As a consequence, eq. (\ref{eq:anti_lilear_checking}) is satisfied for arbitrary $\varepsilon$. In particular, the no boundary saddles satisfy $N_{+}^{\rm nb}=-(N_-^{\rm nb})^*$, and hence the real and imaginary parts of the quantum-corrected lapse action at the two no-boundary saddles satisfy ${\rm Re}[i{\cal A}_{\rm ren}^{\zeta-{\rm reg}}(N^{\rm nb}_+)]={\rm Re}[i{\cal A}_{\rm ren}^{\zeta-{\rm reg}}(N^{\rm nb}_-)]$ and ${\rm Im}[i{\cal A}_{\rm ren}^{\zeta-{\rm reg}}(N^{\rm nb}_+)]=-{\rm Im}[i{\cal A}_{\rm ren}^{\zeta-{\rm reg}}(N^{\rm nb}_-)]$. It can be explicitly seen from eq. (\ref{eq:Afin_ReIm_fixed_k}). As a result, when we add the two saddles, we get the real no-boundary wave function \cite{DiazDorronsoro:2017hti,DiTucci:2019bui,Ailiga:2025fny}.
On the other hand, when we deform $\Lambda$ by $i\ep$, the saddles lying on the real axis are moved in different directions in the complex $N_c$ plane; see eq. (\ref{eq:Lam-defo}). This explicitly breaks the antilinear symmetry \cite{Ailiga:2025fny}. As a consequence, the wave function in the complexified de Sitter case is complex rather than real.

\subsection{$\zeta$-regularized one-loop action with fixed final size}
\label{sec:wave_function_other_regime}
In this appendix, we write the exact expression for zeta regularized finite action (${\cal A}_{\rm ren}^{\zeta-{\rm reg}}$), as mentioned in eq. (\ref{eq:Nc_act_total_nb_fin_at_saddle}), evaluated at the no-boundary saddles at the fixed finite $q_f$. In the Euclidean regime, we have $q_f<3/\Lam$, while in the Lorentzian regime, $q_f>3/\Lam$ (see eq. (\ref{eq:NNB_sad})). Similar exact one-loop expressions are given in \cite{Ailiga:2024wdx}, where some other regularization method was chosen. Here, we prove the expressions for zeta regularization. \\

\begin{figure}
    \centering
    \includegraphics[width=0.58\linewidth]{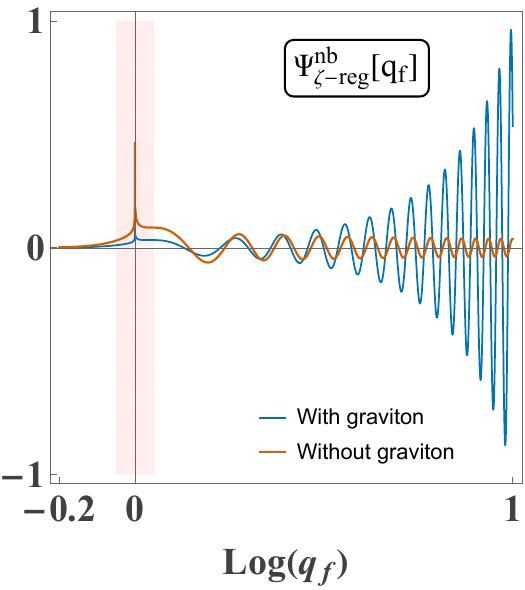}
    \caption{The plot showing $\Psi^{\rm nb}_{\zeta-{\rm reg}}[q_f]$ as a function of final size of the universe ($q_f$). The blue curve shows the growth of the wave function arising from the one-loop gravity fluctuation. The orange curve is the background. The pink shaded region is where the WKB approximation breaks down. The behaviour is qualitatively similar to \cite{Ailiga:2024wdx}, obtained with a different regularization.}
    \label{fig:wave_function_fixed_qf}
\end{figure}
For $q_f>3/\Lam$, the renormalized and zeta-regularized action is given by
\begin{eqnarray}
\label{eq:Log_parts_fin_zeta_regu_lorentzian}
&&\mathcal{A}_{\rm ren}^{\zeta-{\rm reg}}(N^{\rm nb}_\pm)\biggr|_{q_f>3/\Lam}
=-\frac{3V_3}{4\pi G \Lambda}\left[i\pm (\bold{q}_f-1)^{3/2}\right]+\frac{i\hbar}{2}\biggl\{\ln\biggl[1 - x \mp i x\sqrt{\bold{q}_f-1}\biggr]\notag\\
&& - \frac{4}{3} \ln q_f-\frac{8}{3}\ln(2)+\frac{191}{120}\ln\biggl[\frac{12(1\pm i\sqrt{\bold{q_f}-1})}{\Lambda(1\mp i\sqrt{\bold{q_f}-1})}\biggr]+\log \left(2 \pi ^7 A^4\right)\notag\\
&&-2\zeta_H'(-2,\mp i\sqrt{\bold{q_f}-1}+3)\mp 4 i\sqrt{\bold{q_f}-1}\zeta_H'(-1,\mp i\sqrt{\bold{q_f}-1}+3)\notag\\
&&+2(\bold{q_f}+3) \zeta_H'(0,\mp i\sqrt{\bold{q_f}-1}+3)-\frac{\zeta (3)}{\pi ^2}-\frac{1}{3}\biggr\},
\end{eqnarray}

where we define $\bold{q}_f=\Lambda q_f/3$. For $q_f<3/\Lam$, it is given by
\begin{eqnarray}
\label{eq:Log_parts_fin_zeta_regu_euclidean}
&&\mathcal{A}_{\rm ren}^{\zeta-{\rm reg}}(N^{\rm nb}_\pm)\biggr|_{q_f<3/\Lam}
=-\frac{3iV_3}{4\pi G \Lambda}\left[1\mp(1-\bold{q}_f)^{3/2}\right]+\frac{i\hbar}{2}\biggl\{\ln\biggl[1 - x\pm x\sqrt{1-\bold{q}_f}\biggr] \notag \\
&& - \frac{4}{3} \ln q_f-\frac{8}{3}\ln(2)+\frac{191}{120}\ln\biggl[\frac{12(1\mp\sqrt{1-\bold{q_f}})}{\Lambda(1\pm\sqrt{1-\bold{q_f}})}\biggr]+\log \left(2 \pi ^7 A^4\right)\notag\\
&&-2\zeta_H'(-2,\pm\sqrt{1-\bold{q_f}}+3)\pm 4 \sqrt{1-\bold{q_f}}\zeta_H'(-1,\pm\sqrt{1-\bold{q_f}}+3)\notag\\
&& +2(\bold{q_f}+3) \zeta_H'(0,\pm\sqrt{1-\bold{q_f}}+3)-\frac{\zeta (3)}{\pi ^2}-\frac{1}{3}\biggr\}.
\end{eqnarray}
Note that the above expression is completely imaginary.
When $\bold{q}_f\rightarrow 0$, one requires an additional counterterm as mentioned in eq. (\ref{eq:Nc_act_total_nb_ct}).
The expression for $q_f=3/\Lam$ can be obtained from either eq. (\ref{eq:Log_parts_fin_zeta_regu_lorentzian}) or (\ref{eq:Log_parts_fin_zeta_regu_euclidean}). It is explicitly given by
\begin{eqnarray}
    \label{eq:qf=3/lam_wave_function}
   &&\mathcal{A}_{\rm ren}^{\zeta-{\rm reg}}(N^{\rm nb}_\pm)\biggr|_{q_f=3/\Lambda} =-\frac{3iV_3}{4\pi G\Lambda}-\frac{i\hbar}{240} \biggl[\biggl\{-120\ln(1-x)-480 \log (A)+\frac{60 \zeta (3)}{\pi ^2}\biggr\}\notag\\
   &&+40-22 \log (2)-360 \log (\pi )-31\log \left(\frac{3}{\Lambda }\right)\biggr]
\end{eqnarray}
The wave function with fixed size can be obtained using eq. (\ref{eq:Gtrans_qf>3/lam_gen}) in different regimes. We plot the expression in fig. (\ref{fig:wave_function_fixed_qf}). Although the expression for the wave function differs across the various regimes from that obtained in \cite{Ailiga:2024wdx}, its qualitative infrared behaviour remains the same. In particular, we observe a similar growth in the infrared limit to that reported in \cite{Ailiga:2024wdx}. This also establishes that the leading IR behaviour is independent of the details of the regularization method.


\end{document}